\documentclass[a4paper,11pt]{article}
\pdfoutput=1 % if your are submitting a pdflatex (i.e. if you have
\usepackage{lineno}
\usepackage{jheppub} % for details on the use of the package, please
\usepackage[T1]{fontenc} % if needed
\usepackage{xspace}
\usepackage{booktabs}
\usepackage{topcapt}

\usepackage[
backend=bibtex,
            backref=true,
            style=numeric-comp,
            maxcitenames=300,
            maxbibnames=500,
            block=none,
            sorting=none
            ]{biblatex}
\title{\boldmath Combination of Run-1 Exotic Searches in Diboson Final States at the LHC}
\preprint{CERN-TH-2016-004}

\author[a]{F. Dias,}
\author[b]{S. Gadatsch,}
\author[c]{M. Gouzevich,}
\author[a]{C. Leonidopoulos,}
\author[d]{S.F. Novaes,}
\author[e]{A.~Oliveira,}
\author[b]{M. Pierini,}
\author[d]{T. Tomei}

\affiliation[a]{University of Edinburgh, Edinburgh, UK}
\affiliation[b]{CERN, Geneva, Switzerland}
\affiliation[c]{Institut de Physique Nucleaire de Lyon, 
Universite de Lyon, Universite Claude Bernard Lyon 1, 
CNRS-IN2P3, Villeurbanne, France}
\affiliation[d]{Universidade Estadual Paulista, Sao Paulo, Brazil}
\affiliation[e]{Universita e INFN, Padova, Italy}

\emailAdd{Flavia.Dias@cern.ch}
\emailAdd{Stefan.Gadatsch@cern.ch}
\emailAdd{Mgouzevi@ipnl.in2p3.fr}
\emailAdd{Christos.Leonidopoulos@cern.ch}
\emailAdd{Sergio.Novaes@cern.ch}
\emailAdd{Alexandra.Oliveira@cern.ch}
\emailAdd{Maurizio.Pierini@cern.ch}
\emailAdd{Thiago.Tomei@cern.ch}

\newcommand {\ie}{\mbox{\sl i.e.}\,\,}     %i.e.
\newcommand {\eg}{\mbox{\sl e.g.}\,\,}     %e.g.
\newcommand{\PV}{\ensuremath{\mathrm{V}}\xspace}
\newcommand{\PZ}{\ensuremath{\mathrm{Z}}\xspace}
\newcommand{\PW}{\ensuremath{\mathrm{W}}\xspace}
\newcommand{\PWW}{\ensuremath{\mathrm{WW}}\xspace}
\newcommand{\PWZ}{\ensuremath{\mathrm{WZ}}\xspace}
\newcommand{\PZZ}{\ensuremath{\mathrm{ZZ}}\xspace}
\newcommand{\PWH}{\ensuremath{\mathrm{WH}}\xspace}
\newcommand{\PZH}{\ensuremath{\mathrm{ZH}}\xspace}
\newcommand{\PH}{\ensuremath{\mathrm{H}}\xspace}
\newcommand{\PVL}{\ensuremath{\PV_\mathrm{L}}\xspace}
\newcommand{\PVT}{\ensuremath{\PV_\mathrm{T}}\xspace}
\newcommand{\PZp}{\ensuremath{\mathrm{Z^\prime}}\xspace}
\newcommand{\PWp}{\ensuremath{\mathrm{W^\prime}}\xspace}
\newcommand{\PGbulk}{\ensuremath{\mathrm{G}_\mathrm{bulk}}\xspace}

\newcommand{\VVJJ}{\ensuremath{\mathrm{VV}\rightarrow\mathrm{JJ}}\xspace}
\newcommand{\VVLNJ}{\ensuremath{\mathrm{WV}\rightarrow\mathrm{\ell\nu J}}\xspace}
\newcommand{\VVLLJ}{\ensuremath{\mathrm{ZV}\rightarrow\mathrm{\ell\ell J}}\xspace}
\newcommand{\JJ}{\ensuremath{\mathrm{JJ}}\xspace}
\newcommand{\LNUJ}{\ensuremath{\mathrm{\ell\nu J}}\xspace}
\newcommand{\LLJ}{\ensuremath{\mathrm{\ell \ell J}}\xspace}
\newcommand{\GtoZZ}{\ensuremath{\PGbulk \rightarrow \PZ_L\PZ_L}\xspace}
\newcommand{\GtoWW}{\ensuremath{\PGbulk \rightarrow \PW_L\PW_L}\xspace}
\newcommand{\WptoWZ}{\ensuremath{\PWp \rightarrow \PW_L\,\PZ_L}\xspace}

\newcommand{\MJJ}{\ensuremath{m_\mathrm{JJ}}\xspace}
\newcommand{\PT}{\ensuremath{p_\mathrm{T}}\xspace}
\newcommand{\ZLZL}{\ensuremath{\PZ_\mathrm{L}\,\PZ_\mathrm{L}}\xspace}
\newcommand{\WLZL}{\ensuremath{\PW_\mathrm{L}\,\PZ_\mathrm{L}}\xspace}
\newcommand{\WLWL}{\ensuremath{\PW_\mathrm{L}\,\PW_\mathrm{L}}\xspace}
\newcommand{\MVV}{\ensuremath{m_\mathrm{VV}}\xspace}
\newcommand{\MX}{\ensuremath{m_\mathrm{X}}\xspace}

\abstract{
We perform a statistical combination of the ATLAS and CMS results for the search of a heavy
resonance decaying to a pair of vector bosons with the $\sqrt{s}=8$ TeV
datasets collected at the LHC. We take into account six searches in hadronic and
semileptonic final states carried out by the two collaborations. We
consider only public information provided by ATLAS and CMS in the HEPDATA
database and in papers published in refereed journals. We interpret the
combined results  within the context of a few
benchmark new physics models, such as models predicting the existence of a
{\ensuremath{\mathrm{W^\prime}}\xspace}  or a bulk Randall-Sundrum 
spin--2 resonance, for which we present exclusion limits, significances,
$p$-values and best-fit cross sections.
A heavy diboson resonance with a production cross
section  of $\sim$4-5 fb and mass between 1.9 and 2.0 TeV is the
exotic scenario most consistent with the experimental results. Models in which 
a heavy resonance decays preferentially to a WW final state are disfavoured.
}

\begin{document}
\maketitle
\flushbottom
\newpage
\section{Introduction}
Searches for new heavy resonances are one of the major components of the
ATLAS and CMS physics programmes at the Large Hadron Collider (LHC) at CERN. Of
particular interest is the coupling of new resonances to pairs of vector bosons.
%since these types of interactions are predicted in many extensions of the
%Standard Model (SM) and directly challenge the consistency of the Electroweak
%Symmetry-breaking mechanism with the discovered Higgs boson. 
Models with Vectorial heavy resonances (\ie \PWp-like and \PZp-like bosons) are
commonly considered as possible extensions
of the SM, either in weakly coupled (see \cite{Pati:1974yy,Mohapatra:1974gc,Senjanovic:1975rk})
%\cite{Barger:1980ix,Schmaltz:2010xr,Accomando:2010fz}) 
or strongly coupled versions, the so-called {\it composite Higgs}
scenarios~\cite{Dugan:1984hq,Georgi:1984af}. In these scenarios, the
existence of new resonances is introduced to alleviate the hierarchy
problem in the SM. Another common SM extension is the Warped
Extra Dimensions or Randall--Sundrum (RS) model ~\cite{Randall:1999vf},
which is an example of a class of models predicting neutral spin--2
resonances as Kaluza--Klein (KK) excitations of the graviton field 
($G^*$). Two types of models 
are usually considered: the original version, in which only gravity is
allowed to propagate into the extra-dimensional bulk (``RS1'' models, see
Ref.~\cite{Randall:1999ee}) and variants of the original model, in which the
SM fields are also allowed to propagate into the extra dimensional bulk (``bulk
RS'' models, see for example Ref.~\cite{Davoudiasl:2000wi}). RS1 models
favour the decay of $G^*$ to $q \bar{q}$, $\ell^+ \ell^-$ and $\gamma
\gamma$ final states, 
whereas in bulk RS models its decay to vector bosons. 

After a number of direct and indirect bounds from previous experiments, and
in particular, the stringent constraints from the electro-weak precision
measurements carried out at 
LEP~\cite{ALEPH:2010aa}\,\footnote{For recent analyses, including the
  LHC discovery of the Higgs boson, see for
  instance~\cite{Ciuchini:2013pca,Baak:2014ora}.}, nowadays searches for heavy
exotic resonances decaying to pairs of vector bosons typically focus on
resonance masses above 1 TeV. When produced and decayed at the LHC, these
particles would generate vector 
bosons with $\cal{O}$(1~TeV) transverse momenta, requiring special
reconstruction strategies. In particular, the quarks from a
hadronically-decaying 
vector boson are very close to each other in the $\eta-\phi$
space. In their showering and hadronisation process they produce highly
overlapping jets, in 
a so-called {\em 
  boosted topology}. ATLAS and CMS handle this experimental signature
 by reconstructing the two partially
overlapping jets as a single massive (or ``fat'') jet, noted in this paper
as ``J''. 
One then exploits the jet mass $m_J$ and the momentum flow around the jet axis to
distinguish these special jets from those originating from quark or gluon production~\cite{Butterworth:2008iy,
  Ellis:2009me, Stewart:2010tn, 
  Thaler:2010tr, Thaler:2011gf, Gouzevitch:2013qca}.  A typical
  boosted longitudinally polarised and hadronically-decaying \PV
  boson\,\footnote{In this 
    paper we refer to a vector boson (\PW or \PZ) decaying hadronically by
    the generic label \PV.} can be
identified by a tagger with an efficiency of $\sim$ 50\% and with a
false-positive rate for light quarks or gluons of $\lesssim 2\%$
\cite{CMSWTAG, ATLASWTAG}.

The ATLAS and CMS collaborations have employed hadronic boson taggers in
searches for heavy resonances in diboson final states with the
proton-proton collision data collected in 2012 at a centre-of-mass
energy of 8 TeV.  In particular, the ATLAS search in the
fully hadronic final state~\cite{ATLASVV} has generated significant
interest due to an excess of diboson events with invariant mass
mass around 1.9 TeV.
%\,\footnote{See for instance
%  Refs.~\cite{Dobrescu:2015qna,Thamm:2015csa,Brehmer:2015cia,Sanz:2015zha,}
 % for possible beyond-the-SM (BSM) interpretations of this excess.}.
Small deviations in the same mass region are observed in other channels as
well, \eg 
the CMS search in the $Z(\ell^+\ell^-)\,V(q\bar{q})$ channel with 
$\ell=e,\,\mu$~\cite{CMSZVWV}, and the CMS search in the fully hadronic
$V(q\bar{q})\,V(q\bar{q})$ final state \cite{CMSVV}. Other analyses, \eg
the ATLAS and CMS searches 
in the $W(\ell \nu)\,V(q \bar{q})$ channel see no evidence of a deviation,
indicating a possible tension between these experimental results in the 
scenario of a heavy exotic resonance. Additional
results with potentially interesting deviations in the same mass region
include a moderate 
excess ($\approx 1-2\sigma$ of local significance) reported in the
ATLAS~\cite{ATLASjj, ATLASjj13TeV} and  
CMS~\cite{CMSjj, CMSjj13TeV} searches in the dijet channel, 
as well as in the CMS search
in the dilepton channel~\cite{CMSll}. In addition, a search for
right-handed \PWp (and heavy neutrinos) \cite{WR} by CMS has
reported a small excess in the electron channel~\cite{CMSwr} (however, this
excess is not confirmed by a similar ATLAS analysis
\cite{Aad:2015xaa}). Finally, a 
CMS search for $\PW (\ell \nu) \,\PH (b \bar{b})$ resonances reported an
excess of $\approx 2\sigma$, originating from a stronger excess in the
electron channel and no evidence of a deviation in the muon
channel~\cite{CMS:2016yji}. At the same time, the CMS searches for \PWH
or \PZH resonances in the fully 
hadronic channel were inconclusive,  with a mild upward fluctuation around
1.8 TeV and a lack of events around 2~TeV~\cite{CMSVH}. The dedicated
searches for $\PZ(q\bar{q}) \,\PH(\tau^+ \tau^-)$ and $\PH(b \bar{b}) \,\PH(\tau^+ \tau^-)$, 
$\PH(b \bar{b}) \,\PH(b \bar{b})$ final states showed no excess 
\cite{Khachatryan:2015ywa, CMS:2015zug}. 

Several attempts to provide a possible interpretation for this excess have
been made during the 
last months. The deviation has been associated to possible signatures of
various beyond-the-SM models, \eg models with new $W^\prime$ and $Z^\prime$
vector bosons (see for example \cite{Hisano:2015gna, Cheung:2015nha,
  Dobrescu:2015qna, Gao:2015irw, Brehmer:2015cia, Abe:2015jra, Dev:2015pga,
  Coloma:2015una,Cao:2015lia,Das:2015ysz}), models involving new resonances
with different spins (see for example
\cite{Aguilar-Saavedra:2015rna,Cacciapaglia:2015eea, Allanach:2015hba,
  Abe:2015uaa, Fukano:2015uga, Liew:2015osa,
  Collins:2015wua,Sierra:2015zma}), composite and technicolor models (see
for example \cite{Fukano:2015hga, Franzosi:2015zra, Thamm:2015csa,
  Chiang:2015lqa, Low:2015uha,Fukano:2015zua}) and new and composite Higgs
states (see for example \cite{Carmona:2015xaa, Dobrescu:2015yba,
  Sanz:2015zha, Chen:2015xql, Omura:2015nwa, Chen:2015cfa, Kim:2015vba,
  Deppisch:2015cua,Alves:2015vob}). A review of the different models
offering an interpretation of the deviations reported in 
the ATLAS and CMS searches has been made in
Ref. \cite{Brehmer:2015dan}. 

A natural next step would be to carry out a systematic comparison of the
results reported by ATLAS and CMS in various channels, and examine if the
apparent deviations work in a synergistic way towards a coherent picture.
In particular, the goal is to quantify the level of agreement among the different 
results, and by using an exotic signal 
hypothesis for the interpretation of these deviations, to calculate the
corresponding production cross section. We hereby present the first step in
addressing this question, starting with the statistical combination of the
results of the ATLAS and CMS Run-1 searches for vector boson pair
resonances. The exotic models considered by the
experiments are usually connected with the electroweak sector, with the
predicted resonances mainly coupling to longitudinally polarised vector bosons
$\PVL$. We consider the experimental results of the searches for heavy
resonances decaying to 
three final states: $\ZLZL$, $\WLWL$ and $\WLZL$. We combine
the results and interpret the derived exclusion limits in the context of a
(\PWp-like)  
spin--1 charged particle decaying to a $\WLZL$ boson pair, and a neutral spin--2
particle (\PGbulk).  For the latter case, we only consider bulk RS scenarios, namely
particles decaying to the  
$\ZLZL$, $\WLWL$ final states\,\footnote{Models in which the exotic
  resonances have stronger couplings to transverse vector bosons ($\PVT$)
  than longitudinal ones ($\PVL$) typically have larger branching fractions
  to dilepton and dijet final states. It should be noted that boosted  boson
  taggers are more efficient with $\PVL$ than $\PVT$ bosons
  \cite{CMSWTAG}. This topic will be addressed in a future publication.}.

The paper is organised as follows: in Section~\ref{sec:method} we present
a general overview of the methodology used to emulate the ATLAS
and CMS analyses; Sections~\ref{sec:JJ} and~\ref{sec:Semilep} discuss
the emulation of the hadronic and semileptonic analyses,
respectively. Each section covers the individual searches by ATLAS
and CMS, and their combination; in Section~\ref{sec:combo} we combine the Run-1
results provided by the two collaborations and discuss their interpretation
in a few benchmark models considered in this study; we present the summary
of the findings, 
along with the conclusions in Section~\ref{sec:conclusion}.  
A brief note on the compatibility of the findings of this study with the preliminary Run-2 search results reported by ATLAS and CMS in December 2015 has been added in v2 of this paper and is presented after the conclusions. Additional information
on the determination of the background 
and signal modelling for the ATLAS search in the fully hadronic channel is given 
in Appendices~\ref{app-Avvjj}, \ref{sec:narrow}.

%\clearpage
\section{General methodology}
\label{sec:method}
All exotic searches considered in this paper are looking for a diboson
mass peak emerging on top of a falling background spectrum. In order to
evaluate the significance of a deviation observed in the data,
we need as input the shapes of the signal and background distributions, the
total 
number of expected background events, the signal efficiency, and the
experimentally measured distribution (data).

This study is based exclusively on the public information provided by
the two experimental collaborations in the HEPDATA database~\cite{HEPDATA}
and the cited papers (published in refereed journals). In
particular, we employ  the expected backgrounds with
their corresponding uncertainties, as they have been estimated
directly by ATLAS and CMS, wherever possible. The
modelled signal distributions  (namely, shapes and
signal efficiencies for a few benchmark models and mass values) are also
taken from the information publicly provided by 
the experiments, when available\,\footnote{The ATLAS
and CMS collaborations usually provide the histograms for a signal
benchmark model at a fixed mass value. Often, these histograms are not
provided in electronic format. In these cases, we had to extract the
information from the publicly available plots.}. In order to emulate 
signal distributions for additional mass values, we
carry out linear interpolations of the available models within the
benchmark mass points. We derive  
exclusion limits on hypothetical signals by performing
binned templated fits of the data distributions with linear combinations of the
signal and background distributions. These calculations are carried out with the
open-source statistical  framework \texttt{THETA} \cite{theta} which uses
the asymptotic  approximation~\cite{CLSasymp} of the CLs method
\cite{CLS1,CLS2}.

In a few cases, the information published by ATLAS and CMS is not
sufficient for this simple approach to produce satisfactory results. For
example, uncertainty correlations that affect the background
determination, or the mass-dependence of an important systematic
uncertainty are not always properly documented. In these cases, we fit
the data distributions to the functional form documented in the published
analysis, \eg the
function used in the hadronic searches or an exponential function for
the leptonic channels. Details about these fits are given in the
corresponding sections of the paper, where we also discuss the 
agreement achieved in the background modelling. When it is necessary to model
a signal distribution ourselves, we either use a Gaussian
approximation with a resolution inferred from the relevant
experimental paper, or we generate Monte Carlo (MC) samples using the
\texttt{Madgraph5} matrix-element event
generator~\cite{Alwall:2014hca}, matched to
\texttt{Pythia8}~\cite{Sjostrand:2007gs} for the hadronisation process.
For the \PGbulk signal we use the \texttt{Madgraph5} model files as
presented in Ref. \cite{Oliveira:2014kla}, while for the spin--1 signal
\PWp the ones described in Ref. \cite{Pappadopulo:2014qza}.

These approximations are mainly motivated by our
familiarity with the diboson and similar searches by ATLAS and CMS.
The described procedure is validated using the nominal published results as
benchmarks, as well as the comparison of our own calculations of the per-experiment
combinations against the official combination of diboson
searches~\cite{CMSZVWV,ATLAScombo}. We are able to reproduce the exclusion limits of each
analysis individually and their combinations with an agreement of better
than 20\% in the region of interest for all channels, with the exception of
the fully hadronic search in ATLAS (see Appendix~\ref{app-Avvjj}). Our
methodology can be used as a set 
of guidelines for model builders in the absence of official combined
results published by the two experiments. 

All diboson final states considered in this study contain at least one vector boson 
($\PW$ or $\PZ$) decaying hadronically. Because of the limited 
hadronic detector resolution, it is not possible to distinguish between
hadronic $\PW$ and hadronic $\PZ$ jets. When interpreting an
experimental result, special care is needed to account for possible
cross-channel contamination of the final state under
consideration. For example, a neutral heavy resonance decaying to a pair of
vector bosons is 
expected to decay to both $\PW\PW$ and $\PZ\PZ$ final states.  
We consider models in which the relative branching fractions of
neutral particle decays to $\PW\PW$ and $\PZ\PZ$ can vary, in order to
study the relative importance of the different bosonic sub-channels to the
combined result. We quantify this dependence by introducing as a free
parameter the ratio $r$ of
the corresponding branching fractions:
\begin{equation}
r \equiv \frac{\mathcal{B}(X \to \PW\PW)}{\mathcal{B} (X \to \PZ\PZ)}
\label{eq:r}
\end{equation}
with $r=2$ being the default ratio in the baseline bulk RS scenario.

The full list of channels that we consider in this study is as
follows: the fully hadronic searches $X \to V(q\bar{q})\, V(q\bar{q})$
(labelled ``$\JJ$''), searches including a \PW decaying leptonically $X \to \PW(\ell
\nu) \,V(q\bar{q})$ (labelled ``$\LNUJ$''), and searches including a \PZ decaying
leptonically $X \to \PZ(\ell \ell) \,V(q\bar{q})$ (labelled ``$\LLJ$'').
Table~\ref{table:MethodologySummary} summarises the methods
that have been used to emulate each of the analyses considered. Details
of the individual analyses are given in the sections that follow.

%%%%%%%%%%%%%% BEG Vic (run-dependant samples)
\begin{table}[hbt]
\centering
\topcaption{\small Summary of the methods used and the corresponding
  uncertainties for the signal and background modelling per channel and
  experiment. 
\label{table:MethodologySummary}}
%\resizebox{\textwidth}{!}
{\footnotesize
%\begin{tabular}{|c|c|c|c|c|c|c|}
%\hline 
%  {} & {} & Background & Background     & Signal & Signal & Fudge\\ 
% \raisebox{1.5ex}[0cm][0cm]{Experiment} &
% \raisebox{1.5ex}[0cm][0cm]{Channel} &  modelling  &  uncertainties &
% modelling & efficiency & factor \\ \hline 
%  & $\JJ$ \cite{ATLASVV}& Fit & Fit &  Paper \&
% extrap. & Public plots & Yes \\ 
% ATLAS      & $\LNUJ$ \cite{ATLASWV} & HEPDATA & HEPDATA & Gauss. approx. & Public
% plots & Yes \\
%       & $\LLJ$ \cite{ATLASZV} & Fit & HEPDATA & Gauss. approx. &
% Public plots & Yes \\ \hline 
%  & $\JJ$  \cite{CMSVV} & HEPDATA & HEPDATA & HEPDATA &  HEPDATA & No \\
%CMS  & $\LNUJ$ \cite{CMSZVWV} & Fit & Fit & MC &
%Public plots \& MC & Yes \\ 
% & $\LLJ$ \cite{CMSZVWV} & Fit & Fit & MC & Public
%plots \& MC & Yes \\\hline  
%\end{tabular}
\begin{tabular}{crccccc}
\toprule 
  {} & {} & Background & Background     & Signal & Signal & Fudge\\ 
 \raisebox{1.5ex}[0cm][0cm]{Experiment} &
 \raisebox{1.5ex}[0cm][0cm]{Channel} &  modelling  &  uncertainties &
 modelling & efficiency & factor \\ \midrule 
  & $\JJ$ \cite{ATLASVV}& Fit & Fit &  Paper \&
 extrap. & Public plots & Yes \\ 
 ATLAS      & $\LNUJ$ \cite{ATLASWV} & HEPDATA & HEPDATA & Gauss. approx. & Public
 plots & Yes \\
       & $\LLJ$ \cite{ATLASZV} & Fit & HEPDATA & Gauss. approx. &
 Public plots & Yes \\ \midrule 
  & $\JJ$  \cite{CMSVV} & HEPDATA & HEPDATA & HEPDATA &  HEPDATA & No \\
CMS  & $\LNUJ$ \cite{CMSZVWV} & Fit & Fit & MC &
Public plots \& MC & Yes \\ 
 & $\LLJ$ \cite{CMSZVWV} & Fit & Fit & MC & Public
plots \& MC & Yes \\\bottomrule  
\end{tabular}
}
\end{table}
%%%%%%%%%%%%%%%%%%%%%

%The full list of analyses included in our
%combination is summarised in Table~\ref{table:AnalisisSummary}.

%%%%%%%%%%%%%% BEG Vic (run-dependant samples)
%\begin{table}[h]
%\centering
%%\resizebox{\textwidth}{!}
%\footnotesize{
%\begin{tabular}{|c|c|c|c|}
%\hline 
% & $\JJ$ & $\LNUJ$ & $\LLJ$ \\\hline
%ATLAS & \cite{ATLASVV} & \cite{ATLASWV} & \cite{ATLASZV} \\
%CMS  & \cite{CMSVV} & \cite{CMSZVWV} & \cite{CMSZVWV} \\\hline 
%\end{tabular}
%}
%\caption{\footnotesize List of the analyses used in this papers with the corresponding references. 
%\label{table:AnalisisSummary}}
%\end{table}
%%%%%%%%%%%%%%%%%%%%%

%Finally a short study on the relative impact of each experiment is provided in the Appendix \ref{sec:PerExperiment}.

%Also, in theoretical
%models with spin--1 particles, it is not uncommon to have contributions from
%both neutral (e.g. $\PW\PW$) and charged ($\PW\PZ$) resonances (degenerated in
%mass) to the same final state containing hadronic bosons. These
%complications are discussed in more detail in the corresponding sections.

\section{Fully hadronic searches: \VVJJ}
\label{sec:JJ}
In this Section we discuss the analysis of the ATLAS and CMS searches in
the \VVJJ channel. We first present the results of our analysis for the two
searches separately, followed by their combination and a summary of the
findings. 
\subsection{Emulation of ATLAS search}
%We start with a summary of the selection for the ATLAS analysis. We then
%present an overview of the statistical analysis carried out in this channel
%with the major uncertainties. Finally, we discuss the interpretation of
%results for three signal hypotheses and tagging selections: \PWW, \PWZ and
%\PZZ.  

\subsubsection{Description of the ATLAS analysis}

\par The ATLAS fully hadronic search analyses calorimetric dijet
events. The main irreducible background
%in this analysis 
is dijet production in QCD, which is dominated by $2 \to
2$ $t$-channel processes involving quarks and gluons. The contribution of
these processes is minimised by restricting the jet acceptance to $|\eta| <
2.0$ and the rapidity difference between those two jets to $|\Delta \eta| <
1.2$. The events are required to have low missing transverse momentum and a
rather symmetric dijet topology (similar \PT{} for the two leading jets) to
reduce the detector noise. After this selection, the efficiency is
approximately 70-80\% for a heavy
vector boson signal, and above 80\% for a
\PGbulk signal.  

\par To further reduce the multijet background, two fat jets are
reconstructed using the Cambridge-Aachen algorithm
\cite{Dokshitzer:1997in,Wobisch:1998wt} with radius parameter $R = 1.2$. The
mass-drop filtering algorithm~\cite{Butterworth:2008iy} is applied to each
of these jets for the identification of the sub-jets and grooming.
Events are kept if each of the two leading jets satisfies the following
conditions: have two sub-jets with similar transverse momentum, have
less than 30 tracks matched to it, and have a pruned mass within a
$\pm$13 GeV window either around 82.4 GeV (for \PW{} tagging) or around
92.8 GeV (for \PZ{} 
tagging). The selection efficiency of the grooming algorithm for fat jets
from a \PWp{} resonance  is between 30\% and 40\%.

The events are
subsequently classified into three 
non-mutually-exclusive categories, based on the jet-mass values:
\PW{}\PW{}, \PW{}\PZ{} and \PZ{}\PZ{}. The overall product of the geometric
acceptance with the signal
efficiency for this analysis is typically 10-20\%.

\subsubsection{Statistical analysis}

\par
\begin{figure*}[htb]\begin{center}
\includegraphics[width=0.32\textwidth, angle =0 ]{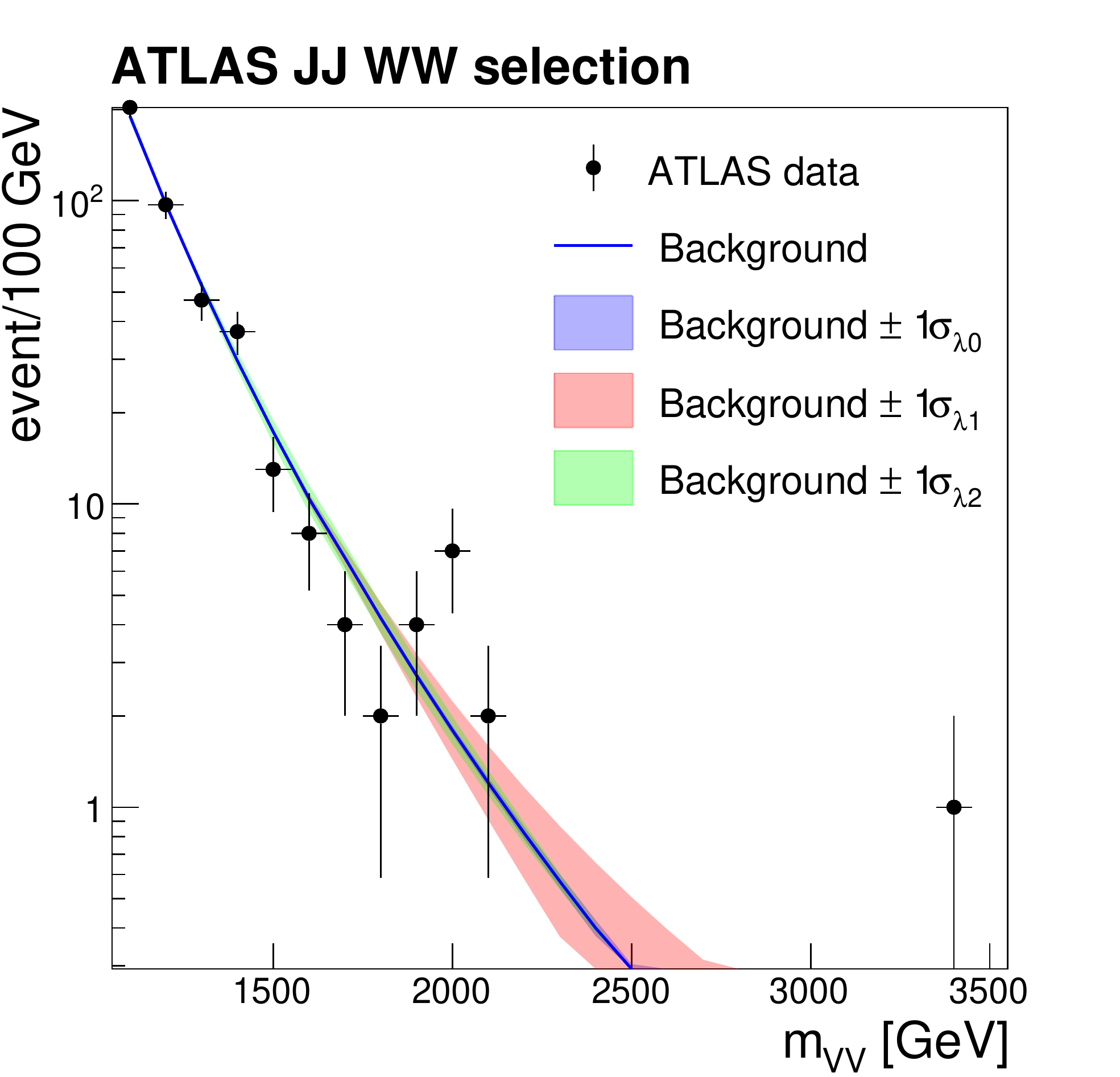}
\includegraphics[width=0.32\textwidth, angle =0 ]{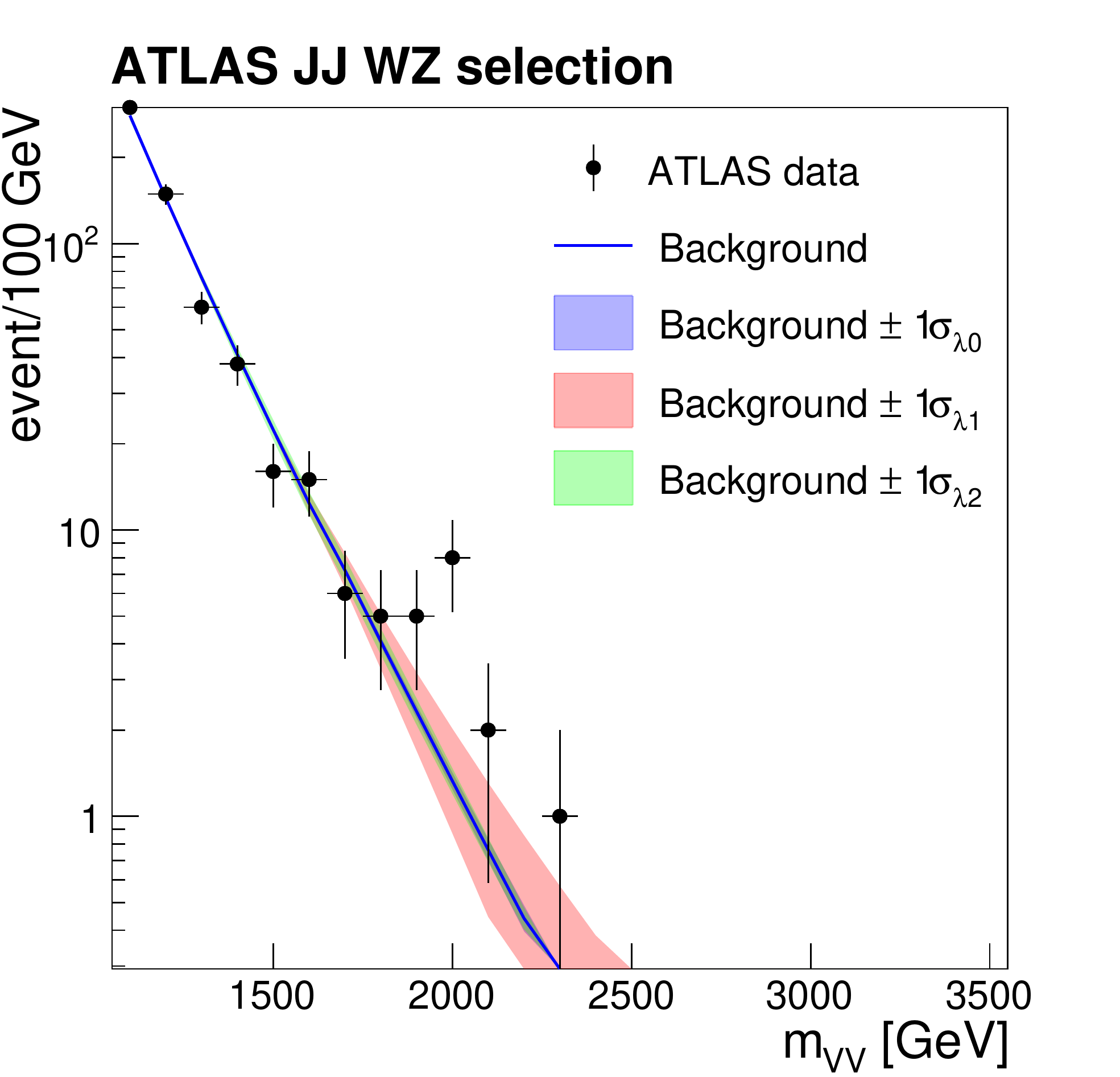}
\includegraphics[width=0.32\textwidth, angle =0 ]{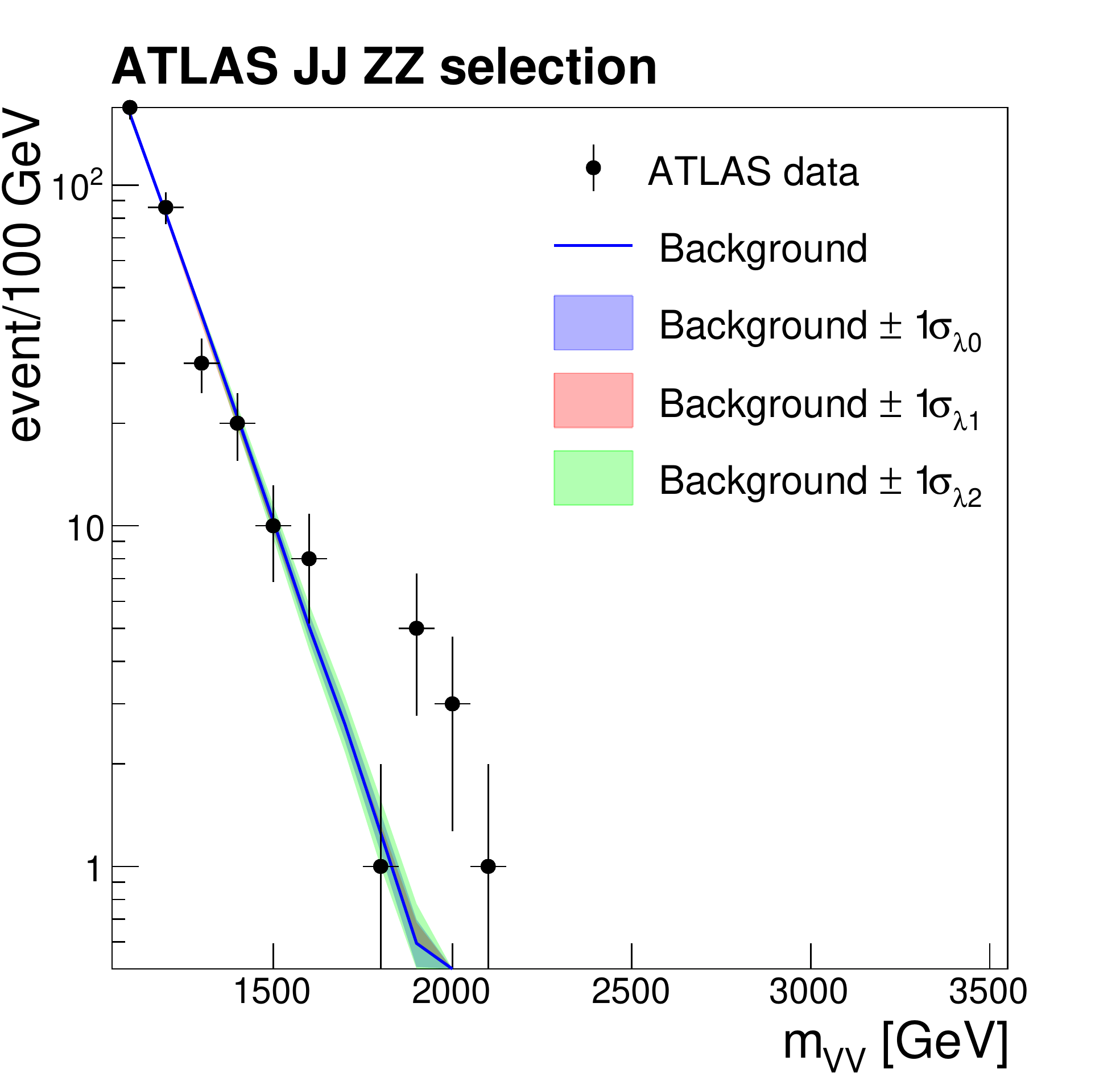}
\caption{\small ATLAS hadronic search: Comparison between the
  official ATLAS fit (blue line) and the fit of this study with uncertainties as
  described in the text (coloured bands), with the
  overlaid  data of the \MJJ spectrum for the \PW{}\PW{} (left), \PW{}\PZ{} (middle) and
  \PZ{}\PZ{} (right) tagging selections.}
\label{fig:ATLASbkg}
%}
\end{center}\end{figure*}

 The analysis uses the smoothness test (``bump search'')
approach: the background is approximated by a steeply falling function,
while the signal template is taken from simulation. The sum of the two components
is then fitted to the data. The background function used by the ATLAS collaboration is:
\begin{equation}
f(\MJJ) = p_0 (1-\MVV)^{p_1 - \xi p_2}\MVV^{p_2}
\end{equation}
where $p_0$, $p_1$ and $p_2$ are free parameters and
\MJJ{} is the dijet invariant mass; ATLAS has also made
the signal templates used in the analysis public.
We employ the same function for the background description, but recalculate the
background uncertainties in order to better account for the large scale correlations in $\MJJ$. 
%%% Is this really supposed to be a new paragraph? I make it not be for the time being.
To this end, we refit the data in each of the three categories above using
the aforementioned background parametrisation. We diagonalise
the uncertainty matrix and obtain three uncertainty eigenvectors
($\sigma_{\lambda_i}$,  with $i = 0,1,2$). 
Our fit result produces a background estimate which agrees with the nominal
background within 10\%, which is well within the uncertainties (see Appendix \ref{app-Avvjj}). 
This background is
subsequently used together with the associated uncertainties in our statistical
analysis (see Fig. \ref{fig:ATLASbkg}). 

We consider the following systematic uncertainties, treated as fully
correlated across \MJJ histogram bins: 
\begin{itemize}
  \item \emph{Background uncertainty}, obtained as described above.
  \item \emph{Signal normalisation uncertainty}, which is separated into
    two further sub-categories: a common-across-channels systematic
    uncertainty corresponding to the luminosity measurement (2.8\%), and an
    additional term applicable to the $\JJ$ channel that covers \PV-tagging
    uncertainties as well as jet systematics. 
  \item \emph{Signal jet energy scale uncertainty}, which includes jet
    transverse momentum and mass uncertainties (with a $\pm 2\%$ and $\pm
    5\%$ impact on $\MJJ$, respectively).
    %The impact of the uncertainties
  %on the likelihood is emulated by shifting the signal shape distribution by $\MJJ$
%  uncertainty. 
  An additional jet energy resolution uncertainty is known to have a
  negligible effect on the signal shape and is ignored in this study.
\end{itemize} 

\par
\begin{figure*}[htb]\begin{center}
%\includegraphics[width=0.39\textwidth, angle =0
%]{../limits_comparison/ATLAS_VV_JJ/VV_ratio.pdf}
\includegraphics[width=0.39\textwidth, angle =0 ]{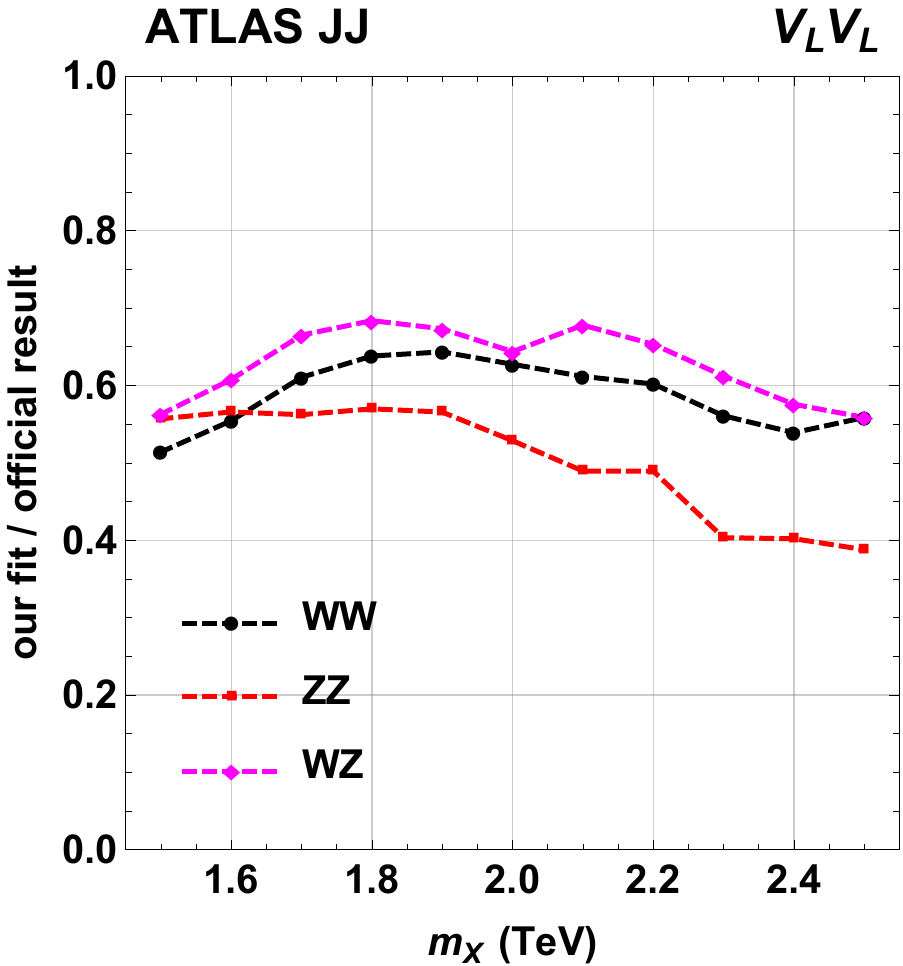}
\caption{\small ATLAS hadronic search: Ratio of observed
  exclusion limits obtained with this study to the ones of the
  official ATLAS result, as a function of the mass $m_X$ of the exotic
  resonance 
  %l to $W' \to WZ$, and bulk-like neutral spin--2 particle decaying exclusively to $WW$ or $ZZ$ pair of bosons in their respective window selections. 
for the \PW{}\PW{} (black), \PZ{}\PZ{} (red)
  and \PW{}\PZ{} (magenta) tagging selections.  
%to  apply the fudge factors. By construction the ration between the expected limits resultant of our fit and nominal are equal to one after apply the fudge factors. 
\label{fig:ATLASfudge}}
\end{center}\end{figure*}

\begin{figure*}[htb]\begin{center}
%\includegraphics[width=0.3\textwidth, angle =0 ]{../limits_comparison/ATLAS_VV_JJ/WW_less.pdf}
%\includegraphics[width=0.3\textwidth, angle =0 ]{../limits_comparison/ATLAS_VV_JJ/WZ_less.pdf}
%\includegraphics[width=0.3\textwidth, angle =0
%]{../limits_comparison/ATLAS_VV_JJ/ZZ_less.pdf}
\includegraphics[width=0.3\textwidth, angle =0 ]{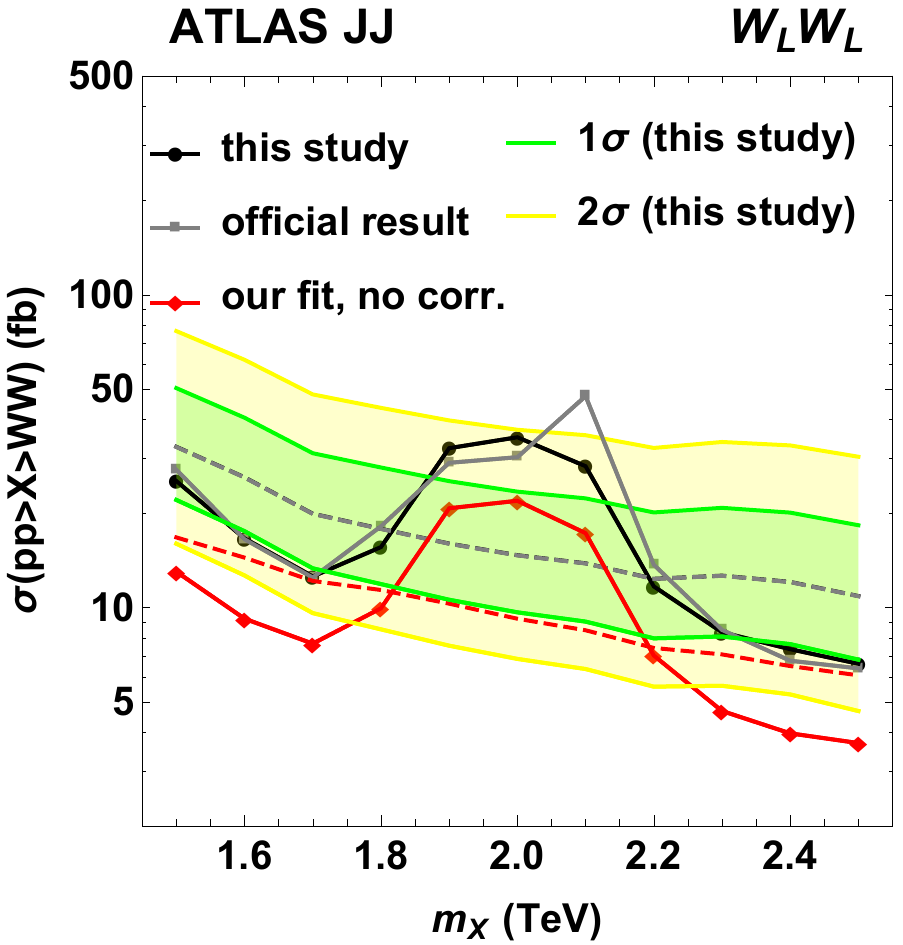}
\includegraphics[width=0.3\textwidth, angle =0 ]{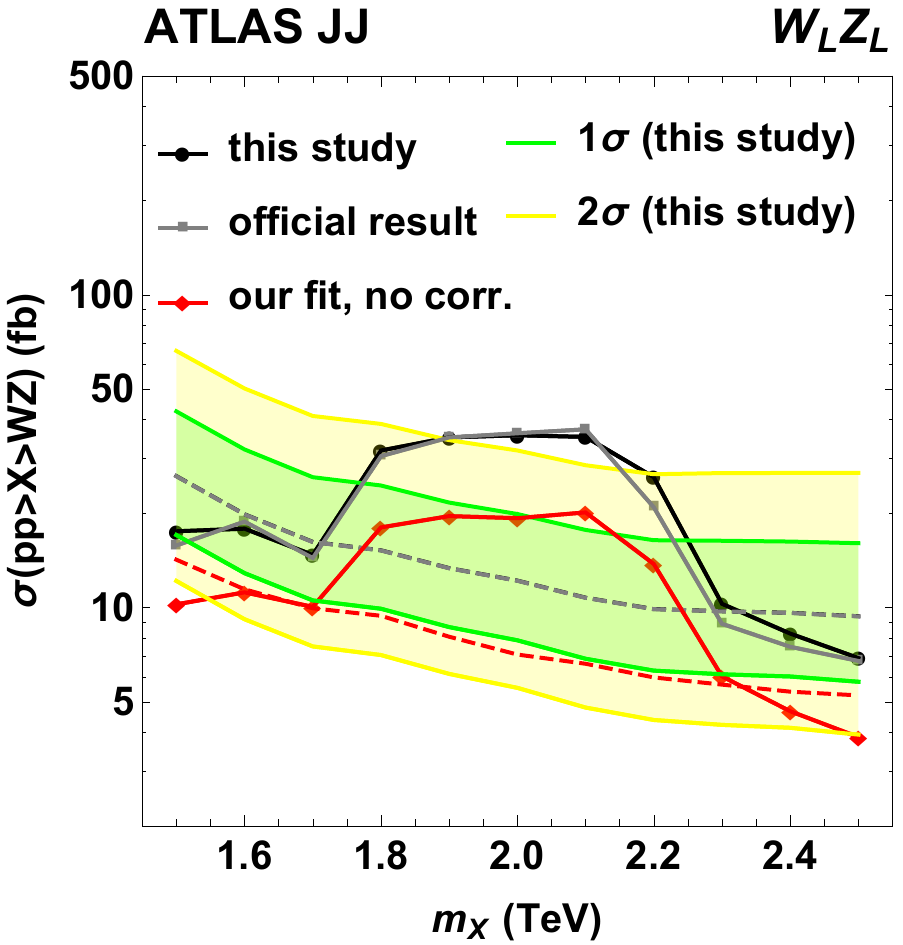}
\includegraphics[width=0.3\textwidth, angle =0 ]{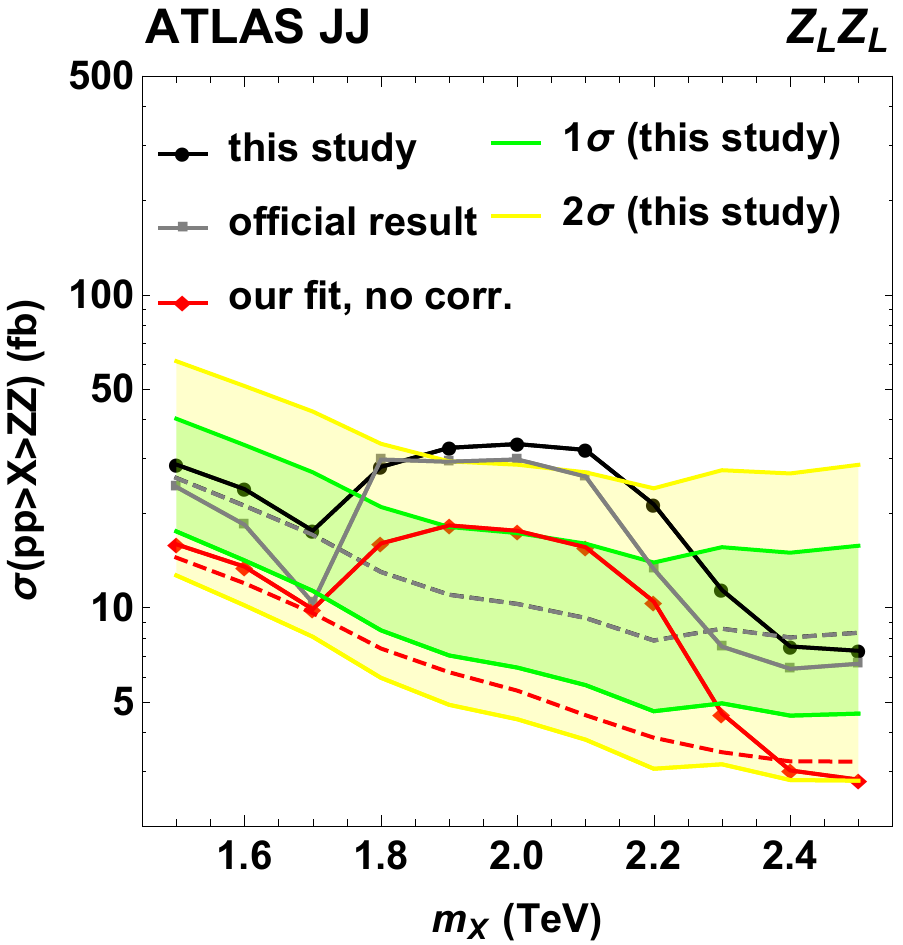}

\caption{\small ATLAS hadronic search: Observed exclusion
  limits on exotic production cross section as a function of the resonance
  mass $m_X$ obtained with this study, with (black) and without (red) the
  correction discussed in the text (``fudge''), and comparison with the
  official ATLAS results (grey)
  for \GtoWW (left), \WptoWZ (middle) and \GtoZZ (right) signal hypotheses
  and tagging selections. The green and yellow bands represent the one and
  two sigma variations around the median expected limits (dashed lines)
  calculated with the same fudge factor.
  \label{fig:ATLASlimit} }
\end{center}\end{figure*}

 Our statistical analysis produces expected exclusion limits that are
typically 
50\% more stringent than the ones publicly provided by ATLAS. This
discrepancy, discussed in detail in Appendix \ref{app-Avvjj}, is corrected
for with the introduction of a {\it fudge} factor, defined as the ratio of
the ATLAS expected exclusion
 limits and the ones from this study obtained with the {\sc THETA}
statistical framework (see Fig.
\ref{fig:ATLASfudge}). With this correction, our calculated exclusion limits
are in good 
agreement with the public ATLAS results (see Fig. \ref{fig:ATLASlimit}).

%%%%%%%%%%%%%% BEG Vic (run-dependant samples)
%\begin{table}[htb]
%\centering
%\resizebox{\textwidth}{!}
%\footnotesize{
%\begin{tabular}{|c|c|c|c|}
%\hline 
%mass (GeV)  &   OW signal l &  OZ signal &  OZ sign\\ \hline
%1500  & 0.534724 &  0.531894 & 0.530061 \\
%1600  & 0.580952 & 0.555135 & 0.569708   \\ 
%1700  & 0.63504 & 0.566784 & 0.61412 \\
%1800  & 0.654515 & 0.601672 & 0.615966 \\ 
%1900  & 0.650031 & 0.630537 & 0.595161  \\
%2000  & 0.626703 & 0.618207 & 0.567389 \\ 
%2100  & 0.607802 & 0.596984 & 0.598222  \\ 
%2200  & 0.598088 & 0.616378 & 0.587551 \\ 
%2300  & 0.559351 & 0.519181 & 0.564537\\ 
%2400  & 0.541227 & 0.520399 &  0.542425 \\ 
%2500 & 0.528914 & 0.495976 & 0.50572 \\\hline
%\end{tabular}
%}
%\caption{\footnotesize Fudge factors used in ATLAS fully hadronic channels. {\bf take care of significative numbers - update to last numbers.}
%The first sample (two prompt photons by SM QCD) is generated with {\tt Sherpa}, all the others are generated with {\tt Pythia6}.
%\label{table:ATLASfudge}}
%\end{table}
%%%%%%%%%%%%%%%%%%%%%
%

%\subsubsection{Signal composition}
\subsubsection{Results with \PWW, \PWZ and \PZZ signal hypotheses}

As discussed above, due to the finite detector resolution, the
$\PV$-tagging tool is not capable to differentiate between
fat jets originating from \PW or \PZ bosons. However, there is a
significant performance difference between \PW and \PZ tagging efficiencies
of up to $\approx 30\%$, mainly as a result of the different boson
masses. By using the mass distribution of longitudinal $\PV$-jets, as
documented in Fig. 1 of Ref.~\cite{ATLASVV}, and by taking into account the
different  \PW and \PZ efficiencies, we can calculate the efficiency of tagging
selections for different signal hypotheses (\PWW, \PWZ, \PZZ).  The
comparison of the tagging selection efficiencies can be found 
in Table~\ref{table:windows}.  
% 
%%%%%%%%%%%%%% BKG cic (run-dependant samples)
\begin{table}[htb]
\centering
%\resizebox{\textwidth}{!}
\topcaption{\small Relative efficiencies for \PWW, \PWZ, \PZZ signal
  hypotheses for tagging selection using different mass windows.
%The first sample (two prompt photons by SM QCD) is generated with {\tt Sherpa}, all the others are generated with {\tt Pythia6}.
\label{table:windows}}
\footnotesize{
\begin{tabular}{cccc}
\toprule 
&  \multicolumn{3}{c}{Signal hypothesis} \\ \cmidrule(lr){2-4}
 Tagging selection        &   \PWW &  \PWZ &  \PZZ \\\midrule
\PWW window   & 1.00      &   0.65 &    0.42 \\
\PWZ window   & 0.84   &   1.00       & 0.65 \\
\PZZ window   & 0.70  &    0.84   &  1.00 \\\bottomrule
\end{tabular}
}
\end{table}
%%%%%%%%%%%%%%%%%%%%%

\begin{figure*}[htb]\begin{center}
\includegraphics[width=0.3\textwidth, angle =0 ]{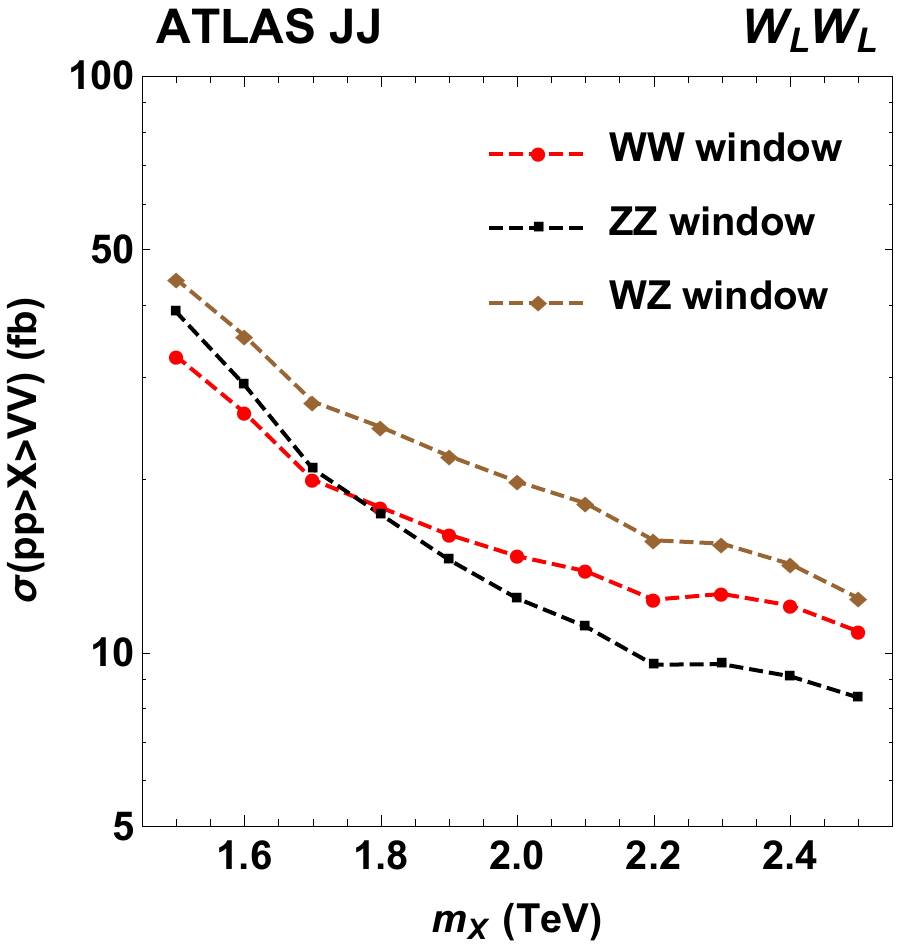}
\includegraphics[width=0.3\textwidth, angle =0 ]{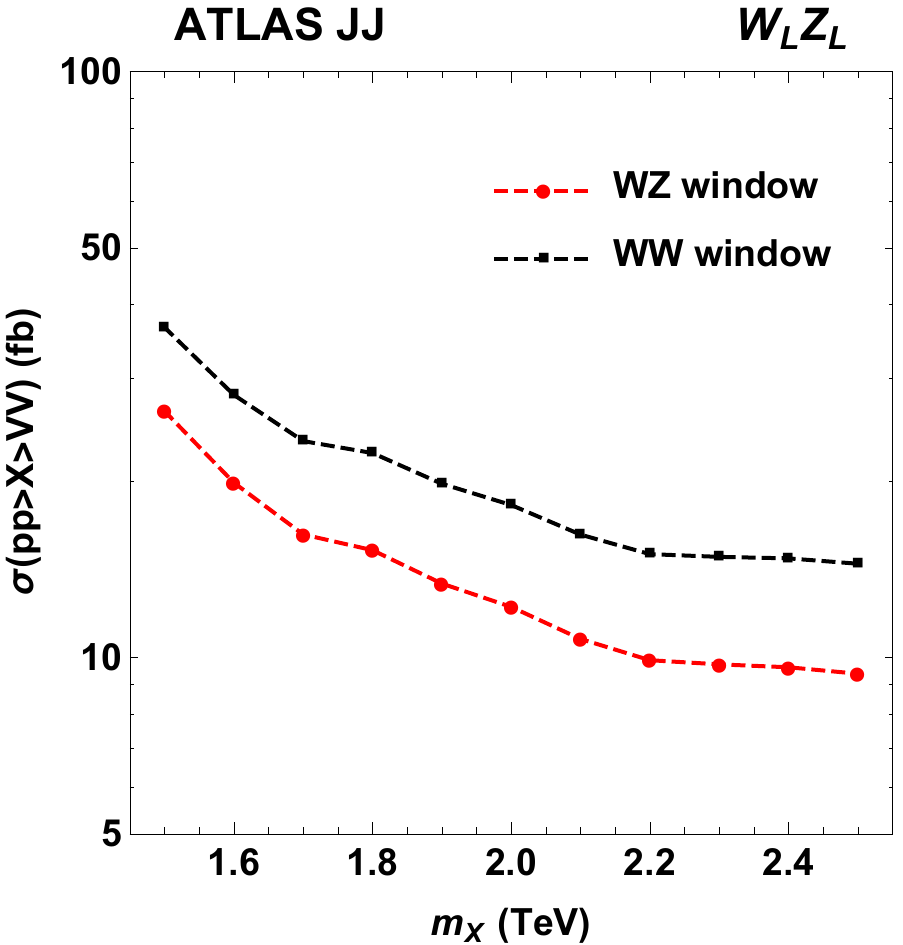}
\includegraphics[width=0.3\textwidth, angle =0 ]{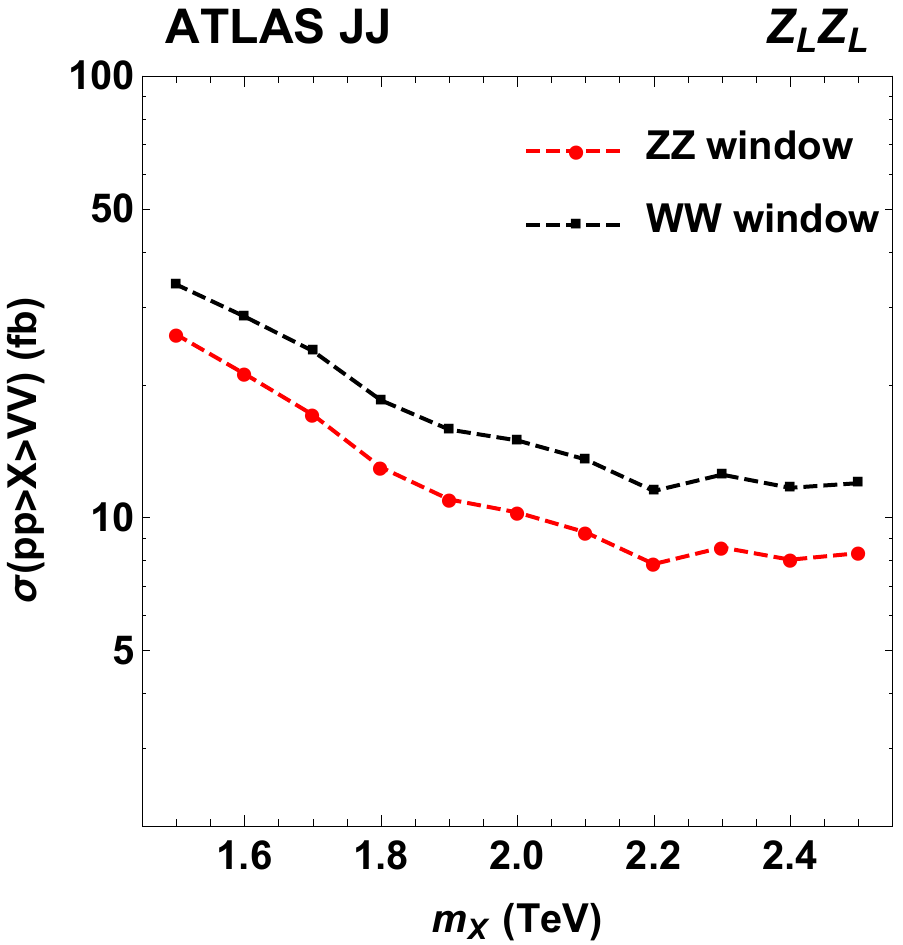}
\caption{\small ATLAS hadronic search: 
Expected exclusion limits for different tagging and mass-window
selections, as a function of the mass $m_X$ of the exotic resonance for
\GtoWW (left), \WptoWZ (middle) and \GtoZZ (right) signal hypotheses. The
results have been obtained with the correction discussed in the
text. \label{fig:windows}} 
\end{center}\end{figure*}

The effect of applying the different tagging selections to the \PWW, \PWZ and
\PZZ signal hypotheses as a function of the resonance mass is shown in
Fig.~\ref{fig:windows}. We assume that 
the $\MJJ$ spectrum is not affected  by the mass window difference in
the tagging selections, \ie that the same distribution describes the three
tagging categories \PWW, \PWZ and \PZZ. Since the three categories have
common events, they cannot be combined as if they were statistically
independent. Instead, for each theoretical model under consideration we
choose the tagging category that gives the best expected exclusion
limits. For the \PWp model the \PWZ tagging selection gives
the best result, whereas for the \PGbulk graviton model in the $\WLWL$ and
$\ZLZL$ final states the \PZZ tagging
selection has the best performance.

\subsection{Emulation of CMS search}
%The strategy followed by the CMS fully hadronic search is very similar to
%the corresponding ATLAS analysis presented above. 

\subsubsection{Description of the CMS analysis}
The  jet acceptance is restricted to $|\eta| < 2.5$ and $|\Delta \eta| <
1.3$ in order to reduce the contamination from multijet events. The detector
noise is removed by requiring tight quality criteria on the jets. 

The pruning algorithm~\cite{Ellis:2009me} is used to clean up the jet from
soft and large-angle radiation. The mass of the resulting fat jet is
constrained in the$70 < m_J < 100$ GeV range. Finally, the
signal-to-background ratio is enhanced by exploiting the jet
\textit{N-subjettiness}~\cite{Thaler:2010tr, Thaler:2011gf, Stewart:2010tn}
variable $\tau_N$. This variable is used to quantify how well the jet
constituents can be arranged into N sub-jets, \ie in a consistency check
with the hadronic \PV boson hypothesis.
The ratio $\tau_{12} = \tau_2/\tau_1$  is built with the two leading jets:
the smaller the ratio, the  larger the probability that the jet consists of
two sub-jets. The analysis considers two categories: the
high purity (HP) one, defined by requiring $\tau_{12} < 0.5$ for both
jets, and the low purity (LP) one, defined by requiring  
one jet with $\tau_{12} < 0.5$ and the other one with $0.5 < \tau_{12} <
0.75$. The HP category is characterised by a smaller background
contamination. The LP category 
captures signal events with asymmetric decays of the  vector-boson
candidates in the laboratory frame. Dividing the event sample into the LP
and HP categories improves the sensitivity of the analysis in the mass
range between 1~TeV and 2~TeV, while avoiding the inefficiency of a tight
$\tau_{12}$  selection at large jet momenta.

The product of the geometrical acceptance with the signal
efficiency is similar to the one in the ATLAS search, ranging between 10\%
and 20\%.  

\subsubsection{Statistical analysis}

\begin{figure*}[htb]\begin{center}
\includegraphics[width=0.40\textwidth, angle =0 ]{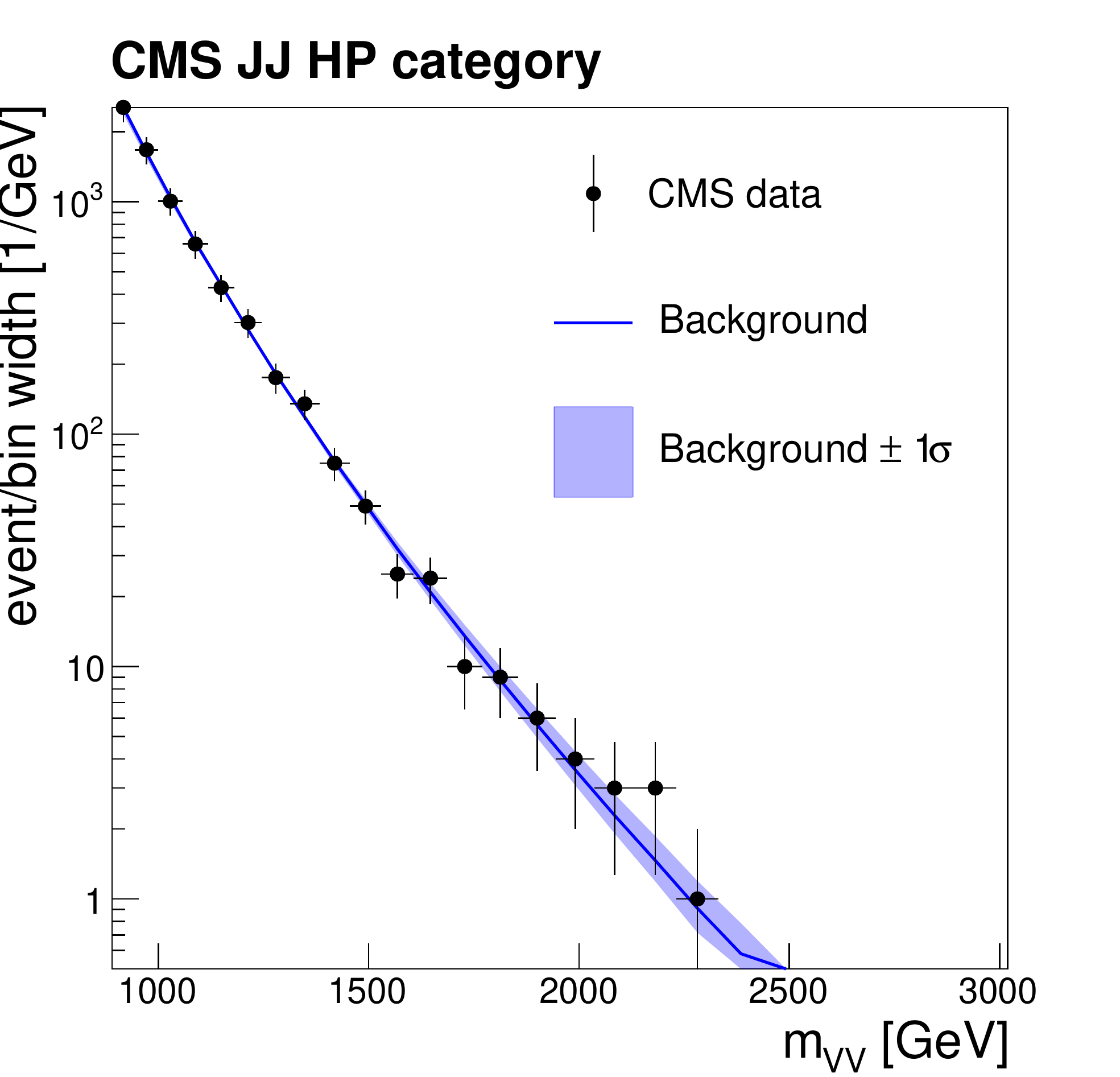}
\includegraphics[width=0.40\textwidth, angle =0 ]{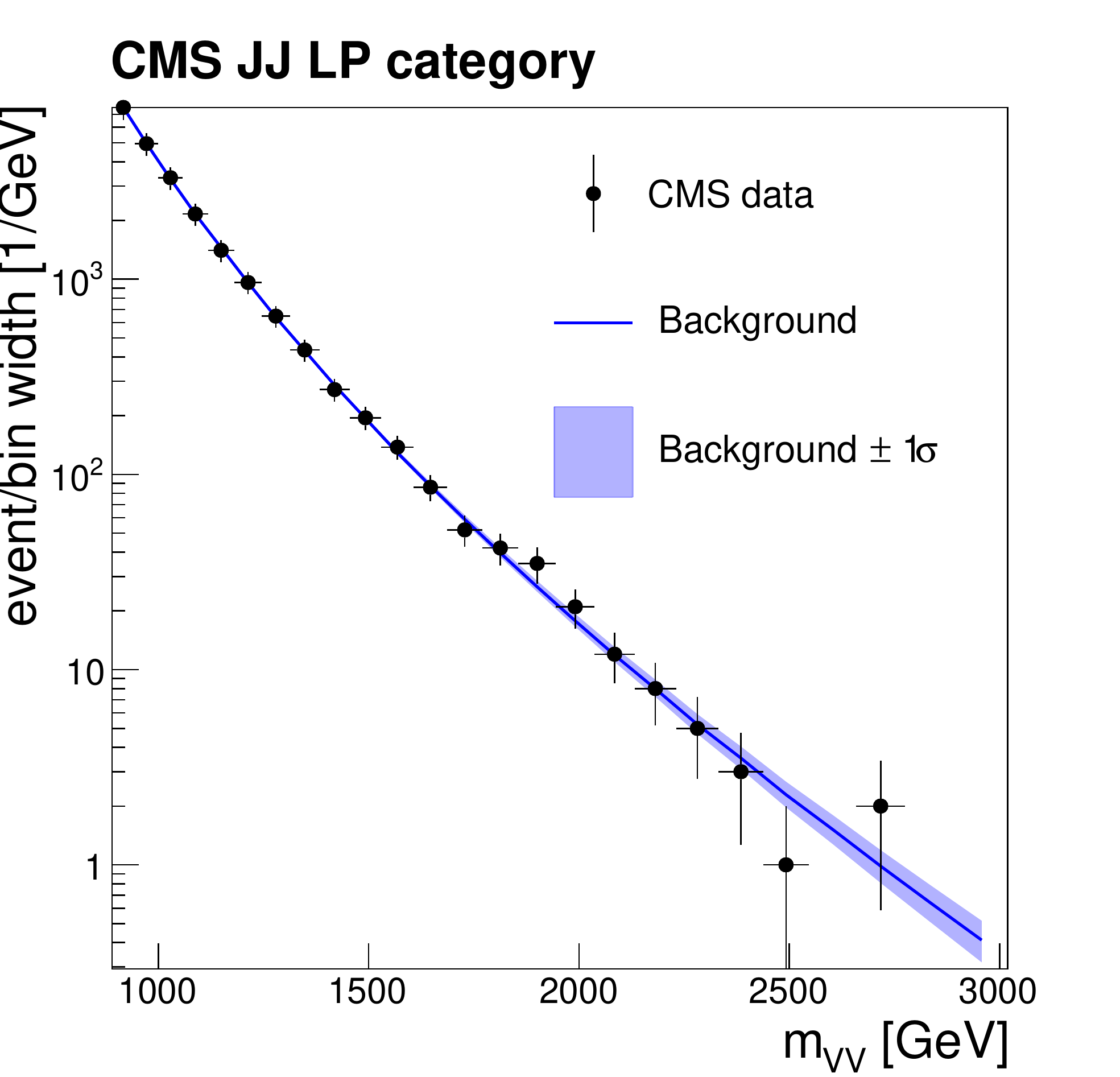}
\caption{\small CMS hadronic search: \MJJ data
  distribution overlaid with the background fit employed in this study with
  uncertainties 
  for High (left) and Low (right) Purity samples. See text for details.
\label{fig:compareCMS} }
\end{center}\end{figure*}

The CMS collaboration provides the binned data and background distributions
with the associated uncertainties in the HEPDATA database (see
Fig. \ref{fig:compareCMS}), as well as the signal distributions for three
different models along with their efficiencies~\cite{CMSVV}: \WptoWZ and
\PGbulk decaying exclusively to \ZLZL or \WLWL. 
We consider the following systematic uncertainties:
\begin{itemize}
  \item {\em Background uncertainty}, provided by CMS (in HEPDATA) and considered as
    fully correlated across the bins of the \MJJ distribution. 
  \item {\em Signal normalisation uncertainty}, which is separated
    further into two sub-categories: a common-across-channels systematic uncertainty
    corresponding to the luminosity measurement (2.2\%), and an additional
    term applicable to the \JJ channel that covers \PV-tagging
    uncertainties, such as \PT{}, pile-up and PDF dependencies (13\%). The
    $\tau_{12}$ uncertainties are treated separately in the category
    below. 
  \item {\em Signal purity category migration uncertainty}, which covers the 
    effects of events ``migrating'' from the HP to the LP category, or
    vice-versa. This uncertainty amounts to 7.5\% and 54 \%, respectively. 
  \item {\em Signal jet energy scale uncertainty}, propagates to $\pm 1\%$ of uncertainty on $\MJJ$;
  It is treated in the same way as in the ATLAS case.
\end{itemize}
All systematic uncertainties are treated as fully correlated across different
\MJJ bins. They are also considered as fully correlated between
the LP and the HP categories,  with the exception of the ``purity
category migration'' uncertainty, which is treated as fully anti-correlated. 

\begin{figure*}[htb]\begin{center}
\includegraphics[width=0.40\textwidth, angle =0 ]{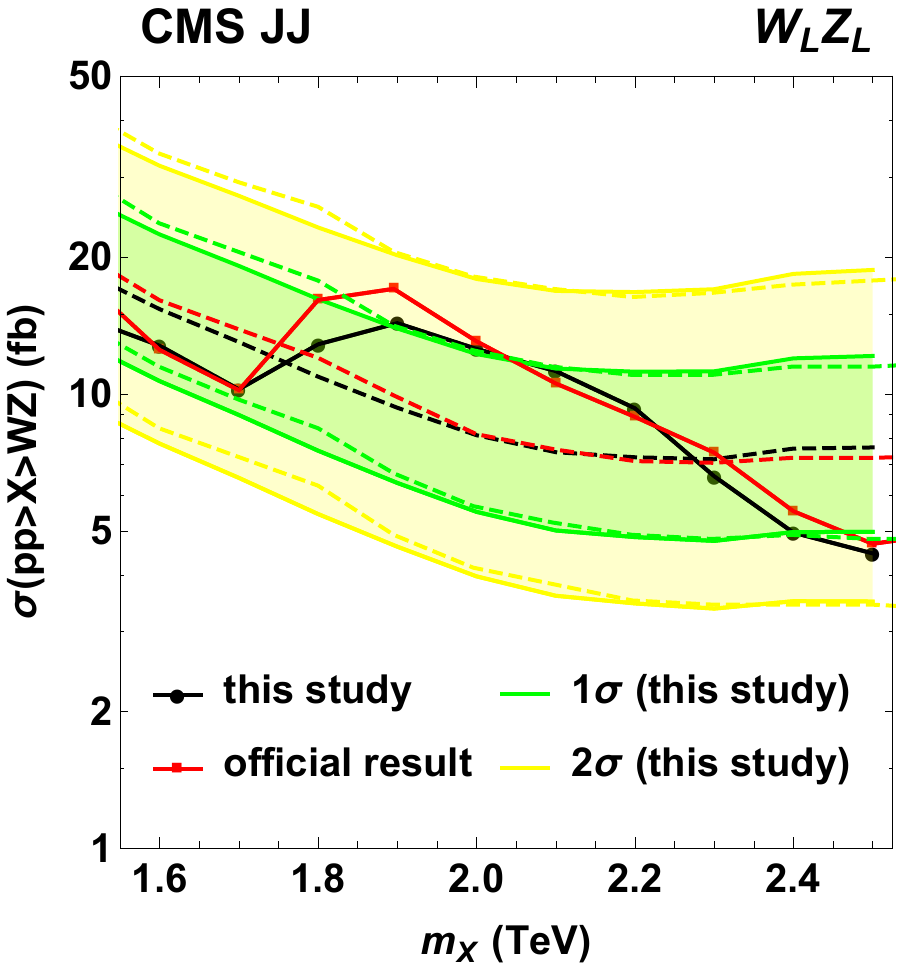}
\includegraphics[width=0.40\textwidth, angle =0 ]{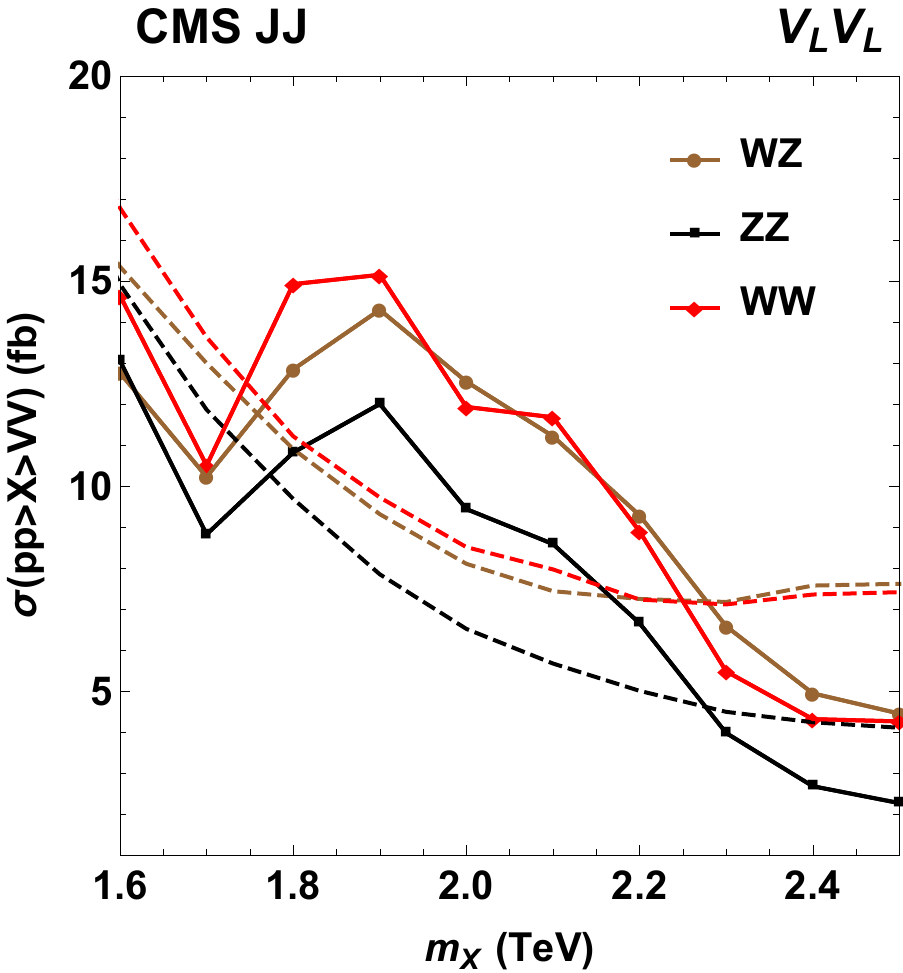}
\caption{\small CMS hadronic search. {\bf Left:} Expected
  (dashed lines) and observed (continuous lines) exclusion
  limits on \WptoWZ production cross sections as a function of the
  resonance mass $m_X$ obtained with this study (black), and comparison
  with the official CMS results (red). The green and yellow bands (dashed
  lines) represent
  the one and two sigma variations around the median expected limits
  calculated in this study (by CMS).
  {\bf Right:} Expected (dashed lines) and
  observed (continuous lines) exclusion limits on exotic production cross
  section as a function of the resonance mass $m_X$ obtained with this
  study for \WptoWZ (brown), \GtoWW (red) 
  and \GtoZZ (black) signal hypotheses.   
\label{fig:checkCMS} }
\end{center}\end{figure*}

Our statistical analysis for \WptoWZ, \GtoWW and \GtoZZ models produces
exclusion limits that are in very good agreement with the ones publicly
provided by CMS. An example of this agreement can been seen in the left plot of
Fig.~\ref{fig:checkCMS}.
The exclusion limits calculated in a few benchmark models can be seen in the
right plot of Fig. \ref{fig:checkCMS}. The most stringent limits are
obtained for the \GtoZZ hypothesis,  thanks to the higher \PV-tagging
efficiency for \PZ bosons.

\subsection{Combined LHC results of hadronic searches}
This section describes the combination of the ATLAS and CMS searches in the
fully hadronic channel \JJ and the interpretation of the results
under different signal hypotheses. 

As a first step we note that ATLAS
assumes a wide resonance in its \JJ searches, whereas CMS assumes a narrow
one. To ensure a consistent treatment of the search in the hadronic channel
between the two experiments we introduce a $+10\%$ scale factor in the
ATLAS selection efficiency. A description of the derivation of the scale
factor and its impact on the search sensitivity is discussed in
Appendix~\ref{sec:narrow}. For every signal hypothesis under consideration
we use the optimal mass selection windows as defined by ATLAS.  
%One shall remember that CMS data and signal was not modified, while a fudge factor was applied on ATLAS data to account for the difference between our expected limits and the ones published by ATLAS collaboration.

\begin{figure*}[htb]\begin{center}
\includegraphics[width=0.32\textwidth, angle =0 ]{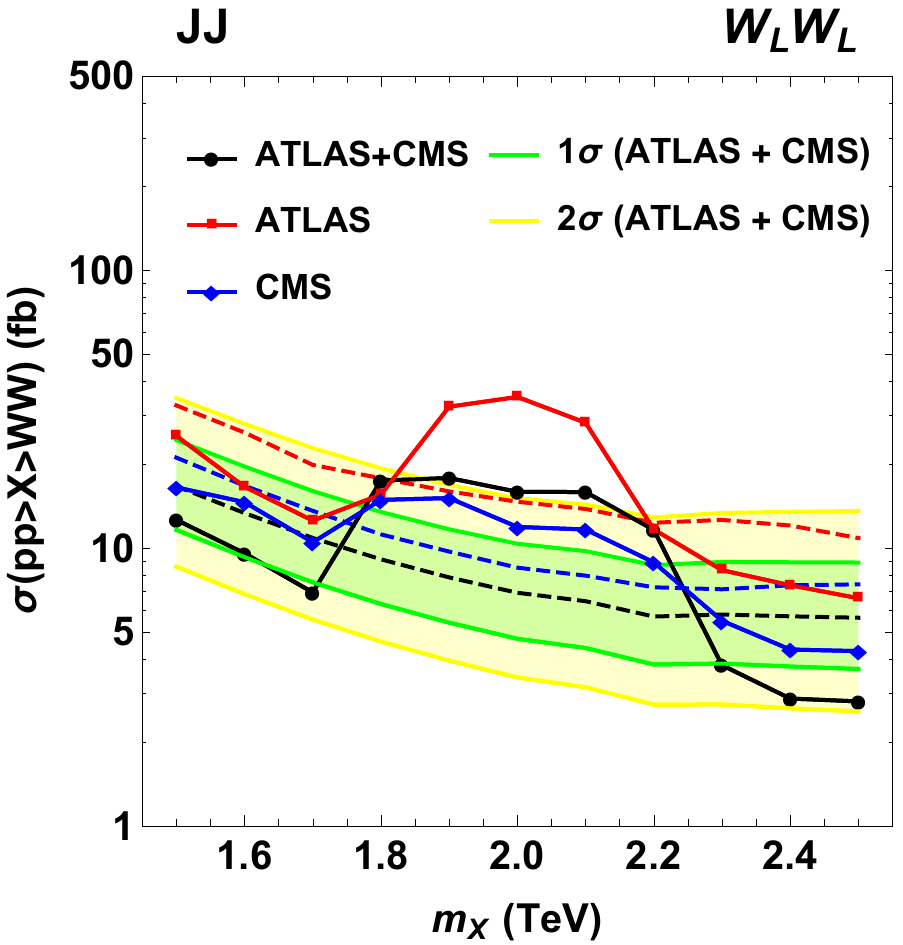}
\includegraphics[width=0.32\textwidth, angle =0 ]{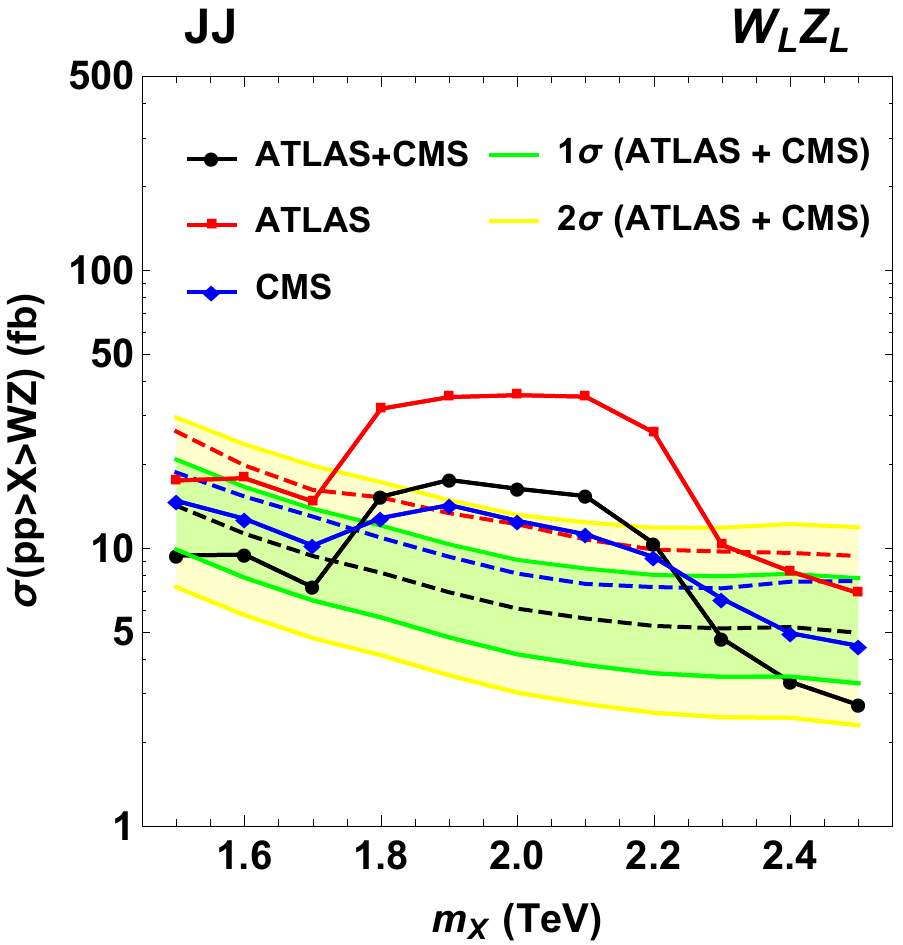}
\includegraphics[width=0.32\textwidth, angle =0 ]{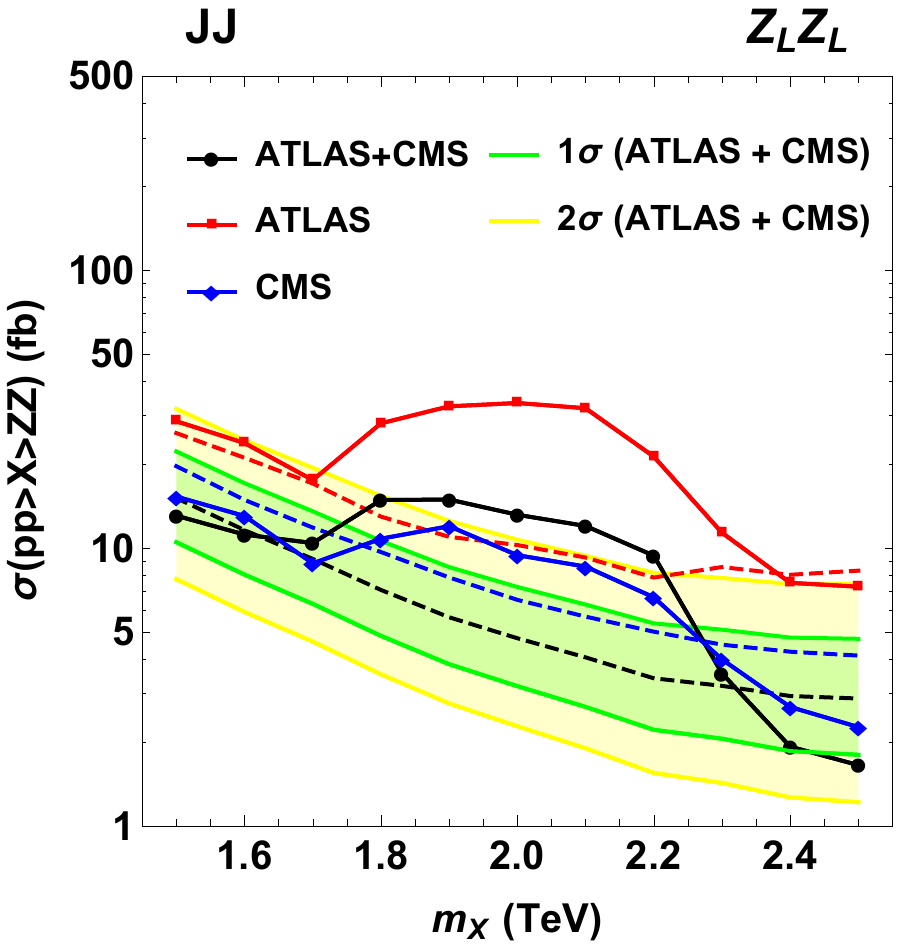}
\caption{\small Combination of hadronic searches: Expected (dashed lines)
  and observed (continuous lines) exclusion limits on exotic production cross
  section as a function of the resonance mass $\MX$ obtained with the
  emulation of the ATLAS (red) and CMS (blue) searches and their
  combination (black) for \WLWL (left),  \WLZL (middle) and \ZLZL
  (right) selections and signal hypotheses. The green and yellow bands
  represent the one and two sigma variations around the median expected
  limits. The results include the 10\% scale factor discussed in the text.
  \label{fig:VVJJ}
}
\end{center}\end{figure*}

We proceed by combining the {\sc THETA} data cards of the individual ATLAS
and CMS searches. The results of the statistical combination for the \WLZL,
\WLWL, and \ZLZL signal hypotheses can be seen in Fig. \ref{fig:VVJJ}. In
the $1.7 < \MX < 2.2$ TeV region we observe the largest discrepancy between
expected and observed exclusion limits due to the presence of the excess in
the \MJJ spectrum. The excess is much smaller in the CMS analysis, which
forces the combined results to lie between the ATLAS and the CMS
curves. The sensitivity of the combined search as we move away from the
deviation region is driven by the CMS analysis. 

\begin{figure*}[htb]\begin{center}
\includegraphics[width=0.32\textwidth, angle =0 ]{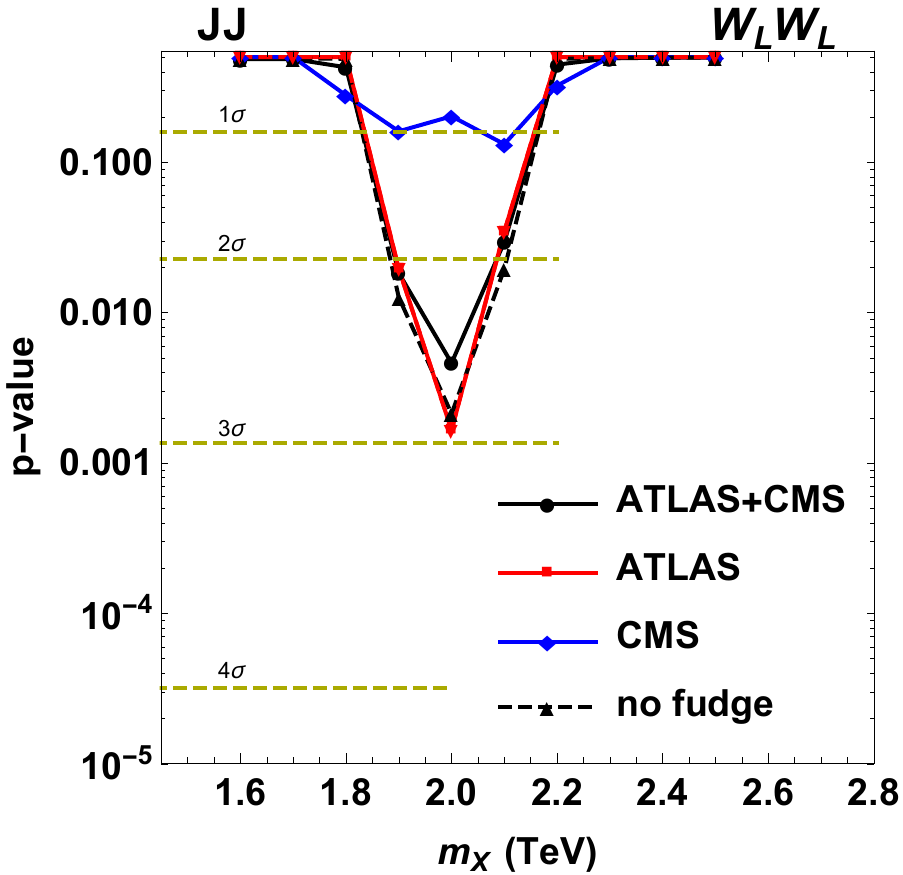}
\includegraphics[width=0.32\textwidth, angle =0 ]{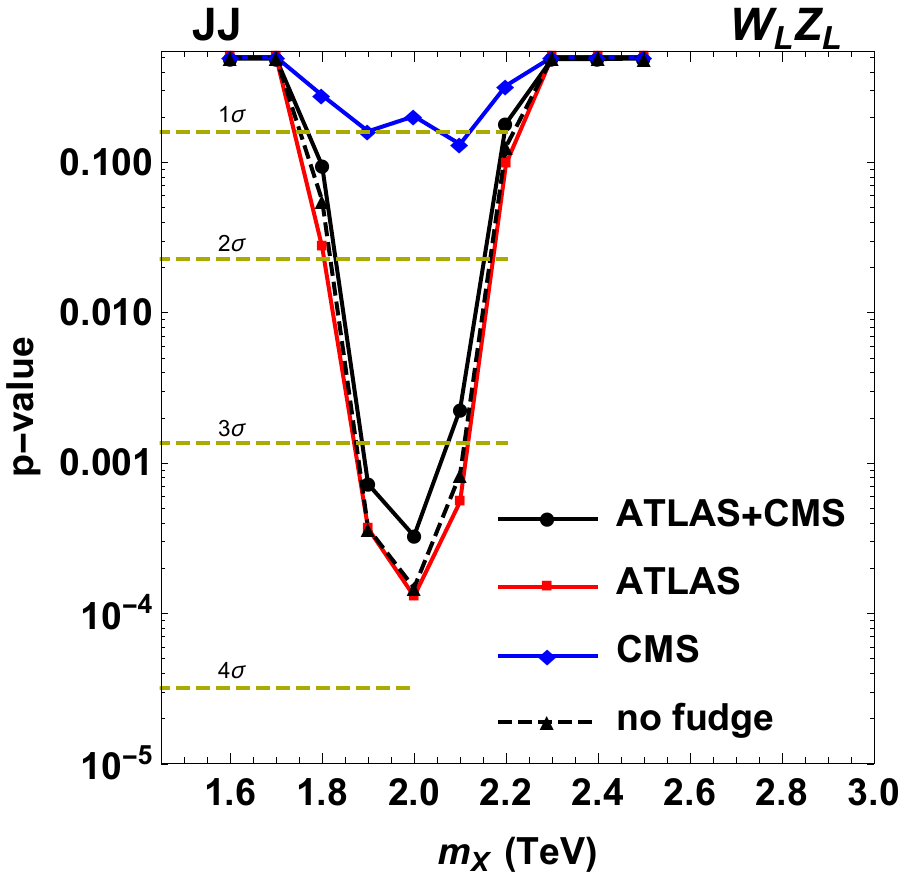}
\includegraphics[width=0.32\textwidth, angle =0 ]{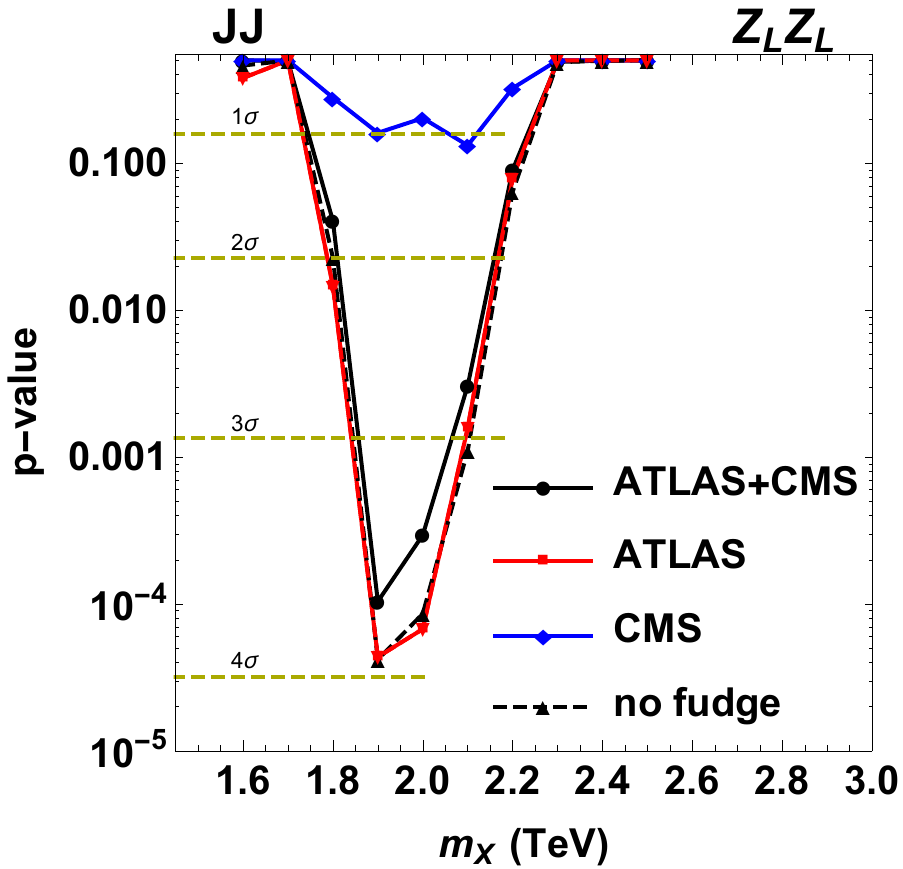}
\caption{\small Combination of hadronic searches: likelihood ratio $p$-values
  as a function of the exotic resonance mass \MX obtained with the 
  emulation of the ATLAS (red) and CMS (blue) searches and their
  combination (continuous black) for \WLWL (left),  \WLZL (middle) and \ZLZL
  (right) selections. The dashed black curve corresponds to the combined
  search without the 10\% scale factor discussed in the text. 
\label{fig:VVJJpval}}
\end{center}\end{figure*}

The impact of the individual experimental results on the combination can be
seen in the distribution of $p$-values (obtained using Wilks' theorem) depicted in
Fig. \ref{fig:VVJJpval}. The CMS $z$-value or significance\footnote{{The statistics community tends to use the term $z$-value or $z$-score, whereas the physics community
prefers to use the term \textit{significance}.}}
in the excess region is of the
order of 1$\sigma$, independently of the considered model and corresponding
selections. The ATLAS significance ranges from less than 3$\sigma$ for the \WLWL
 selection to nearly 4$\sigma$ for the \ZLZL selection, as a
result of the different \PW and \PZ  mass selection windows. The statistical
significance of the combined 
result is very close to the one obtained with the ATLAS result alone, although
slightly reduced. In fact, the ATLAS and CMS results are not contradictory:
due to the small CMS excess observed in the same mass region, the CMS result
cannot exclude the larger ATLAS excess.

\begin{figure*}[htb]\begin{center}
\includegraphics[width=0.32\textwidth, angle =0 ]{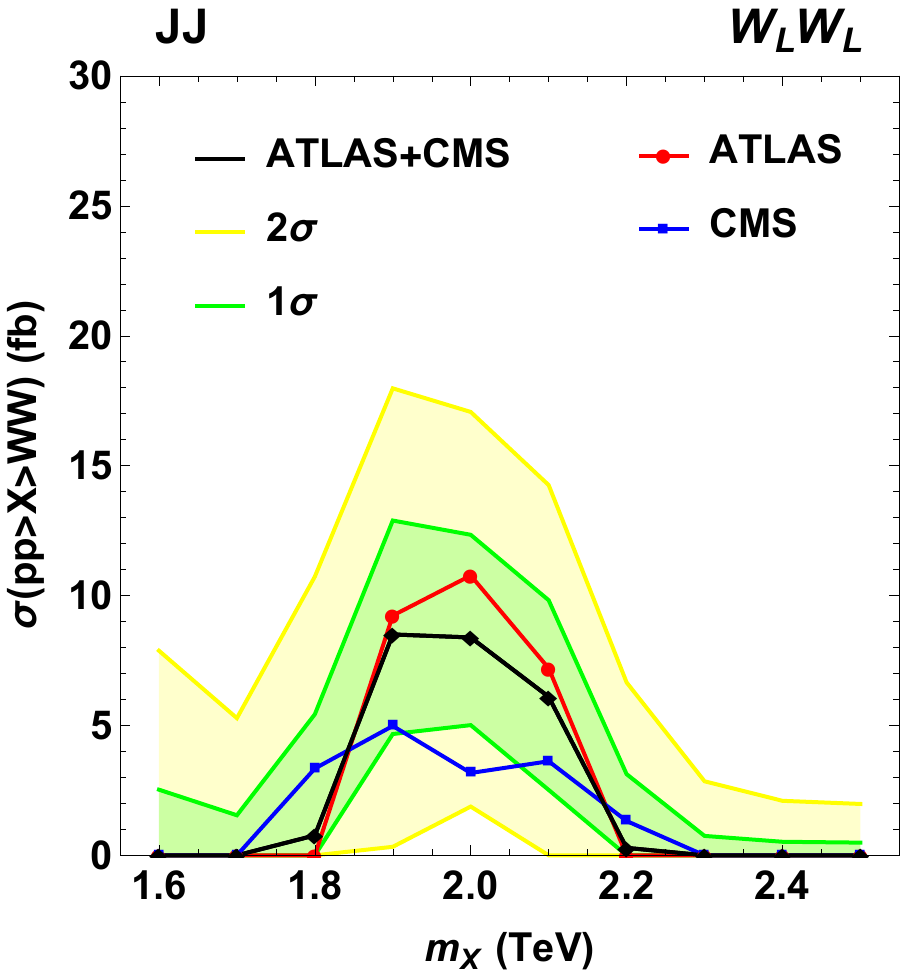}
\includegraphics[width=0.32\textwidth, angle =0 ]{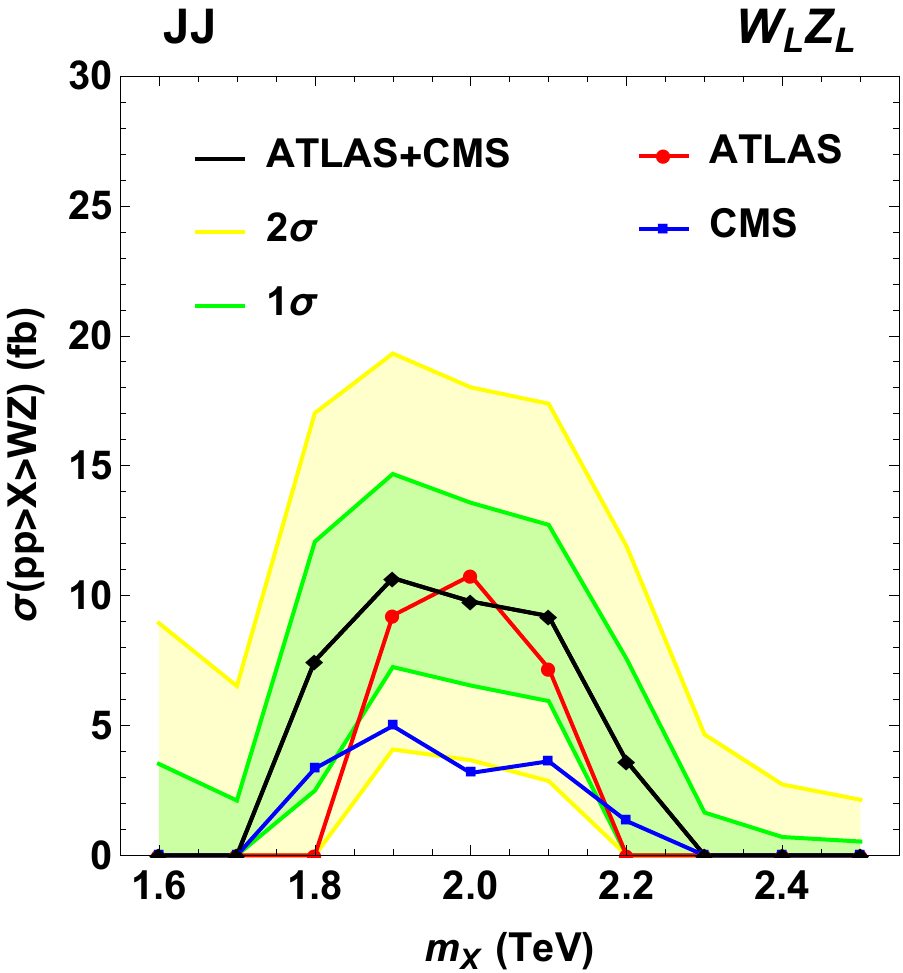}
\includegraphics[width=0.32\textwidth, angle =0 ]{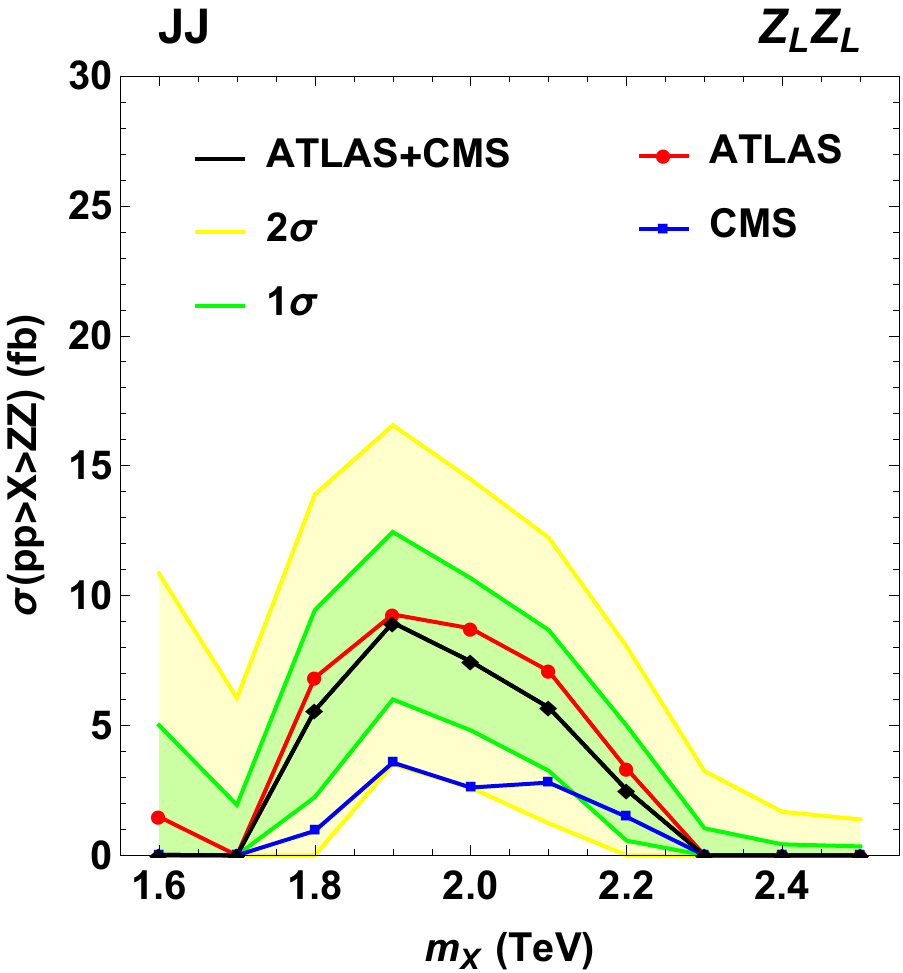}
\caption{\small Combination of hadronic searches: Best fitted exotic
  production cross section as a function of the resonance mass $m_X$
  obtained with the emulation of the ATLAS (red) and CMS (blue) searches
  and their combination (black) for \WLWL (left),  \WLZL (middle) and \ZLZL
  (right) selections and signal hypotheses. The green and yellow bands
  represent the one and two sigma variations around the median 
  values. The results include the 10\% scale factor discussed in the text.
\label{fig:VVJJML}
}
\end{center}\end{figure*}

In order to further characterise the interplay between the ATLAS and the
CMS results in the combination, we show in Fig.~\ref{fig:VVJJML} the
best-fit exotic signal cross section as
a function of the resonance mass $m_X$ value for a few benchmark models and
corresponding selections: \WLZL, \WLWL and \ZLZL. The best-fitted cross
section values are shown separately for the emulation of ATLAS and CMS
searches, and their combination. The largest excess for the \WLZL and
\WLWL signal hypotheses is observed in the $1.9 < \MX < 2.1$ TeV mass
range, while the excess extends down to $\MX = 1.8$ TeV for the 
\ZLZL signal hypothesis. In these mass ranges, the ATLAS data suggests a
production cross section of $\approx 10$~fb, whereas the CMS data favours
smaller values ($\approx 3$~fb) and is more consistent with the no-signal
hypothesis. The \MX profile of the fitted exotic signal cross section is
essential identical to the one obtained from the ATLAS search emulation. 

\begin{figure*}[htb]\begin{center}
\includegraphics[width=0.32\textwidth, angle =0 ]{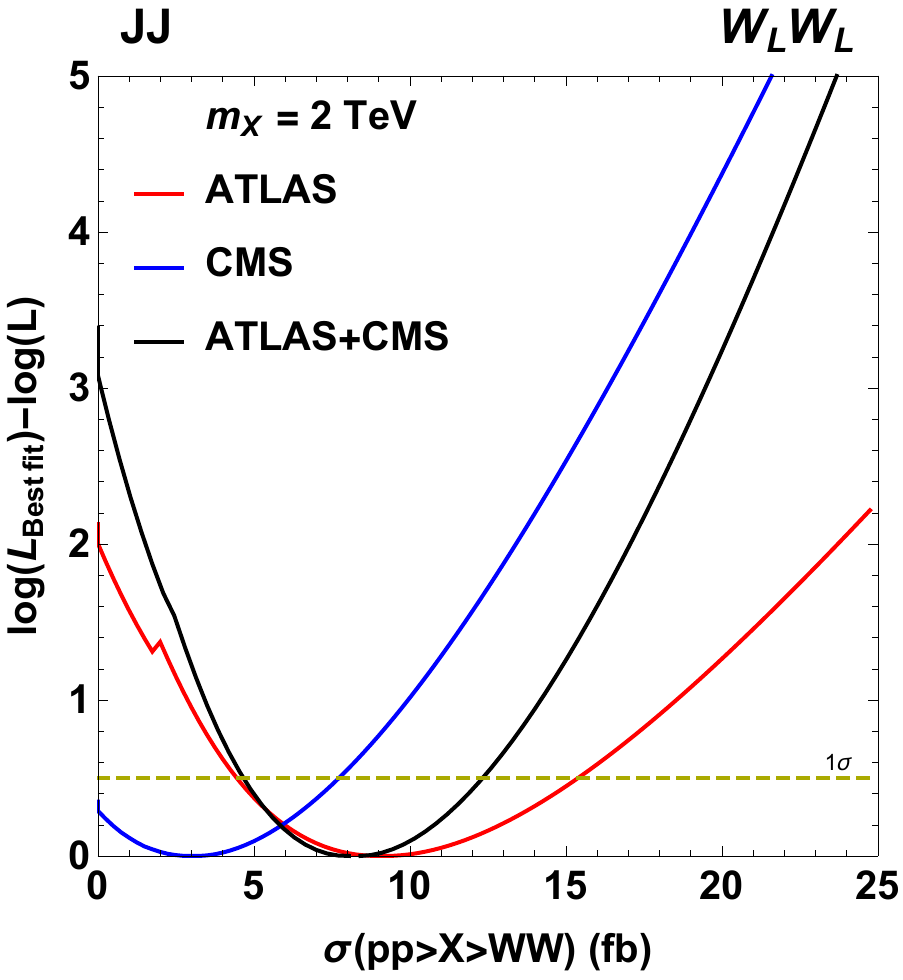}
\includegraphics[width=0.32\textwidth, angle =0 ]{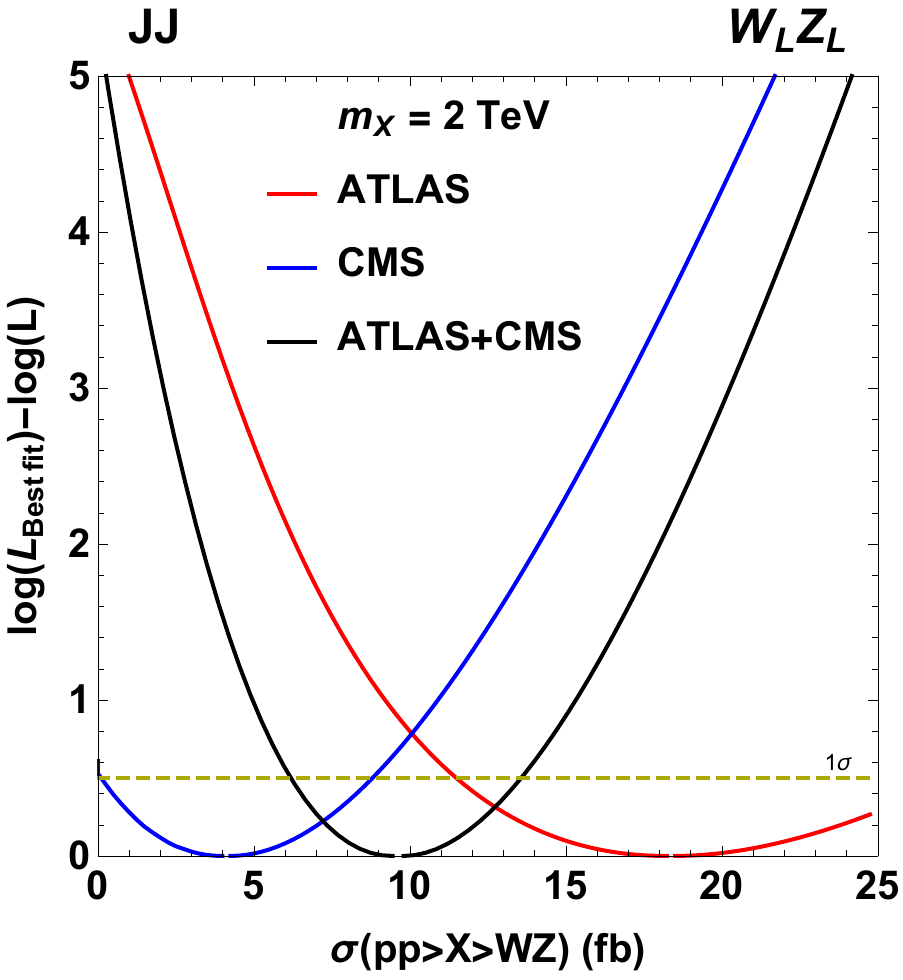}
\includegraphics[width=0.32\textwidth, angle =0 ]{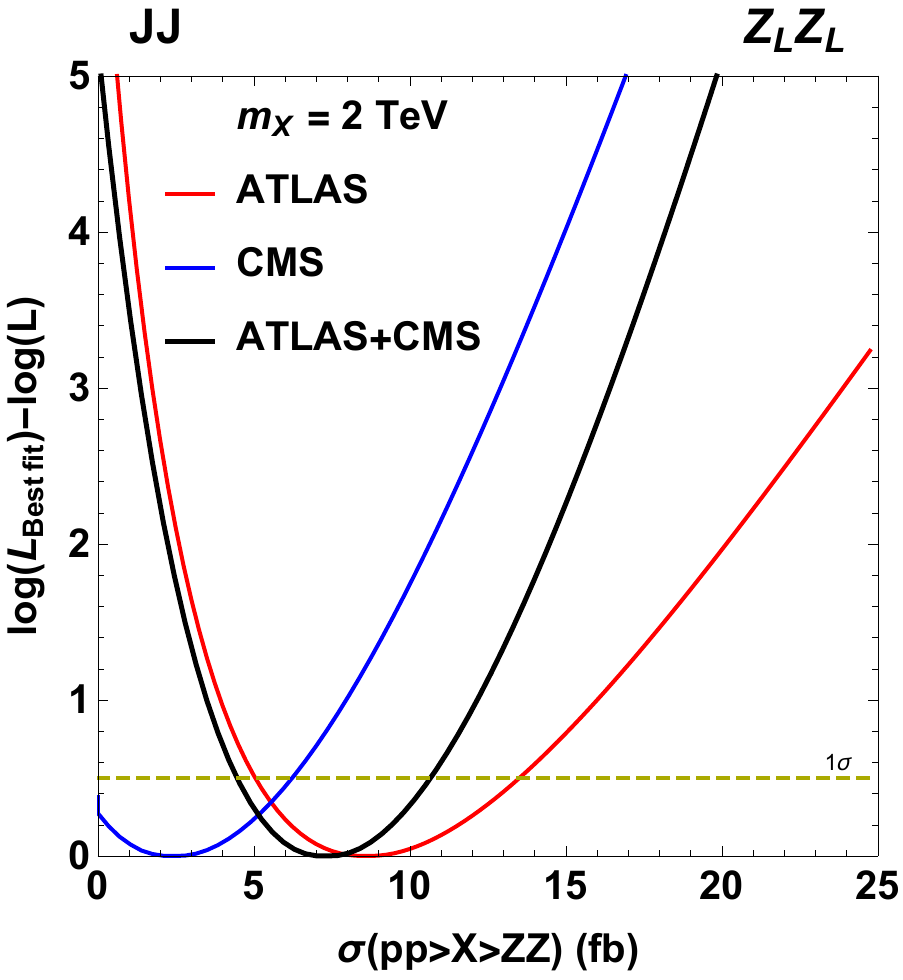}
\caption{\small Combination of hadronic searches: Scans of the profile 
  likelihood as a function of the exotic production cross section for a
  $\MX$ = 2 TeV signal (mass value of largest 
  excess) for the emulation of the ATLAS (red) and CMS (blue) searches
  and their combination (black) for \WLWL (left),  \WLZL (middle) and \ZLZL
  (right) selections and signal hypotheses.
 %From left to right: $m_X = $ 1.8 TeV, 1.9 TeV and 2 TeV.  
 \label{fig:VVJJbf}}
\end{center}\end{figure*}

\begin{figure*}[htb]\begin{center}
\includegraphics[width=0.3\textwidth, angle =0 ]{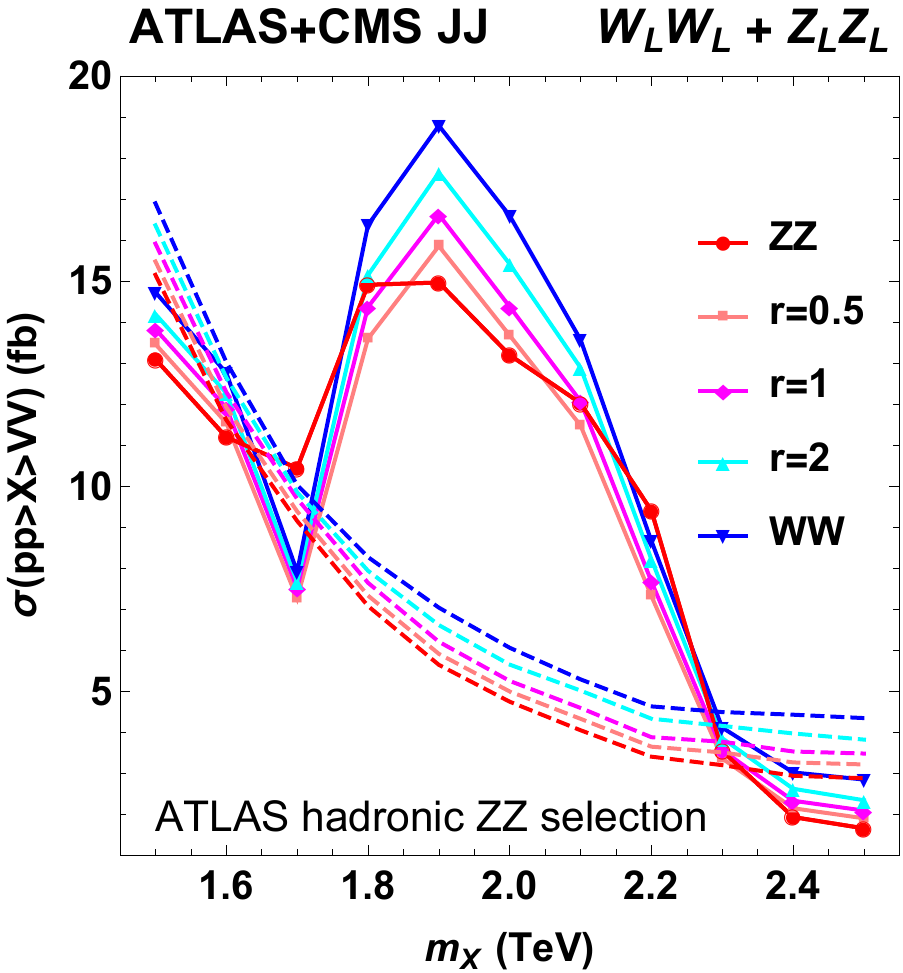}
\includegraphics[width=0.32\textwidth, angle =0 ]{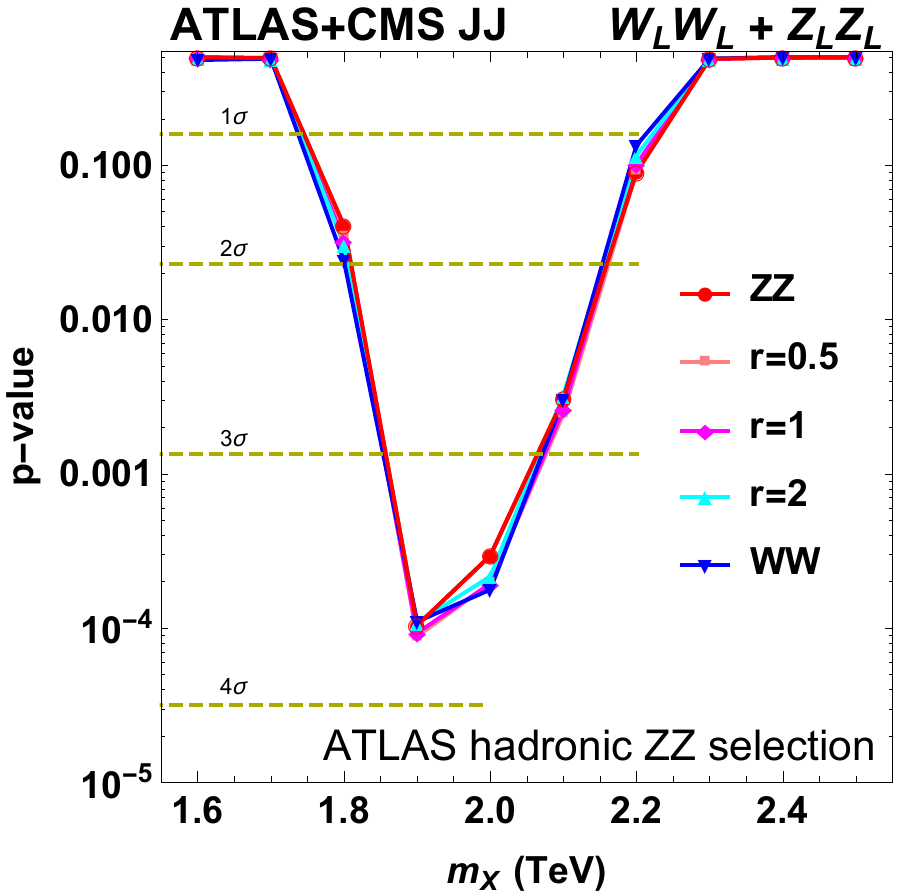}
\includegraphics[width=0.3\textwidth, angle =0 ]{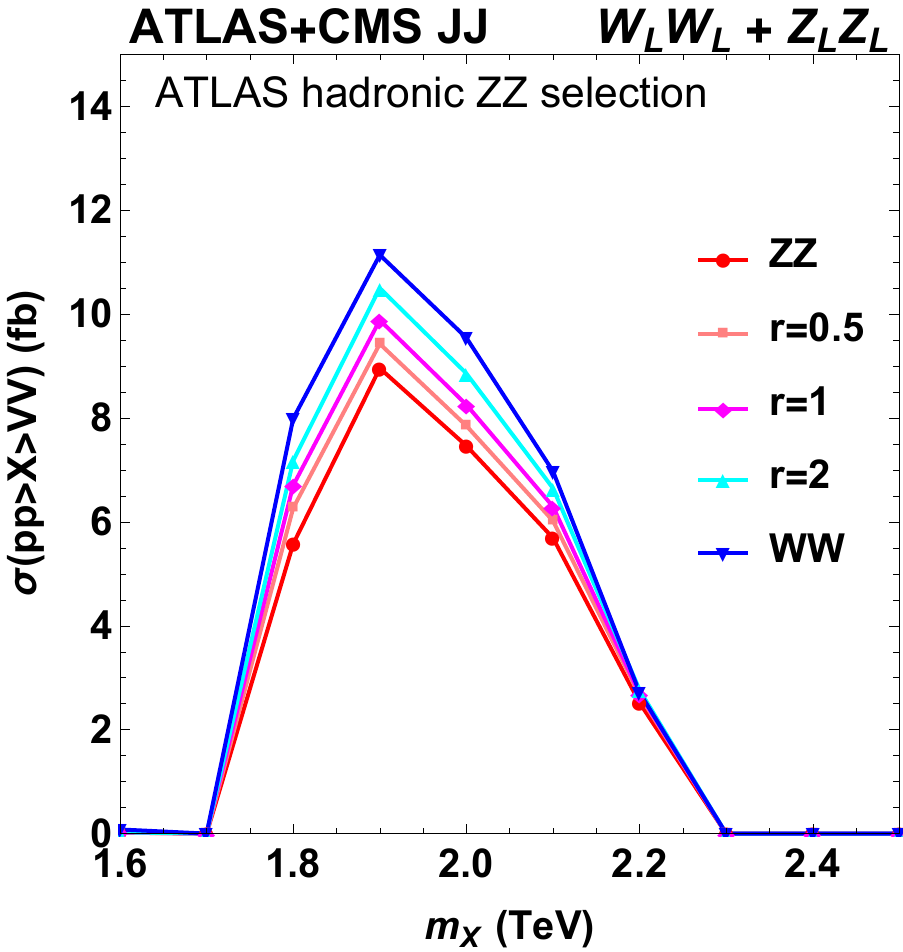}
\caption{\small  Combination of hadronic searches, and dependence of results
  obtained in this study on
  the $r \equiv {\mathcal{B}(X \to \PW\PW)}/{\mathcal{B} (X \to \PZ\PZ)}$
  parameter for a neutral bulk RS-like spin--2
  particle hypothesis, and as a function of the resonance mass \MX. {\bf Left:}
  expected (dashed lines) and  
  observed (continuous lines) exclusion limits on exotic production cross
  section. {\bf Middle:} likelihood-ratio $p$-values. {\bf
    Right:} best fitted exotic 
  production cross section.
\label{fig:VVJJbulk}}
\end{center}\end{figure*}

Further tests of the compatibility between the ATLAS and CMS results can be
seen in Fig.~\ref{fig:VVJJbf}, showing scans of the profiled likelihood as
a function of the exotic production cross
section for $\MX = 2$ TeV (mass value of
largest excess). Due to the large uncertainties of the fit, the
best-fit cross-section values by ATLAS and CMS are compatible within $\pm
1\sigma$ for the \WLZL and \WLWL hypotheses. The compatibility of the
results from the two experiments is slightly reduced in the \ZLZL
scenario. The dependence of these results on $r \equiv {\mathcal{B}(X
  \to \WLWL)}/{\mathcal{B} (X \to \ZLZL)}$ can be seen in
Fig. \ref{fig:VVJJbulk}. The 
conclusions discussed above remain mostly unchanged. 

In summary, in the combination of fully hadronic results the small CMS
excess results in a slight reduction of the larger ATLAS excess. However,
the combined-search statistical significance stays well above 3$\sigma$
for the \WLZL and \ZLZL hypotheses and close to 3$\sigma$ for the
\WLWL hypotheses. The preferred mass range for a hypothetical exotic
signal is 1.9 $< \MX < $ 2.0 TeV, with the corresponding production
cross section in the 8-12 fb region.

%\clearpage
\section{Semi-leptonic searches: \VVLNJ and \VVLLJ}
\label{sec:Semilep}
In this Section we discuss the analysis of the ATLAS and CMS searches in
the \VVLNJ and \VVLLJ channels. We follow the discussion pattern of the
fully hadronic section: we first present the results of our analysis for
the two searches separately, followed by their combination and a summary of our
findings.

\subsection{Emulation of ATLAS search}
%We begin this section with the description of the selection for
%the ATLAS analysis, followed by the overview of the statistical analysis
%and its major uncertainties.

\subsubsection{Description of the ATLAS analysis}
The ATLAS semileptonic search considers both the case in which the two
quarks from the vector boson decay are reconstructed as a single merged jet
(boosted regime), and the case in which they are reconstructed as two
distinct jets (resolved regime). In this study, we focus on resonances
heavier than 1.5~TeV, for which the merged regime largely drives the
sensitivity. Thus we consider only the  Merged Region (MR) categories of
Refs.~\cite{ATLASZV,ATLASWV}. 

In both \VVLLJ and \VVLNJ searches, the boosted jet is identified using the
mass-drop filtering algorithm (as in the $\VVJJ$ search). In addition, two
same-flavour opposite-sign leptons, or one charged lepton and missing transverse
energy (MET) are required. The events are selected online by single- or
double-lepton based triggers. The detector coverage includes the tracker
volume ($|\eta| < 2.5$) and the fiducial region of the electromagnetic
calorimeter (for electrons) or the muon detector. The typical $\PT$
threshold for the charged leptons and for MET is 25 GeV.  The main
backgrounds are inclusive \PV production (\ie \PZ+jets for the \LLJ channel
and \PW+jets for the \LNUJ channel), as well as $t\bar{t}$ production. 

\subsubsection{Statistical analysis}
\begin{figure*}[htb]\begin{center}
\includegraphics[width=0.40\textwidth, angle =0 ]{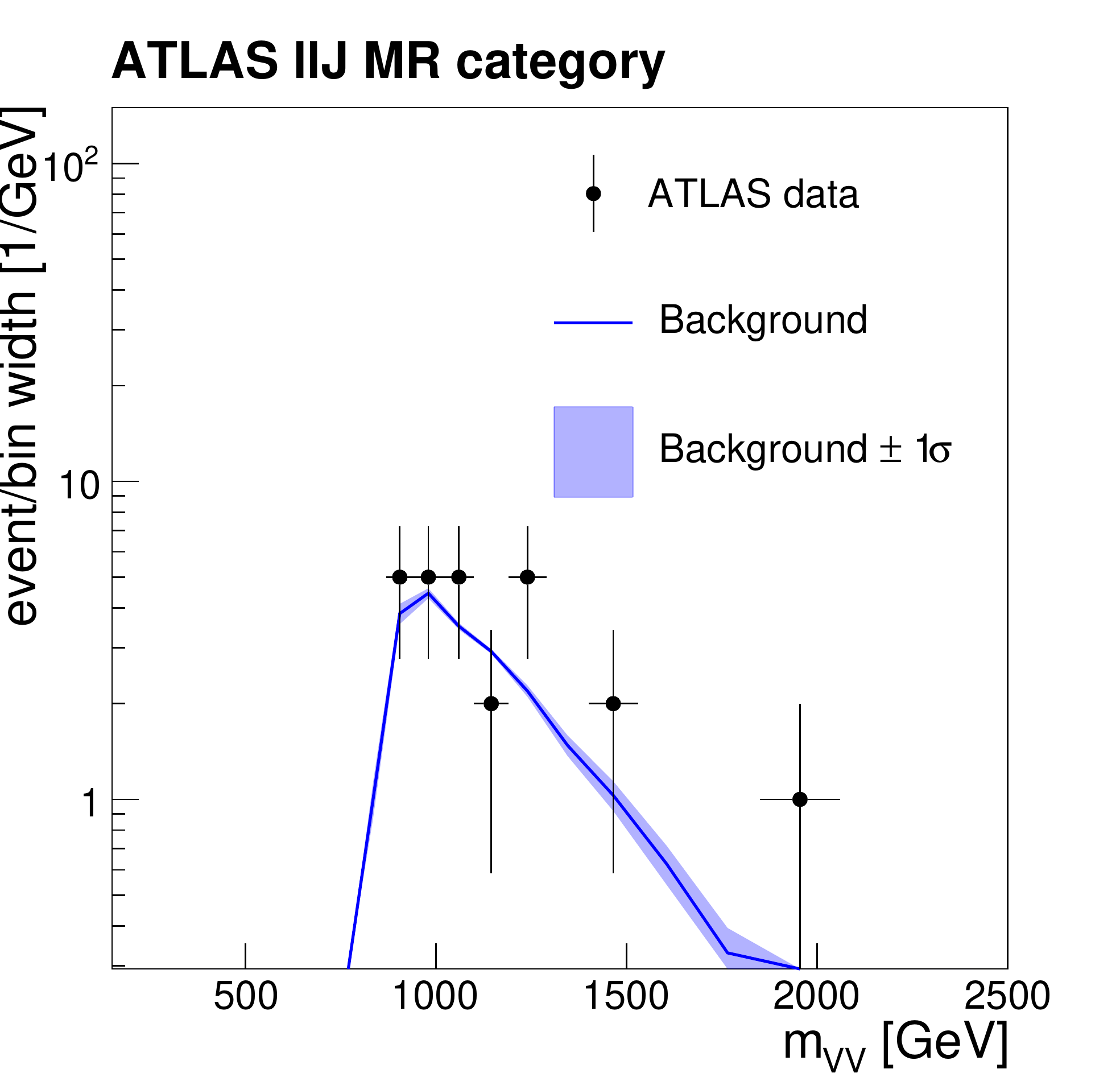}
\includegraphics[width=0.40\textwidth, angle =0 ]{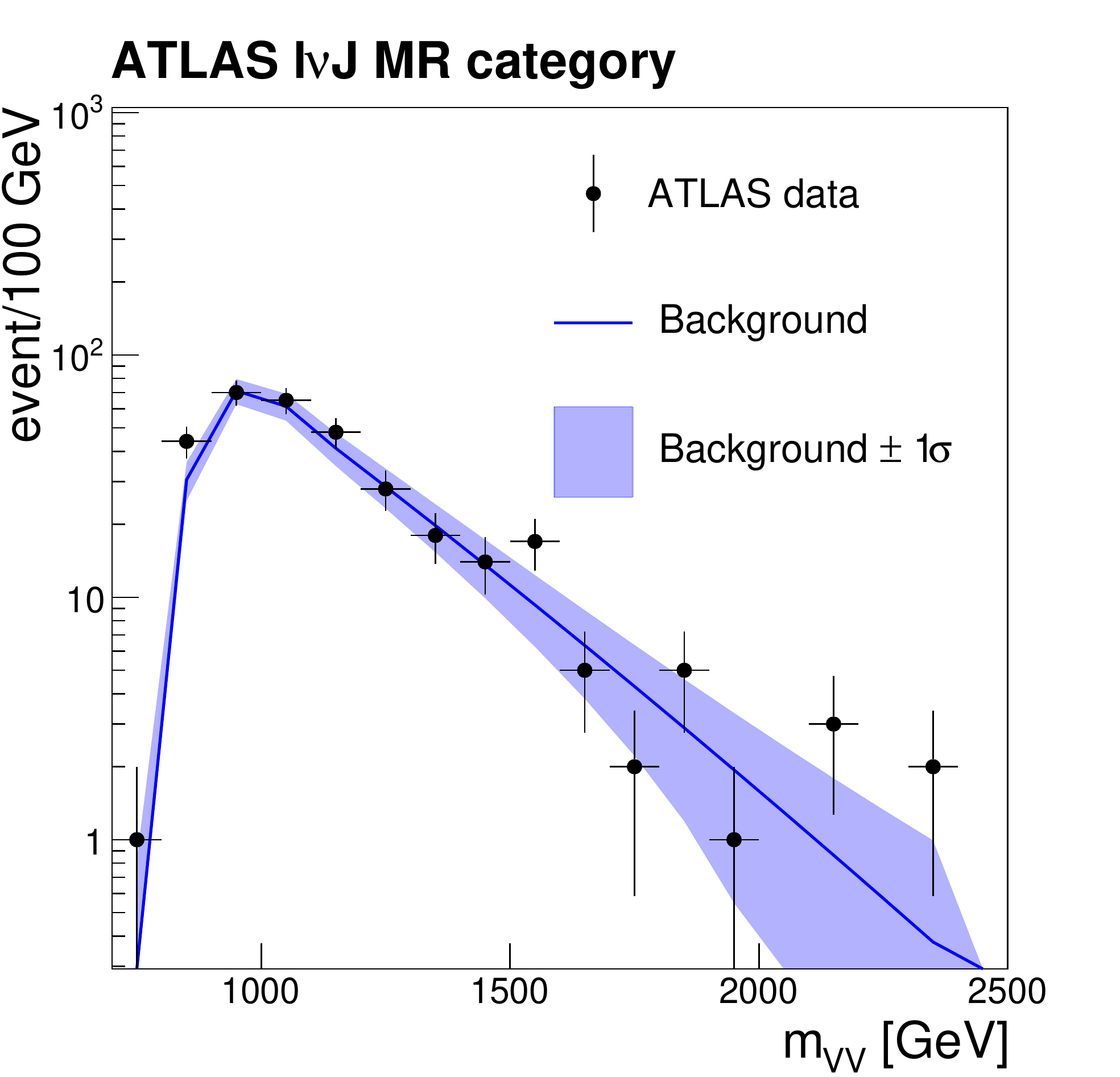}
\caption{\small ATLAS \VVLLJ (left) and \VVLNJ (right)
  searches: Comparison between the official ATLAS background (blue line) and its
  uncertainties (purple band) with the overlaid data of the \MJJ spectrum
  for the Merged Region (of the vector boson hadronic reconstruction) category. 
  \label{fig:check_atlas_vv_llj_lnuj} } 
\end{center}\end{figure*}

We build the likelihood for the ATLAS semileptonic searches using the
information documented in the HEPDATA database. The ATLAS collaboration
estimates the background uncertainties separately for each lepton
category. The electron \PT resolution is better than that of the muon in the high-\PT{}
region.
%\footnote{The
 % energy and momentum of the electrons is reconstructed using calorimetric clusters with
  %the relative resolution scaling as $\sqrt{E}/E$, whereas the momentum of the muons is
  %reconstructed using hits from the tracker and the muon systems with the
  %relative resolution scaling as $\Delta  \PT/\PT$.}. 
The systematic uncertainties associated with different background
sources ($t\bar{t}$ and electroweak components) are
also treated separately. Nevertheless, the background distributions
documented in the HEPDATA database (see
Fig.~\ref{fig:check_atlas_vv_llj_lnuj}) are presented jointly for 
electrons and muons. 
We model the signal distributions in the diboson mass spectrum with a Gaussian
function, centred at the assumed resonance mass and with a width reflecting
the experimental resolution. We assume a fixed value of 4\% resolution in
the \LLJ channel for all mass values\,\footnote{The signal resolution for a
  $\MX = 2$ TeV resonance in the \LLJ channel is 4\%, decreasing to 3\% for
  lower masses \cite{ATLASZV}. We assume a fixed resolution to simplify the
  analysis.}. Similarly, we assume a fixed value of 10\% resolution in the \LNUJ
channel for all mass values\,\footnote{In the case of the \LNUJ channel, the
  reconstruction of the resonance mass requires an assumption on the
  longitudinal momentum of the outgoing neutrino that is not detected. In
  practice, this is estimated from the MET measurement combined with a
  \PW{} mass constraint. The diboson resonance mass is subsequently
  computed using the jet, lepton and calculated neutrino momenta. The
  mass resolution in this channel is degraded compared to
  the \LLJ channel.} (see Fig. 1 in Ref. \cite{ATLASWV}).

The signal distributions are normalised to the expected
yield, as calculated from the theoretical cross section and the selection
efficiency provided by the ATLAS collaboration.  

We consider the following systematic uncertainties, treated as fully
correlated across \MJJ histogram bins: 
\begin{itemize}
  \item {\em Background uncertainty}, provided by the ATLAS experiment (in HEPDATA).
  \item {\em Signal normalisation uncertainty}, which is separated into two
    further sub-categories: a
    common-across-channels systematic uncertainty corresponding to the
    luminosity measurement (2.8\%), and an additional term
    accounting for all types of scale and efficiency systematic effects (10\%). The
    latter is treated as uncorrelated between the $\LLJ$ and $\LNUJ$ channels. 
\end{itemize}

\begin{figure*}[htb]\begin{center}
%\includegraphics[width=0.4\textwidth, angle =0
%]{../limits_comparison/ATLAS_VV_lvJ_semi_paper_fu.pdf}
\includegraphics[width=0.4\textwidth, angle =0 ]{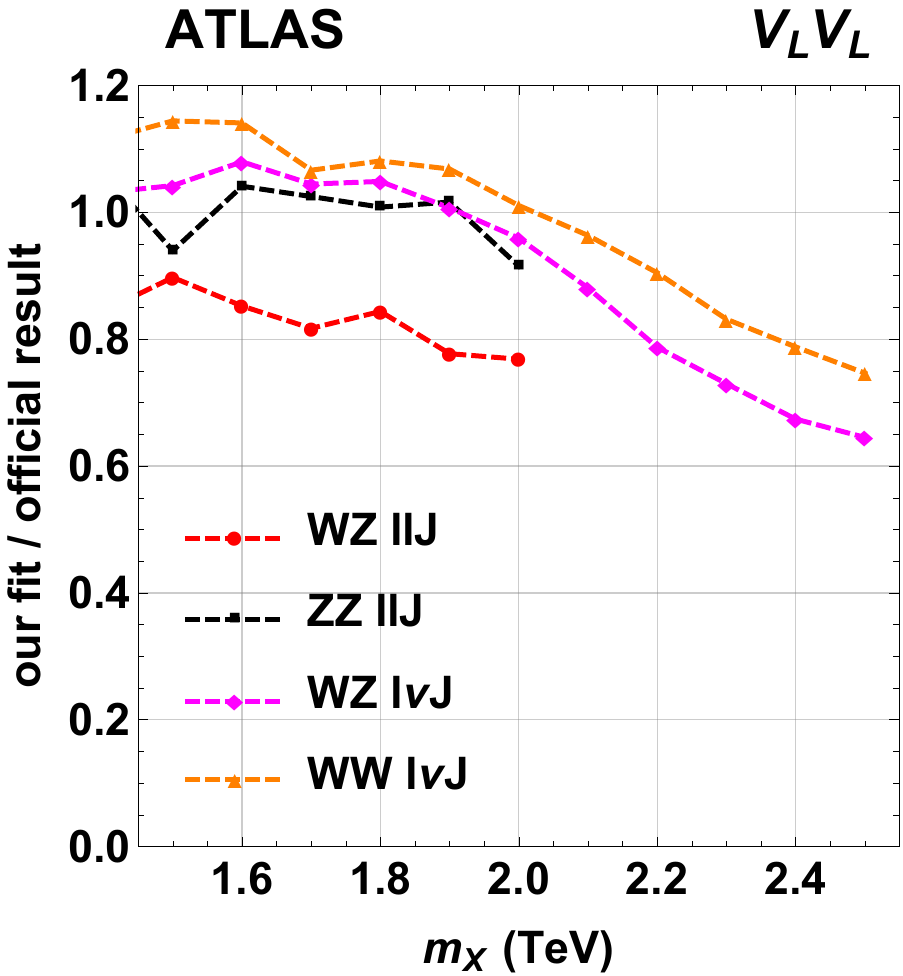}
\caption{\small ATLAS semileptonic searches: Fudge factor as a
  function of the mass $m_X$ of the exotic resonance, calculated via the 
ratio of observed exclusion limits obtained with this study to the ones of
the official ATLAS result, for the \WptoWZ (red) and \GtoZZ (black) signal
hypotheses in the 
\LLJ channel, and for the \WptoWZ (magenta) and \GtoWW (orange) signal
hypotheses in the \LNUJ channel. 
\label{fig:Asemiratio} }
\end{center}\end{figure*}

\begin{figure*}[htb]\begin{center}
\includegraphics[width=0.4\textwidth, angle =0 ]{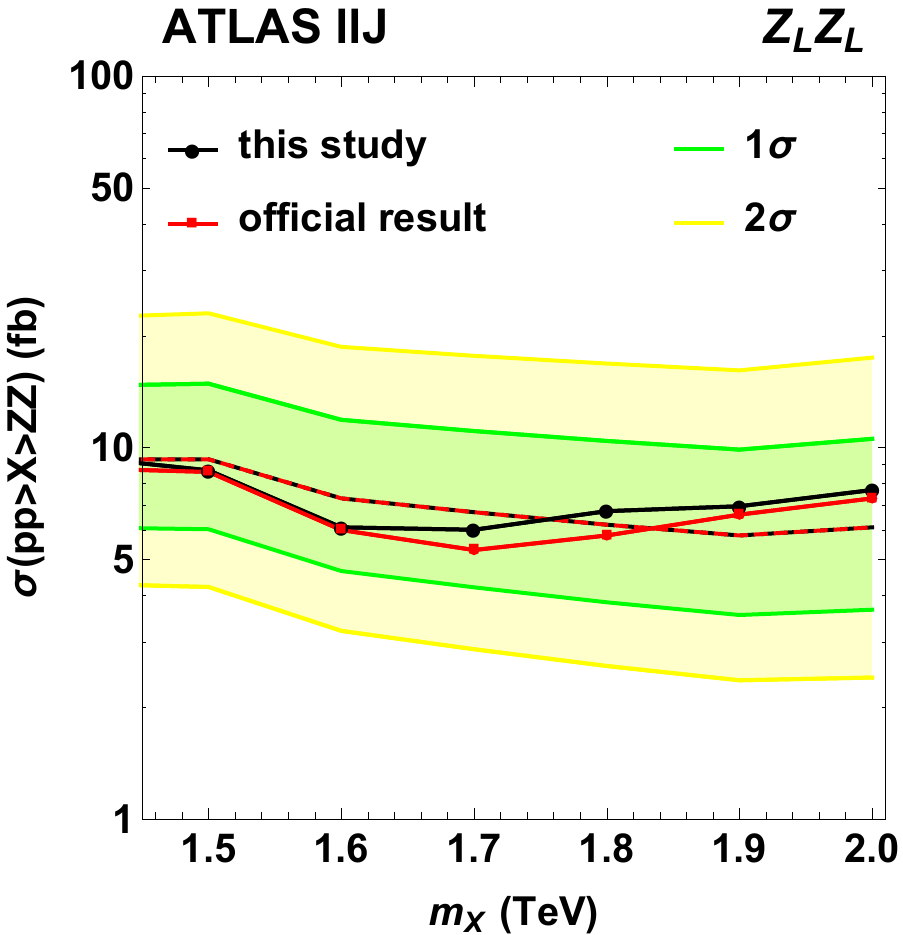}
\includegraphics[width=0.4\textwidth, angle =0 ]{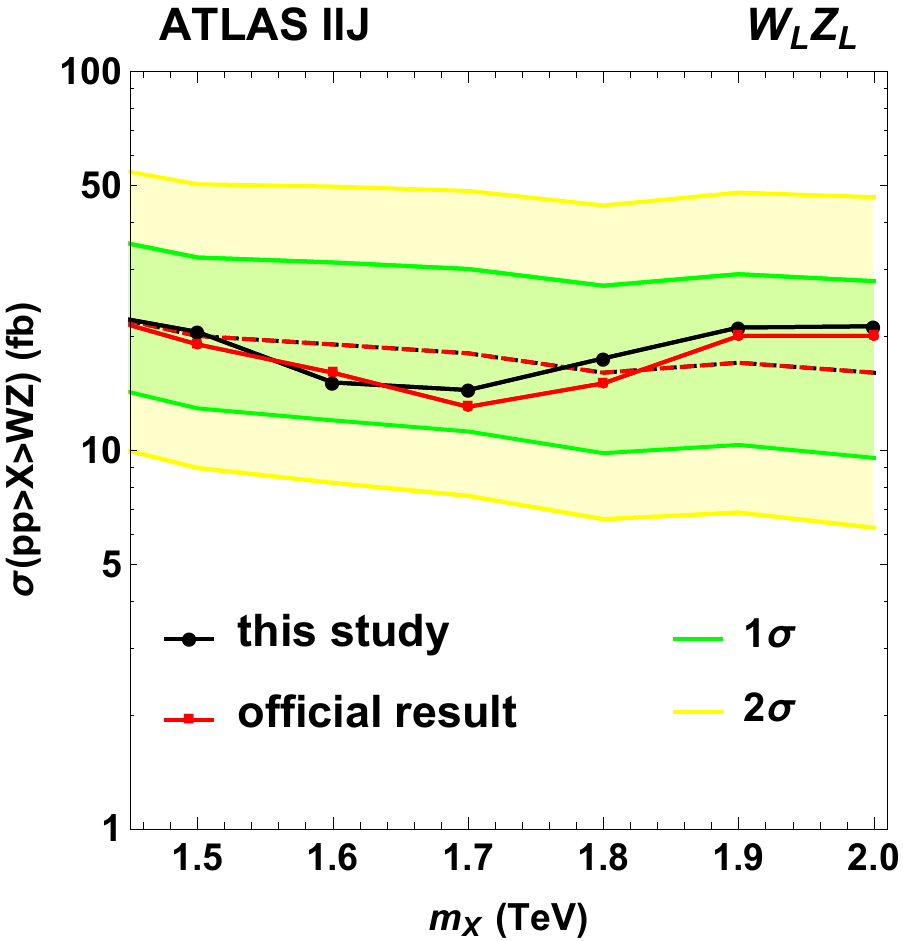}
\includegraphics[width=0.4\textwidth, angle =0 ]{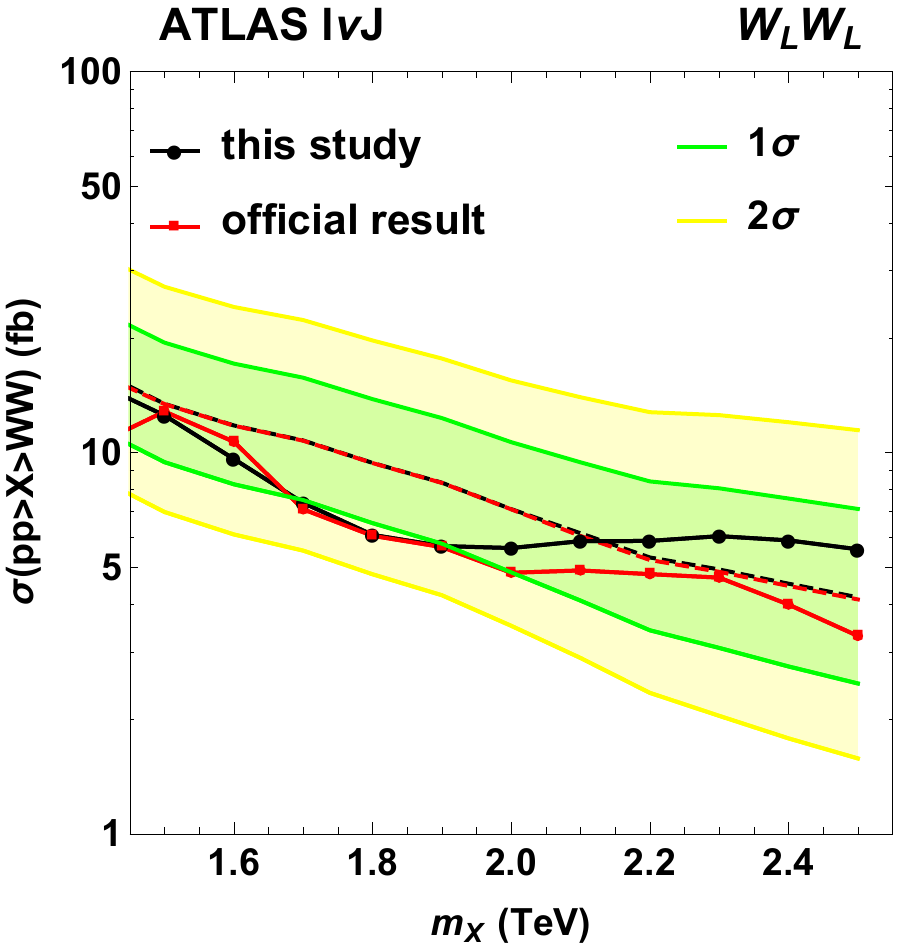}
\includegraphics[width=0.4\textwidth, angle =0 ]{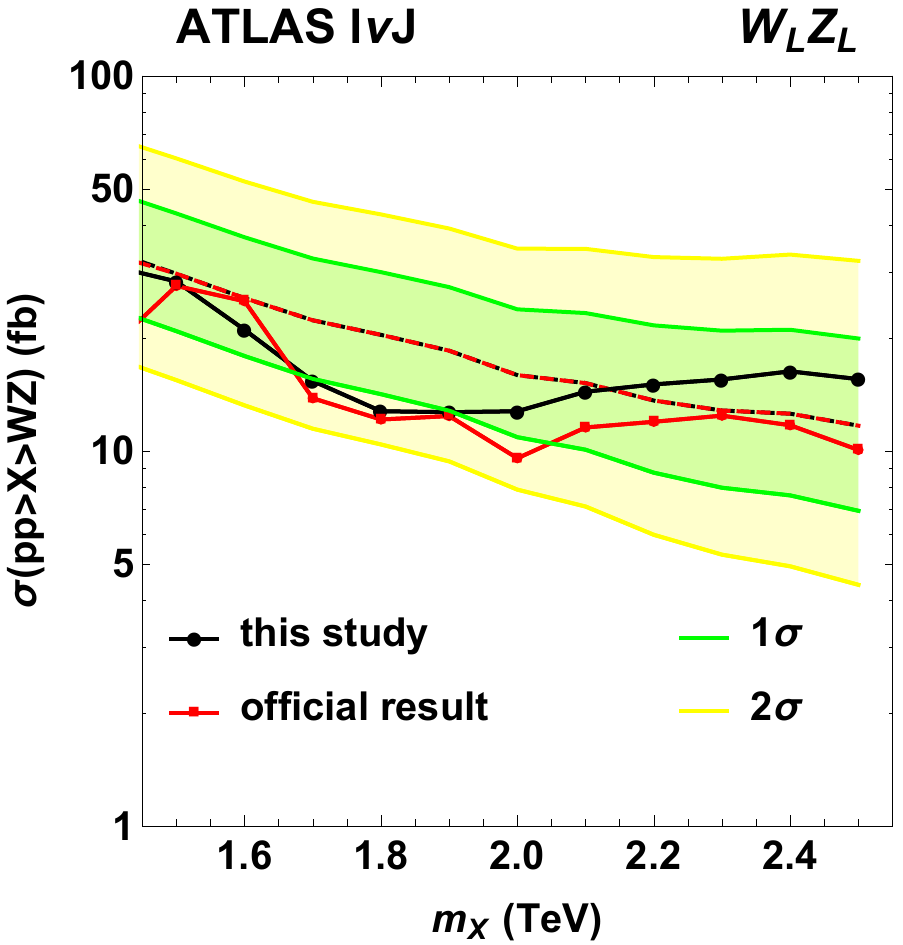}
\caption{\small ATLAS semileptonic searches: Expected
  (dashed lines) and observed (continuous lines) exclusion limits on exotic
  production cross sections as a function of the
  resonance mass $m_X$ obtained with this study (black), and comparison
  with the official CMS results (red) for \GtoZZ (top left), \WptoWZ (top
  right), \GtoWW (bottom left) and \WptoWZ (bottom right) signal hypotheses
  in the \LLJ 
  (top) and \LNUJ (bottom) channels. The green and yellow bands represent
  the one and two sigma variations around the median expected limits
  calculated in this study, with all the corrections described in the text included. 
%{\bf Top}: \LLJ  channel for  ({\bf Left}), and bulk-like neutral spin--2 particle decaying exclusively to $ZZ$ pair of bosons ({\bf Right}).  {\bf Bottom}: $\LNUJ$  channel to $W' \to WZ$, and bulk-like neutral spin--2 particle decaying exclusively to $ZZ$ pair of bosons. 
\label{fig:ll_lnuJ}}
\end{center}\end{figure*}

Given the approximations that we have introduced to model the signal, we do not expect
our statistical analysis to produce results matching with high accuracy the
public ATLAS results. Similarly to the procedure followed for the emulation
of the fully hadronic ATLAS search, we introduce a fudge factor to reduce this
discrepancy. The value of the fudge factor is chosen such that the expected
exclusion limits produced by this study agree
with the official limits by ATLAS. It is found to be between 0.8 and 1.2 in
the resonance mass range of interest, slowly decreasing for larger mass
values (Fig.~\ref{fig:Asemiratio}).
%The potential mis-modelling of the
%signal distribution (presence of non-Gaussian tails for example) also induces
%different corrections to the different benchmarks.
With this correction, our calculated exclusion limits are in good agreement
with the public ATLAS results (Fig. \ref{fig:ll_lnuJ}).

\subsection{Emulation of CMS search}
%CMS uses an analysis strategy very similar to the one presented for ATLAS above.
\subsubsection{Description of the CMS analysis}
The CMS semileptonic analyses~\cite{CMSZVWV} are performed with data
collected by single-lepton triggers for the \LNUJ channel and double-lepton
triggers for the \LLJ channel. Jets are identified as boosted vector bosons
using the same algorithm employed for the fully hadronic search (see
Section~\ref{sec:JJ}). Similarly to the strategy developed in the fully hadronic search, 
 LP and HP categories are introduced, based on the value of $\tau_{21}$, to
 increase the analysis sensitivity.

 The analysis is performed by using a \PGbulk graviton as the benchmark
 signal model. In order to facilitate the interpretation of the search results in
 other theoretical models, the CMS collaboration provides the reconstruction efficiencies
 of leptonic and hadronic $\PW_L$ and $\PZ_L$ in the HP category, as
 function of the boson's \PT{} and $\eta$. 
Those 2D efficiency maps include the effects of the pruned jet mass and
$\tau_{21}$ selections, as well as the resonance mass reconstruction.  
\subsubsection{Statistical analysis}
\begin{figure*}[htb]
\centering
\includegraphics[width=0.4\textwidth, angle =0 ]{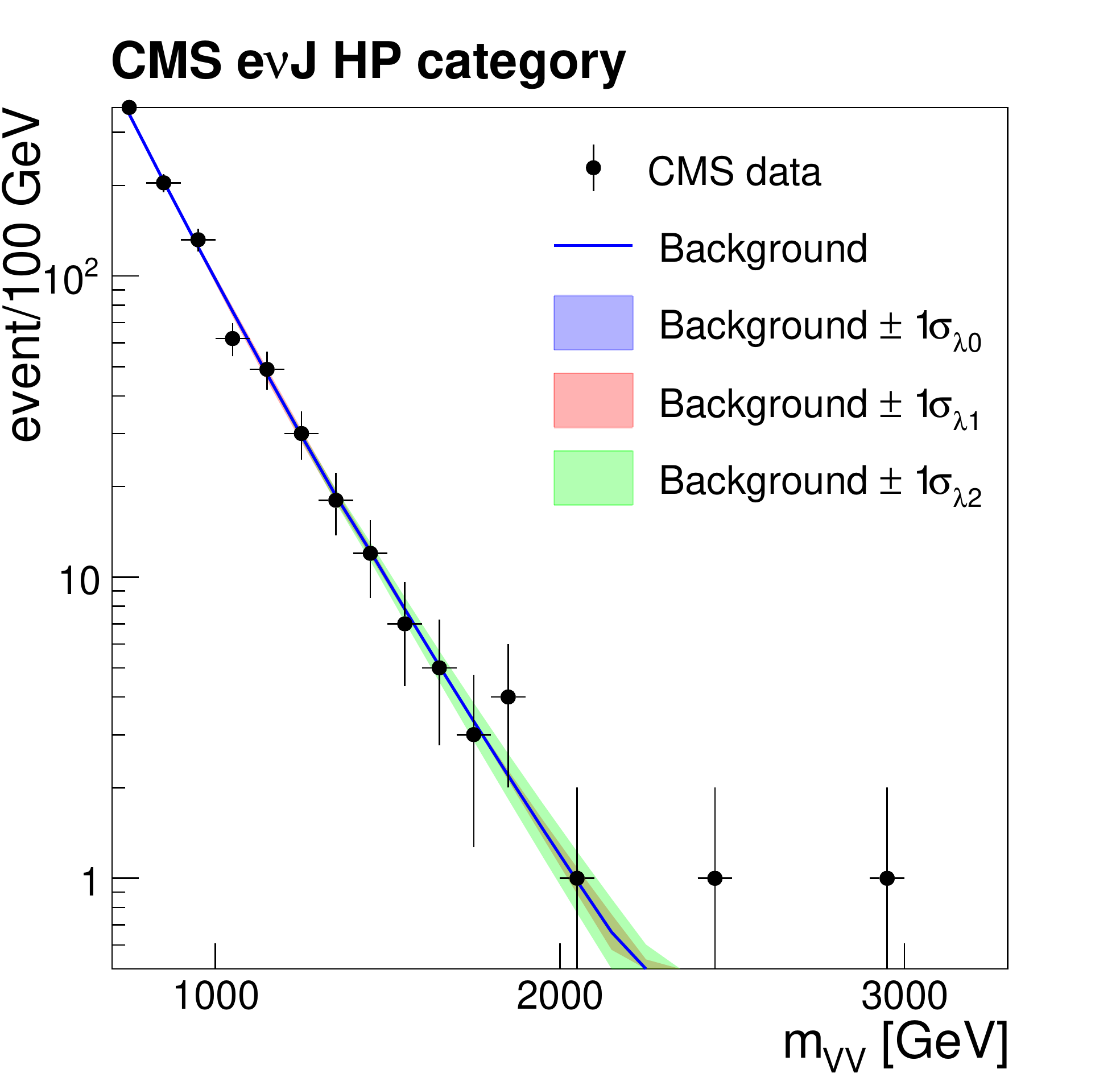}
\includegraphics[width=0.4\textwidth, angle =0 ]{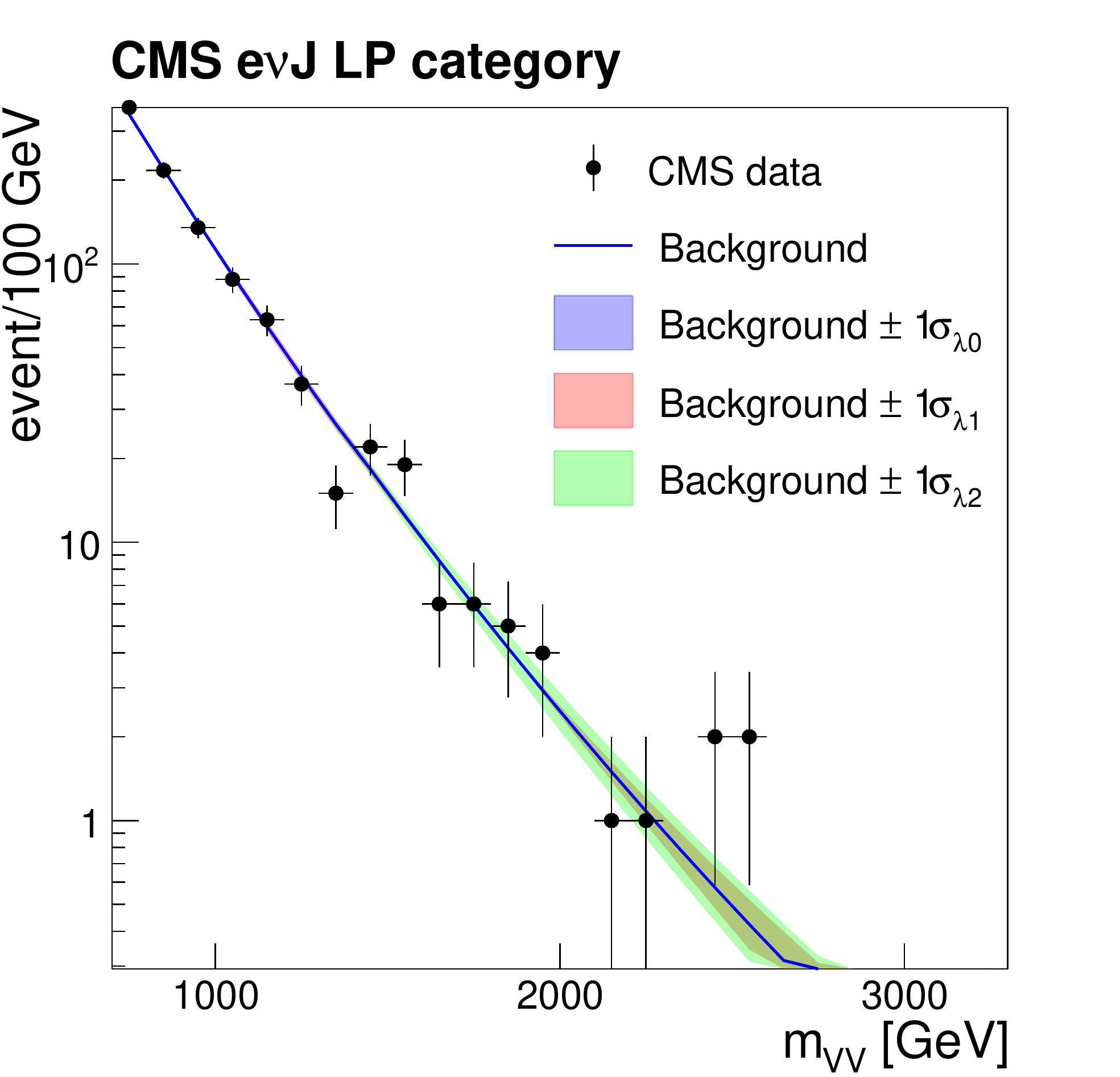}
\includegraphics[width=0.4\textwidth, angle =0 ]{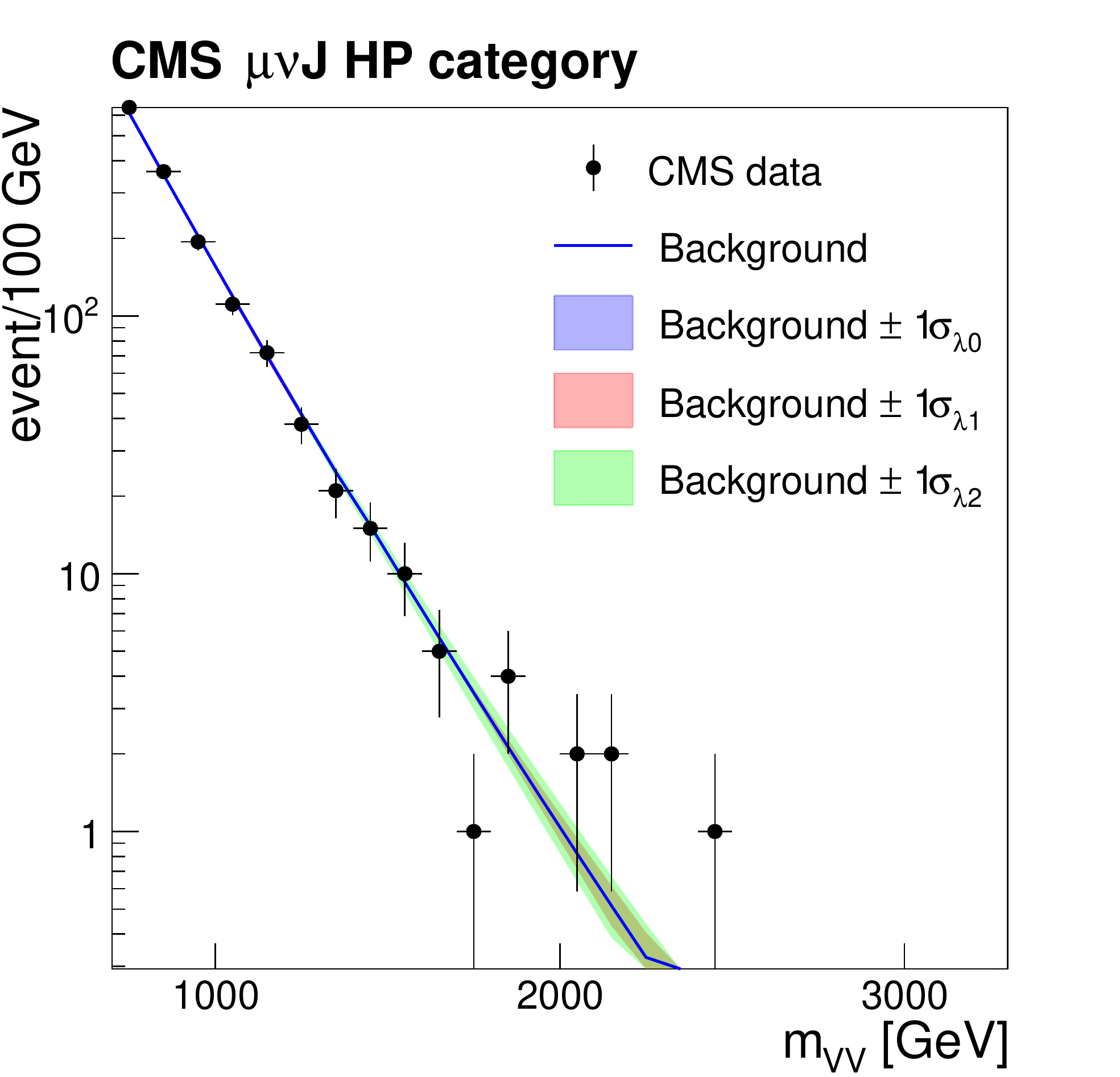}
\includegraphics[width=0.4\textwidth, angle =0 ]{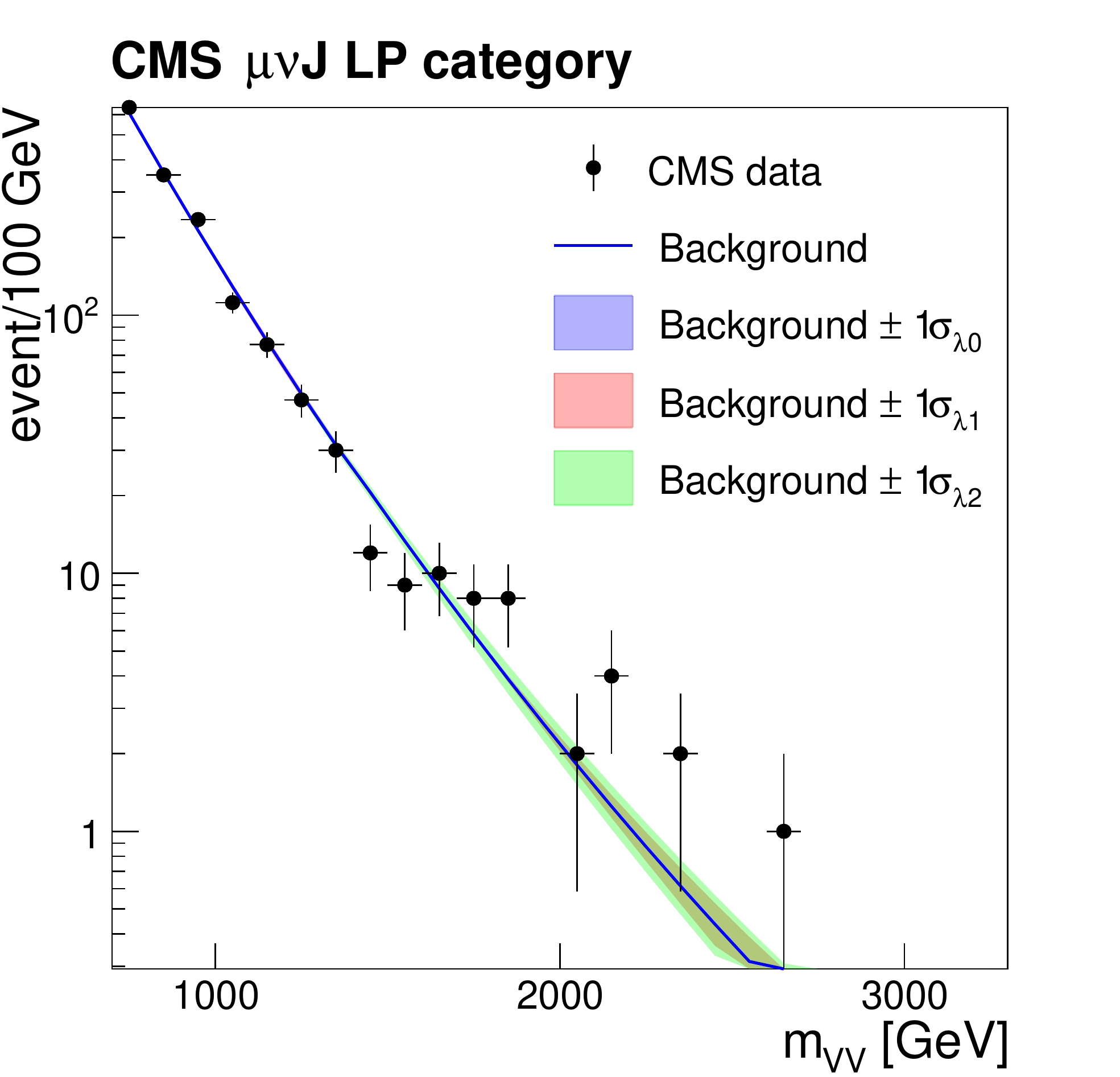}
\caption{\small CMS \VVLNJ search: Comparison between the official CMS background
  (blue line) and the background modelling with uncertainties employed by
  this study (coloured bands), with the overlaid data of the \MJJ spectrum for 
  the HP (left-hand side) and LP (right-hand side) categories, plotted
separately for the electron (top) and the muon (bottom) channels.}
\label{fig:CMSbkgsemiWW}
\end{figure*}

\begin{figure*}[htb]
\centering
\includegraphics[width=0.4\textwidth, angle =0 ]{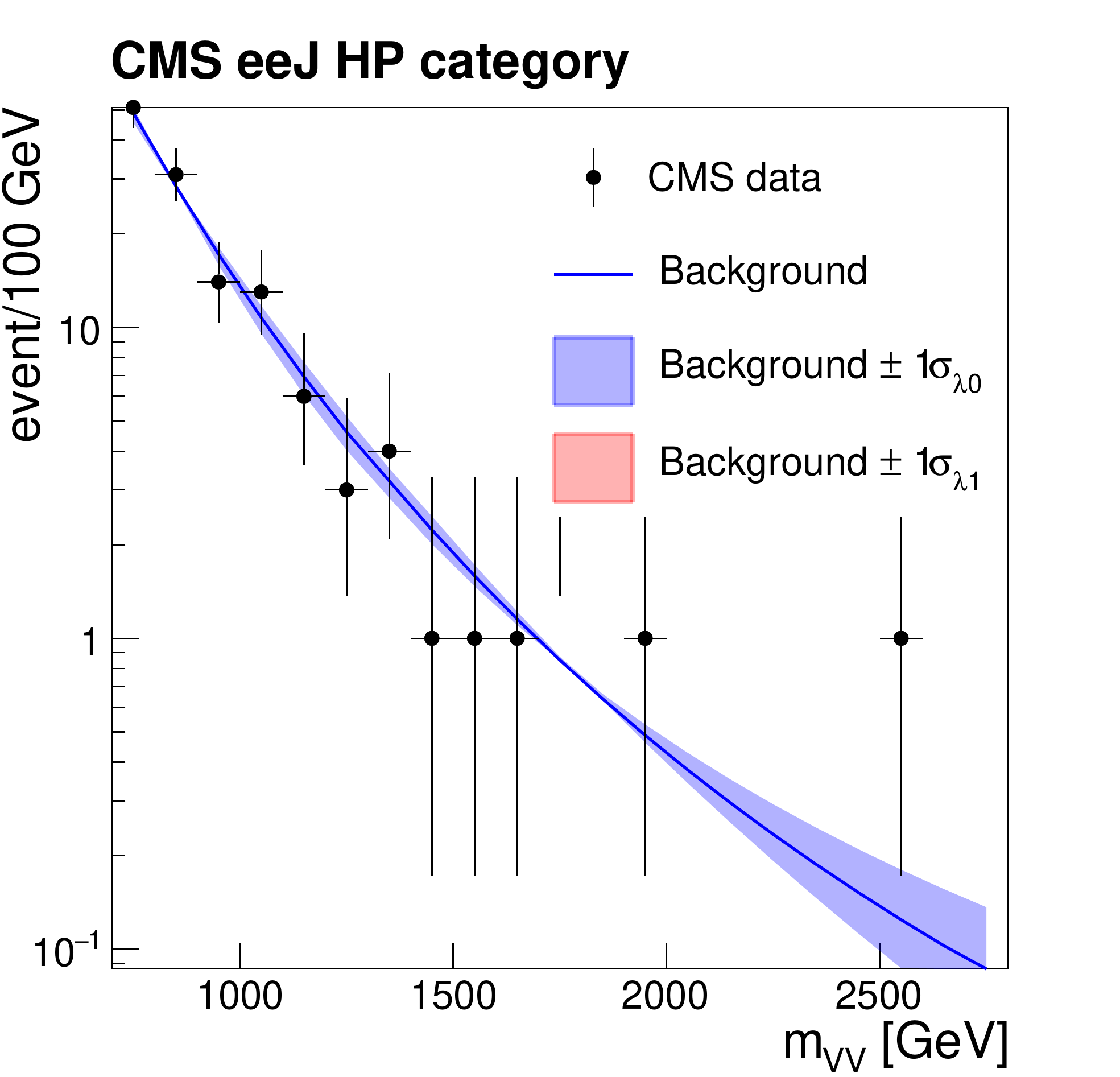}
\includegraphics[width=0.4\textwidth, angle =0 ]{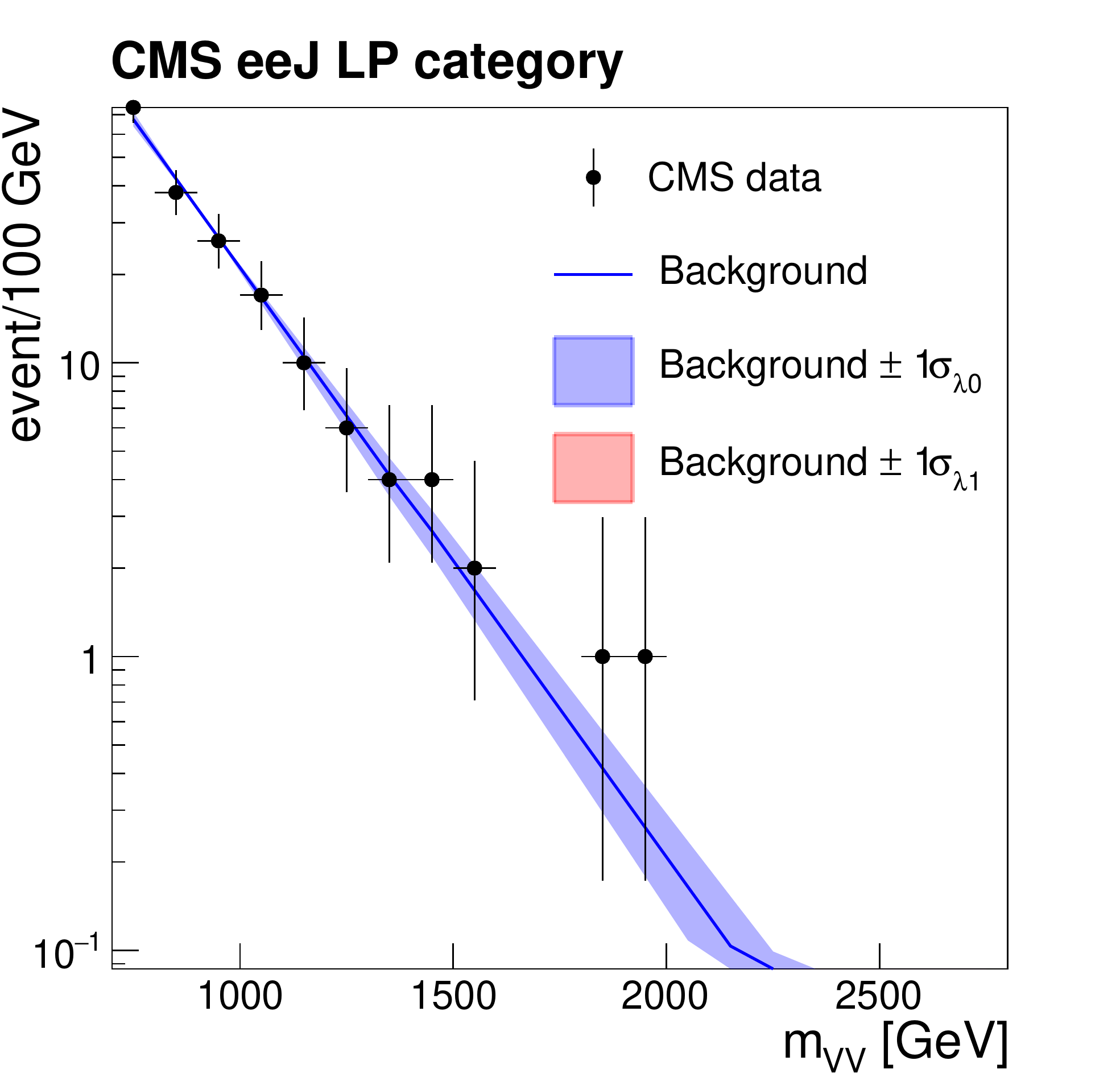}
\includegraphics[width=0.4\textwidth, angle =0 ]{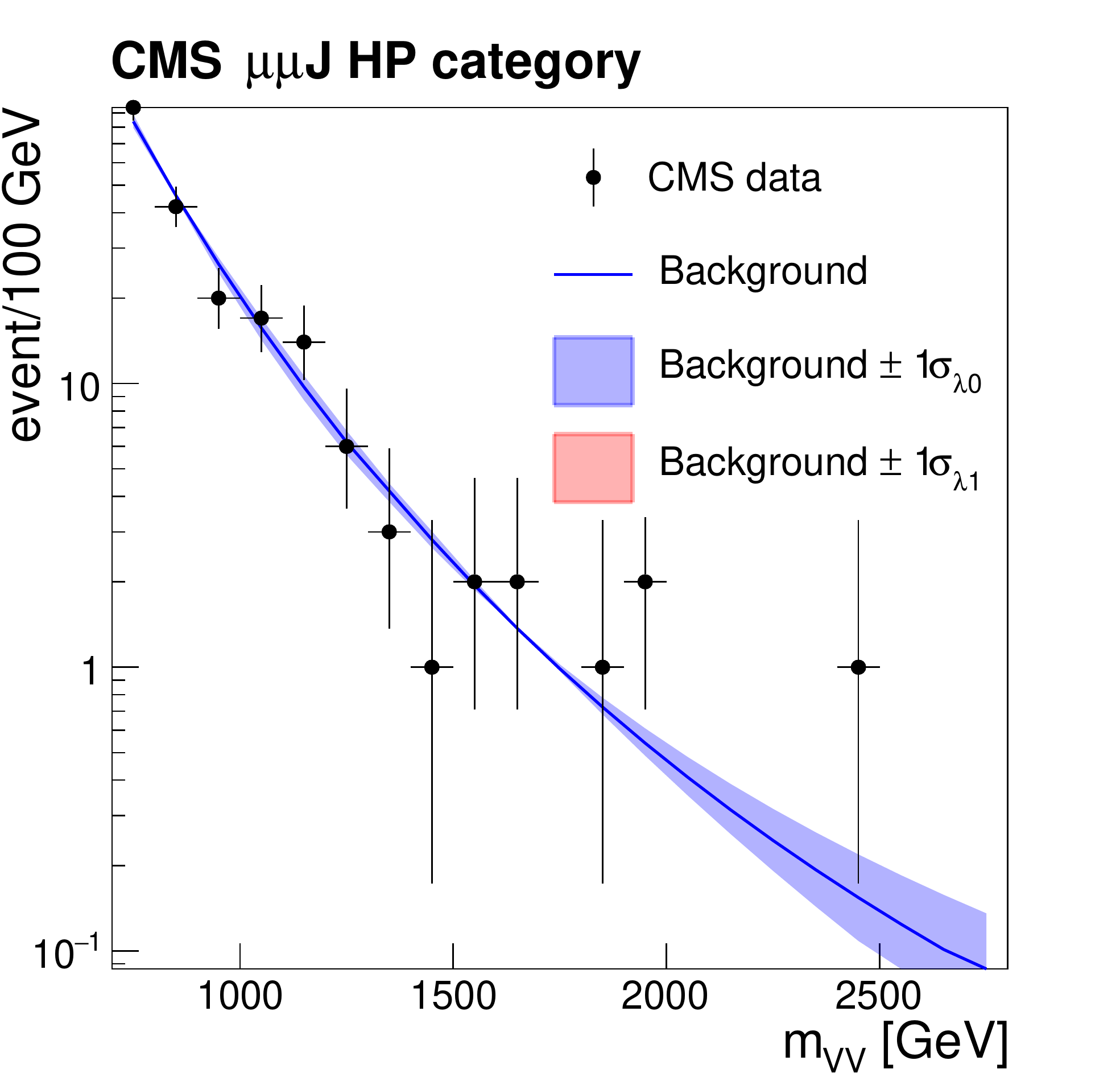}
\includegraphics[width=0.4\textwidth, angle =0 ]{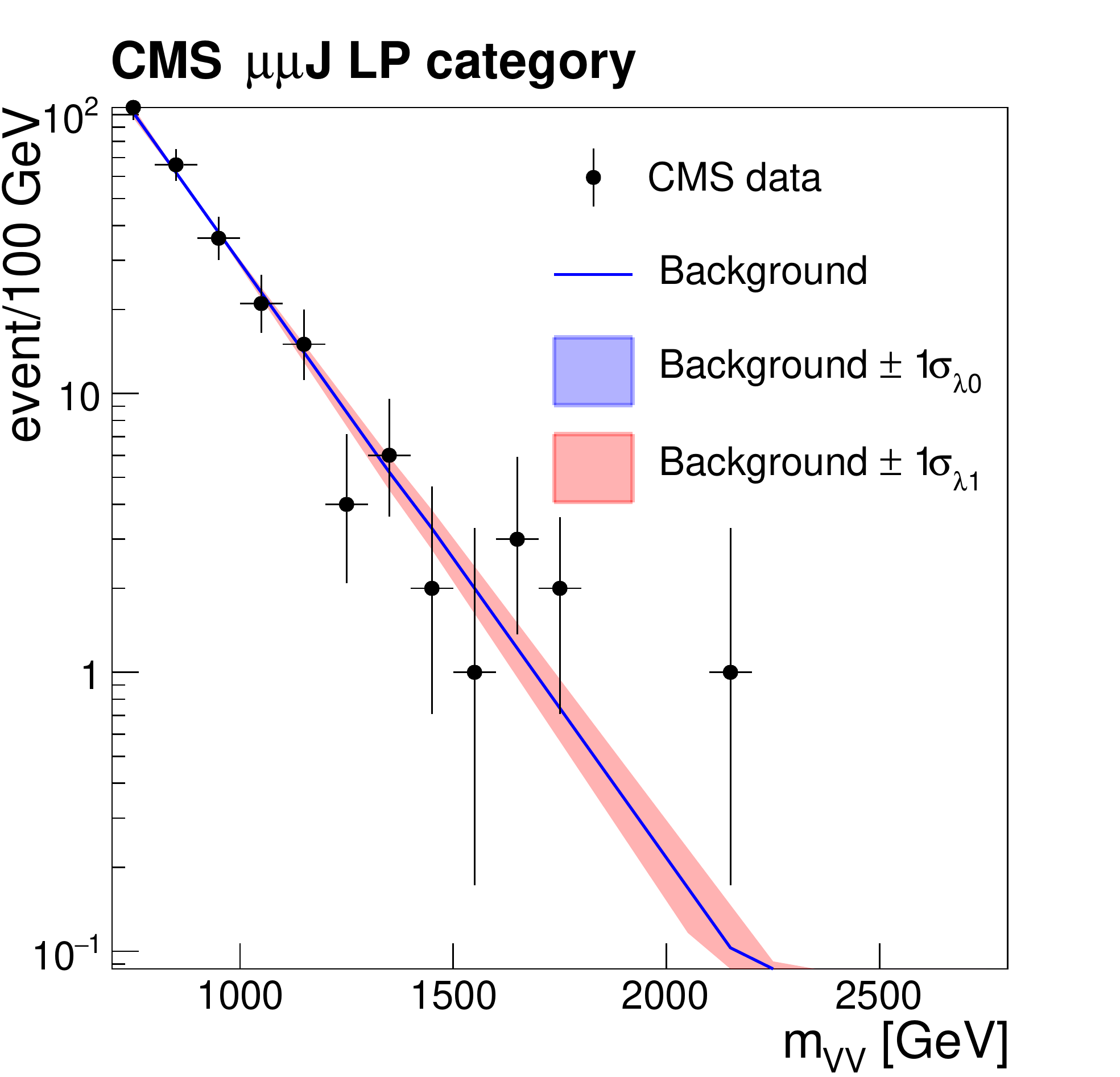}
\caption{\small
CMS \VVLLJ search: Comparison between the official CMS background
  (blue line) and the background modelling with uncertainties employed by
  this study (coloured bands), with the overlaid data of the \MJJ spectrum for 
  the HP (left-hand side) and LP (right-hand side) categories, plotted
separately for the electron (top) and the muon (bottom) channels.}
\label{fig:CMSbkgsemiZZ}
\end{figure*}

The background model is extracted by fitting the \MVV data distributions
for each lepton flavour with a levelled exponential  
\begin{equation}
f(\MVV) = N \, \exp\left [ -\frac{\MVV}{\sigma + k \cdot \MVV} \right ]
\end{equation}
where $N$, $k$ and $\sigma$ are free parameters. This function saturates
in the high $\MVV$ region, and is meant to describe events where $\MVV$ was
significantly mismeasured. For example, this may happen if a high $p_T$
muon leaves a nearly straight track barely bent by the magnetic field, or
if the calculation of the neutrino momentum fails. In practice, this
function is used in Ref. \cite{CMSZVWV} to model the HP category
with $k$ as a free parameter, whereas for the LP category $k$ can be set to 0.
In the $\LLJ$ channel we focus on the $\MVV$ > 700 GeV region, and we
merge the contents of the (publicly available) 50~GeV wide bins to obtain a
uniform, 100-GeV-wide binning for the \MVV distribution.  
We use the diagonalised uncertainties from the fit ($\sigma_{\lambda_i}$,
with $i = 0,1,2$) as
background uncertainties. Figs.~\ref{fig:CMSbkgsemiWW} and
\ref{fig:CMSbkgsemiZZ} show the comparison between the fits produced in this
study and the official CMS fits on the data distributions.

We model the signal distributions in the diboson mass spectrum with a
Gaussian function. The HP signal yield is calculated from the theoretical
cross section and the selection efficiency obtained from the algorithm
described in Ref. \cite{CMSZVWV}. The first step in this process is the
generation of signal samples with the \texttt{Madgraph5} generator as
described in Sec.~\ref{sec:method}. We then apply 
acceptance selections on the leptons and generator-level jets, and use the 2D
efficiency maps to emulate the \PV-boson reconstruction and tagging processes. Finally,
we apply a 90\%  correction to account for $b$-jet
veto inefficiencies. Considering the approximations made, this procedure is
expected to reproduce the official CMS results within a 10\% accuracy. The
HP-category efficiencies that we obtain
are consistent with the nominal \GtoWW efficiencies for $\MX = 1.2$ TeV
within 6\%. 

The LP category signal efficiencies are generally not provided, but
examples of the LP/HP efficiency ratios are given for a \PGbulk signal with
$\MX = 1.2$ TeV. The ratio is 0.47 (0.25) for
%electron and muon categories in - Christos: WHY DO MENTION ELECTRONS AND
%MUONS HERE?
the \LLJ (\LNUJ) channel. The reason for the efficiency difference between
the two cases lies in the different boosted jet selection applied in the
two channels. We make the assumption that we can use the same LP/HP ratio
for all mass points under consideration in this study, and use the values
above to estimate the expected signal yields in the LP category. Finally,
the $\tau_{21}$ categorisation is not 
sensitive to the nature of the resonance\,\footnote{Provided that the
  polarisation of the final state bosons is the same for both models.},
%To understand better the impact of the boosted jet tagger on the efficiency
% to tag the V polarisation},
therefore we use the same LP/HP ratio also for the \PWp signal hypothesis.

We consider the following systematic uncertainties, treated as fully
correlated across \MJJ histogram bins: 
\begin{itemize}
  \item {\em Background uncertainty}, extracted from our fit to the data distributions.
  \item {\em Signal normalisation uncertainty}, which is separated into two
    further sub-categories:  a
    common-across-channels systematic uncertainty corresponding to the
    luminosity measurement (2.2\%), and an additional uncertainty covering
    all lepton-related uncertainties (3.7\%
    for electrons, 3\% for muons), applied separately for the $\LLJ$ and
    $\LNUJ$ channels. 
  \item {\em Signal purity category migration uncertainty}, which covers the
    effects of events ``migrating'' from the HP to the LP category, or
    vice-versa. This uncertainty amounts to 9\% and 24\%, respectively.
\end{itemize}

\begin{figure*}[htb]
\centering
%\includegraphics[width=0.5\textwidth, angle =0
%]{../limits_comparison/CMSsemi_paper_fudge.pdf}
\includegraphics[width=0.5\textwidth, angle =0 ]{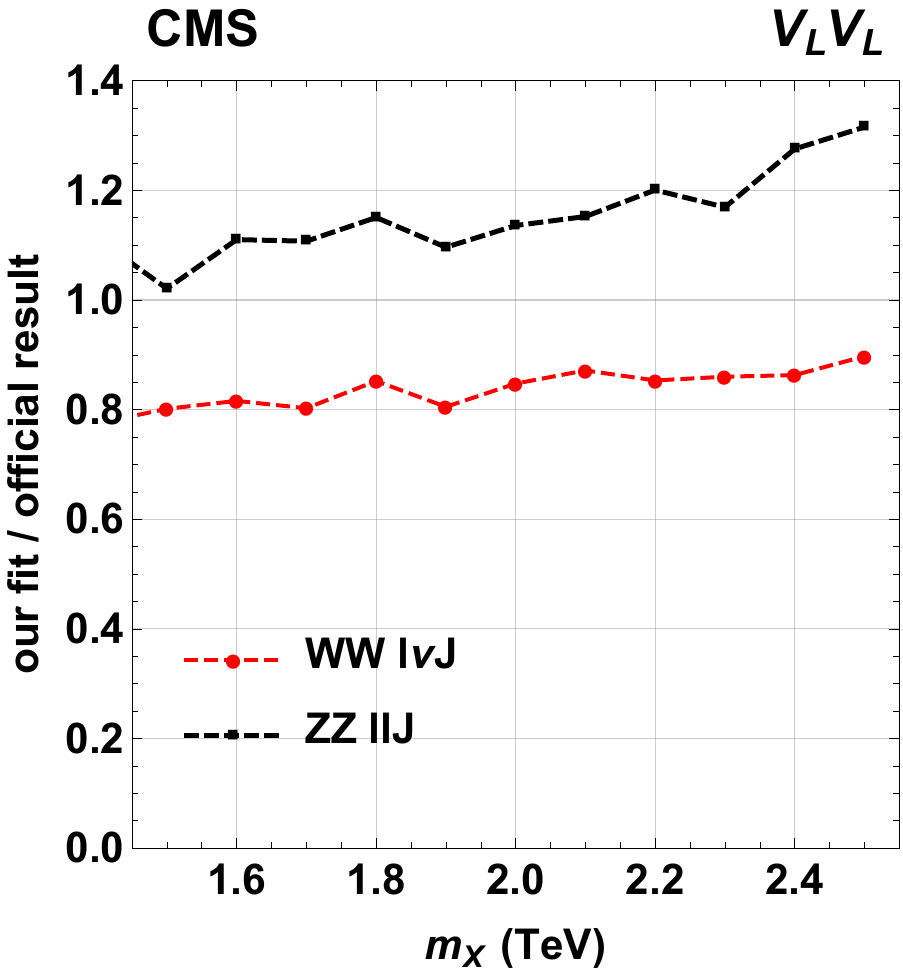}
\caption{\small CMS semileptonic searches: Fudge factor as a
  function of the mass $m_X$ of the exotic resonance, calculated via the 
ratio of observed exclusion limits obtained with this study to the ones of
the official CMS result for the \GtoWW (red) and \GtoZZ (black) semileptonic analyses.
\label{fig:CMSsemiNomA}
}
\end{figure*}

\begin{figure*}[htb]
\centering
\includegraphics[width=0.45\textwidth, angle =0 ]{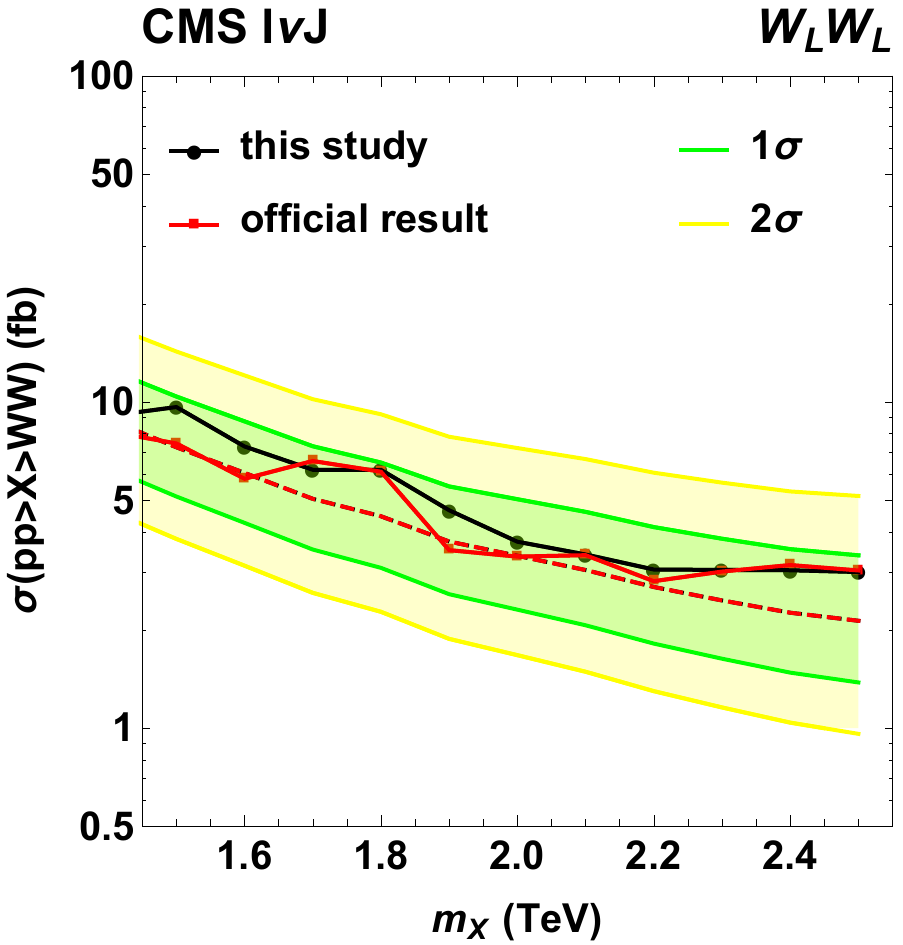}
\includegraphics[width=0.45\textwidth, angle =0 ]{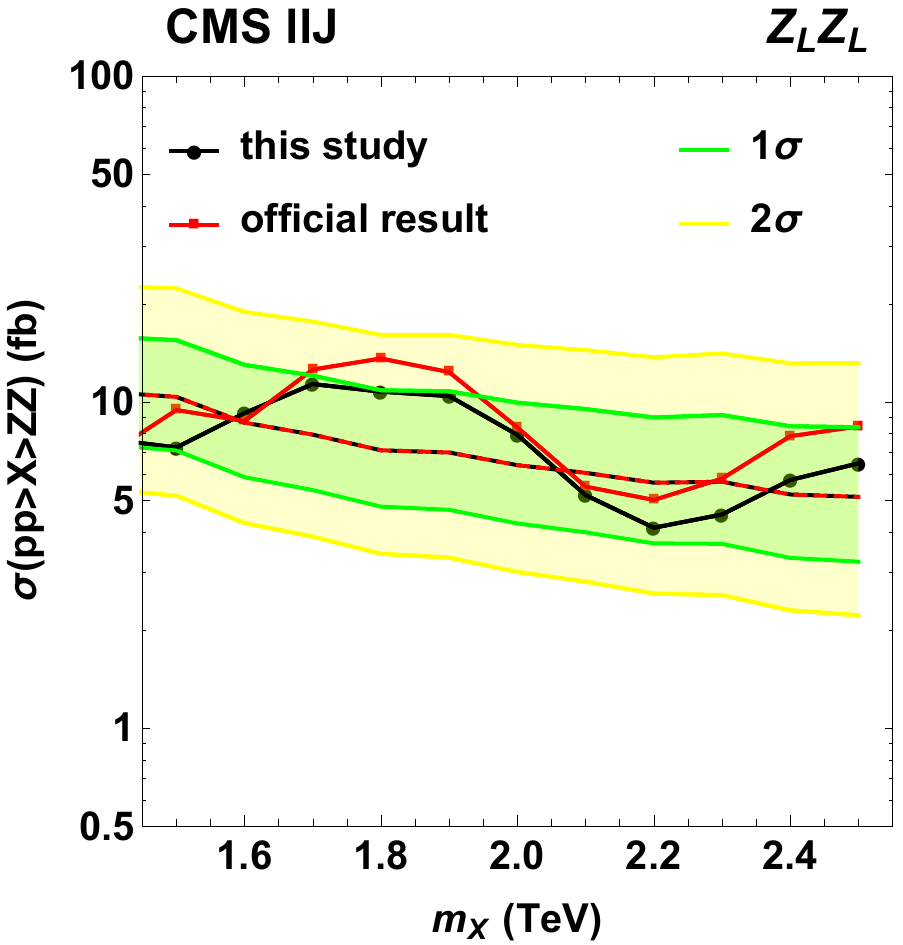}
\caption{\small CMS semileptonic searches:
Expected (dashed lines) and observed (continuous lines) exclusion limits on exotic
  production cross sections as a function of the
  resonance mass $m_X$ obtained with this study (black), and comparison
  with the official CMS results (red) for the \GtoWW search in the \LNUJ
  channel (left) and the \GtoZZ search in the
  \LLJ channel (right). 
The green and yellow bands represent
  the one and two sigma variations around the median expected limits
  calculated in this study, with all the corrections described in the text included. 
\label{fig:CMSsemiNomB}
}
\end{figure*}

As already discussed in previous sections, we apply a fudge factor to
account for differences between our background description and the one from
the public CMS result, as well as for the approximations introduced in the
signal modelling (Fig.~\ref{fig:CMSsemiNomA}). With this correction,
our calculated exclusion limits are in good agreement with the public CMS
results (Fig. \ref{fig:CMSsemiNomB}). The statistical uncertainties
(one- and two-sigma coverage bands) are $\approx 50\%$ smaller than
expected, as they have been calculated with the asymptotic CLs method,
which is known to underestimate uncertainties in tests with small
statistics.  

\begin{figure*}[htb]
\centering
\includegraphics[width=0.4\textwidth, angle =0 ]{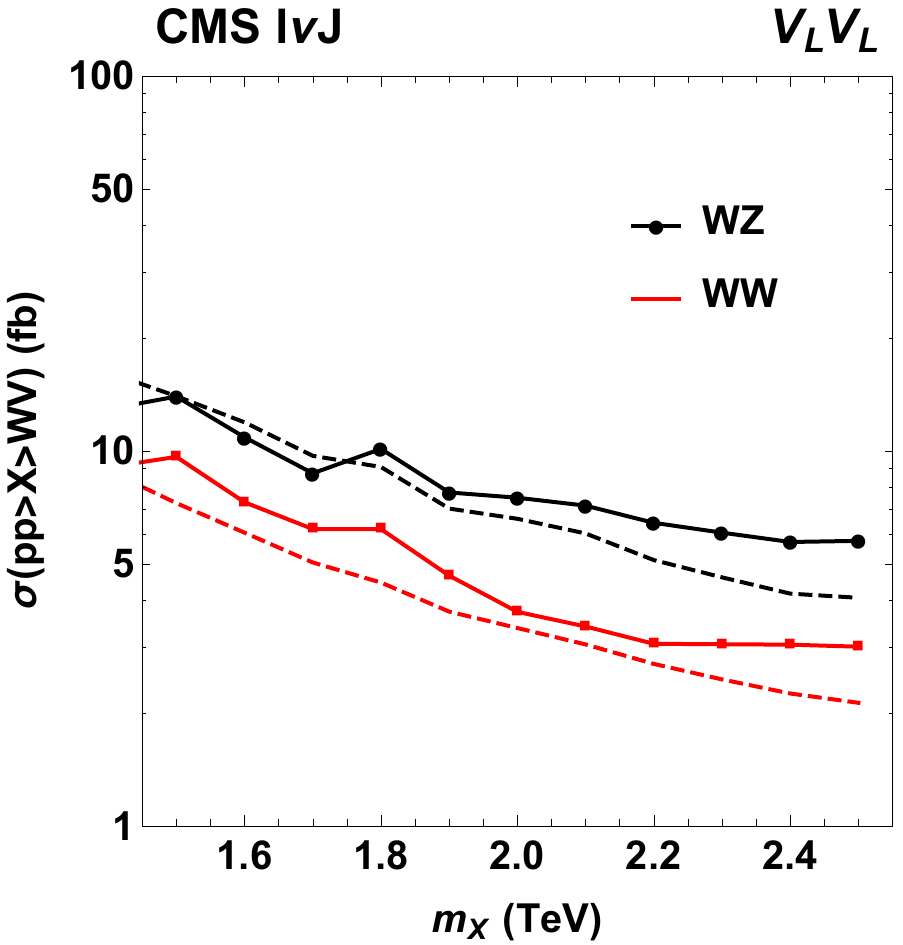}
\includegraphics[width=0.4\textwidth, angle =0 ]{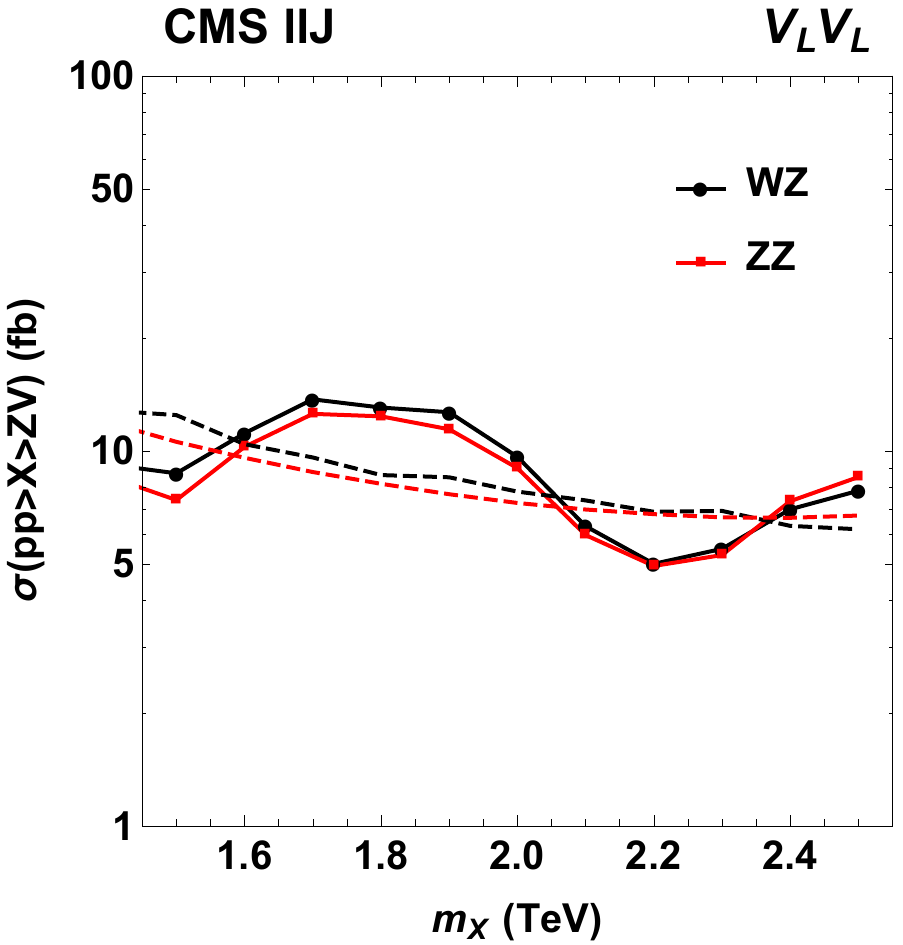}
\caption{\small CMS semileptonic searches:
Expected (dashed lines) and observed (continuous lines) exclusion limits on
exotic production cross section  as a function of the resonance mass $m_X$
obtained with this study for the \GtoWW (red) and \PWp (black) signal
hypotheses in the \LNUJ channel (left) and for the \GtoZZ (red) and \PWp
(black) signal hypotheses in the \LLJ channel (right).
\label{fig:CMSsemiAlter}
}
\end{figure*}

We use the same procedure to recast the results in the context of a \WptoWZ
signal search, with the results presented in Fig.~\ref{fig:CMSsemiAlter}.
%
%As we assumed that the most important effect that contributes to the fudge factors comes from our modelling of the background predictions, we include the respective fudge factors in the limit setting for each channel. 
%
%
The jet mass selection for the \LLJ channel is $70< m_J <110$~GeV, to be
compared with $65< m_J <105$~GeV for the \LNUJ analysis. This choice was
made in order to optimise the search for a neutral resonance (at the expense of
the search for a charged one). Since the \LNUJ channel mass window is shifted to a region with more background, the signal sensitivity for the \LNUJ channel is reduced.

\subsection{Combined LHC results of semi-leptonic searches}
\begin{figure*}[htb]\begin{center}
\includegraphics[width=0.4\textwidth, angle =0 ]{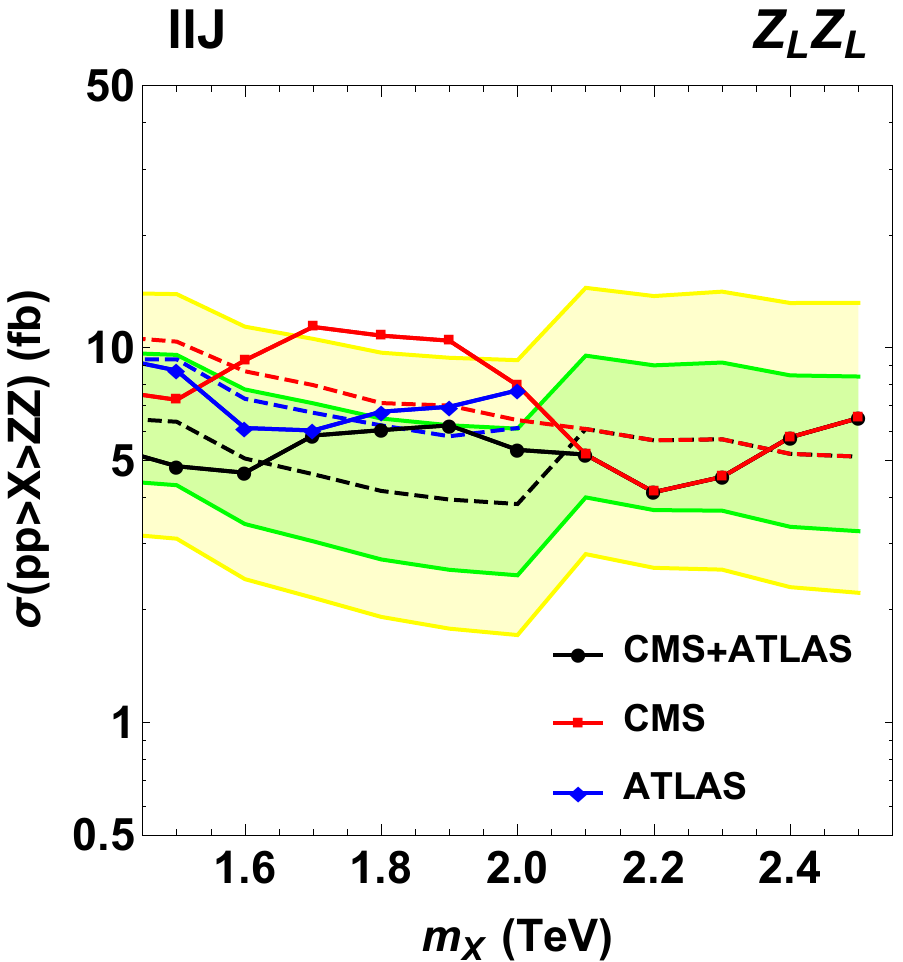}
\includegraphics[width=0.42\textwidth, angle =0 ]{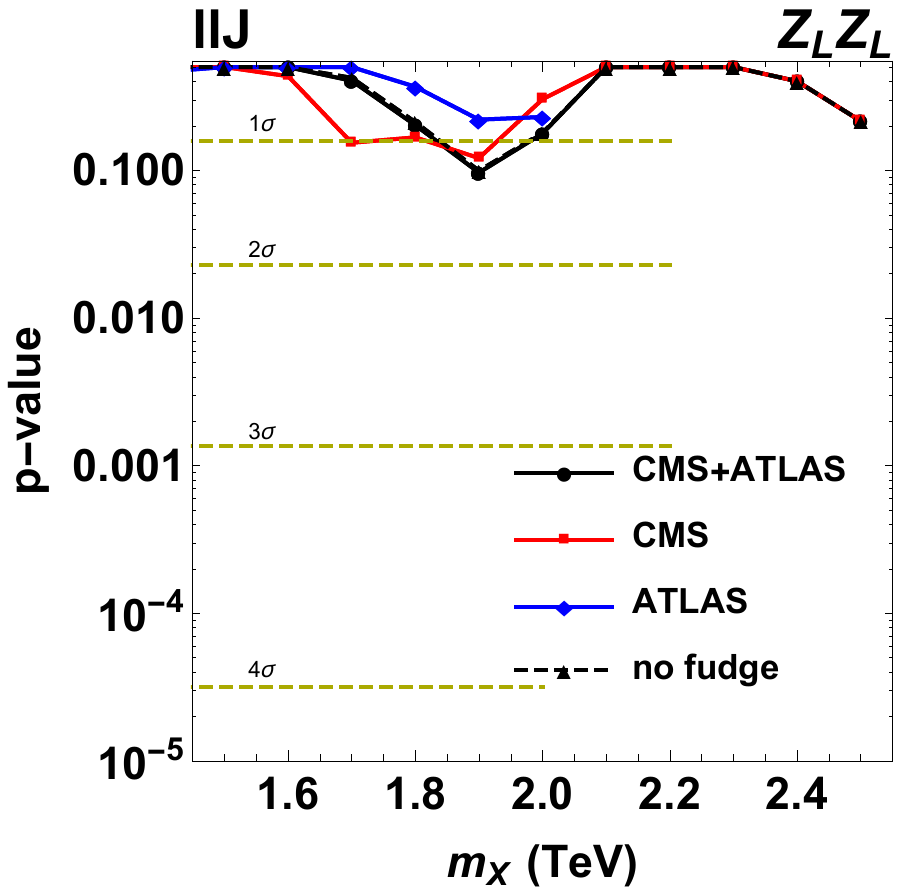}
\includegraphics[width=0.4\textwidth, angle =0 ]{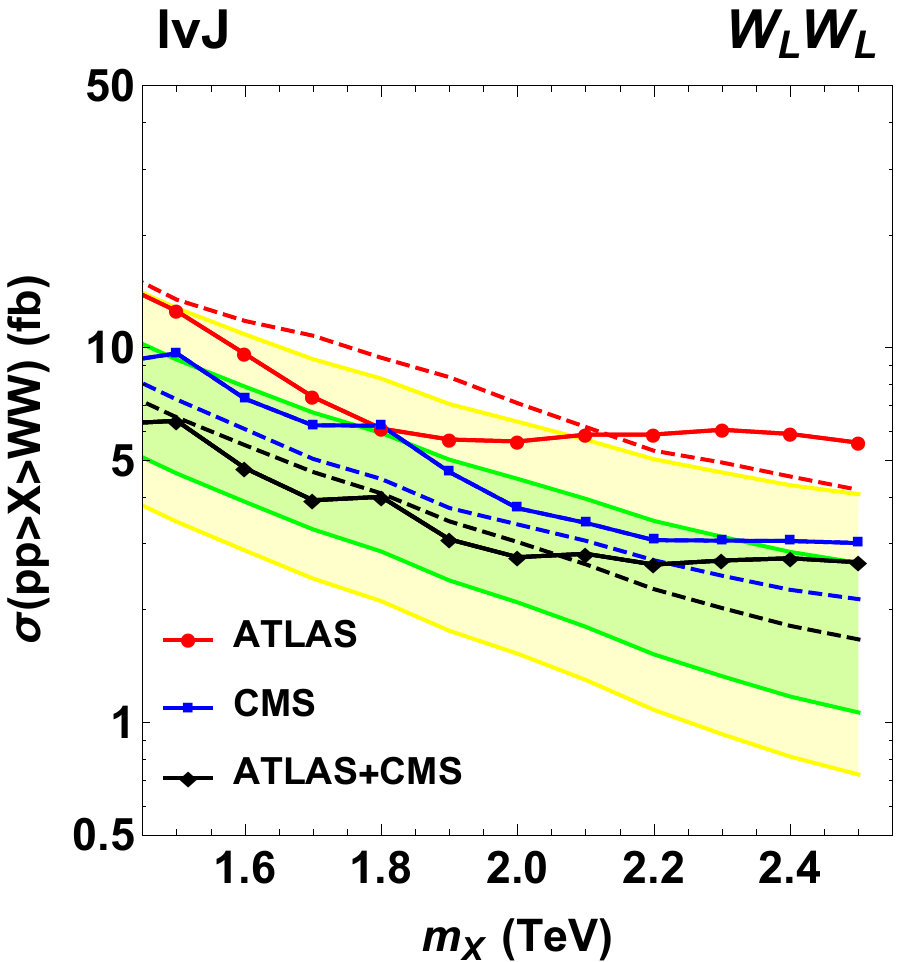}
\includegraphics[width=0.42\textwidth, angle =0 ]{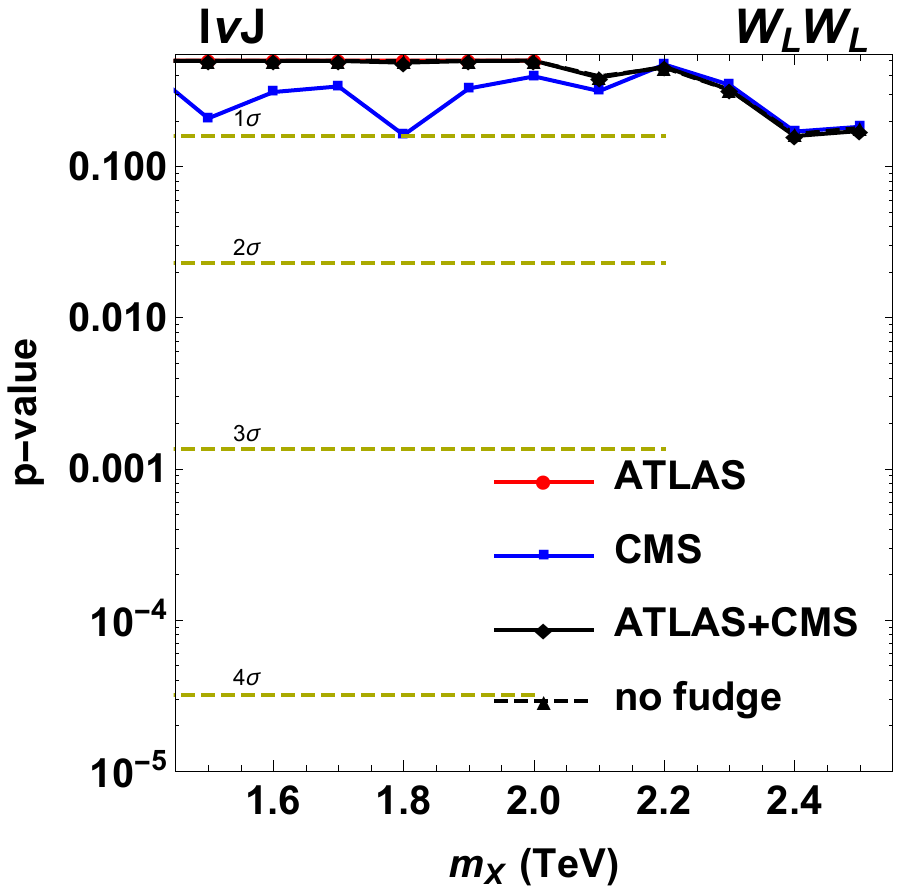}
\includegraphics[width=0.4\textwidth, angle =0 ]{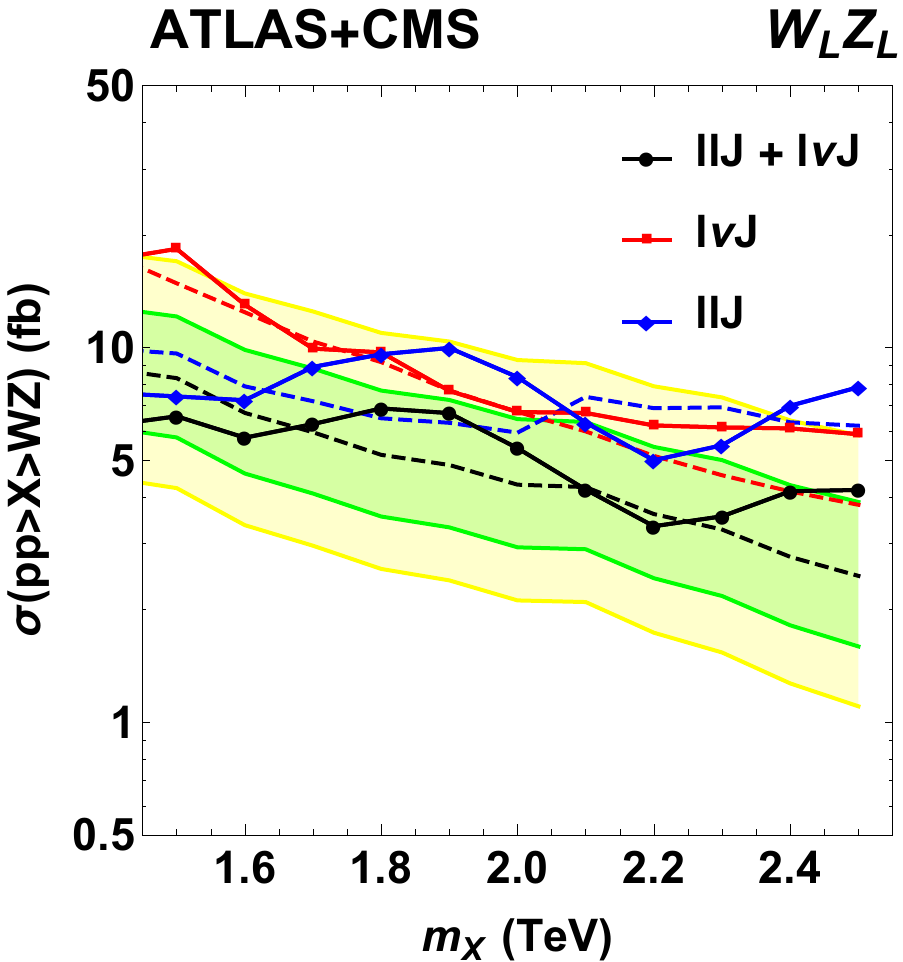}
\includegraphics[width=0.42\textwidth, angle =0 ]{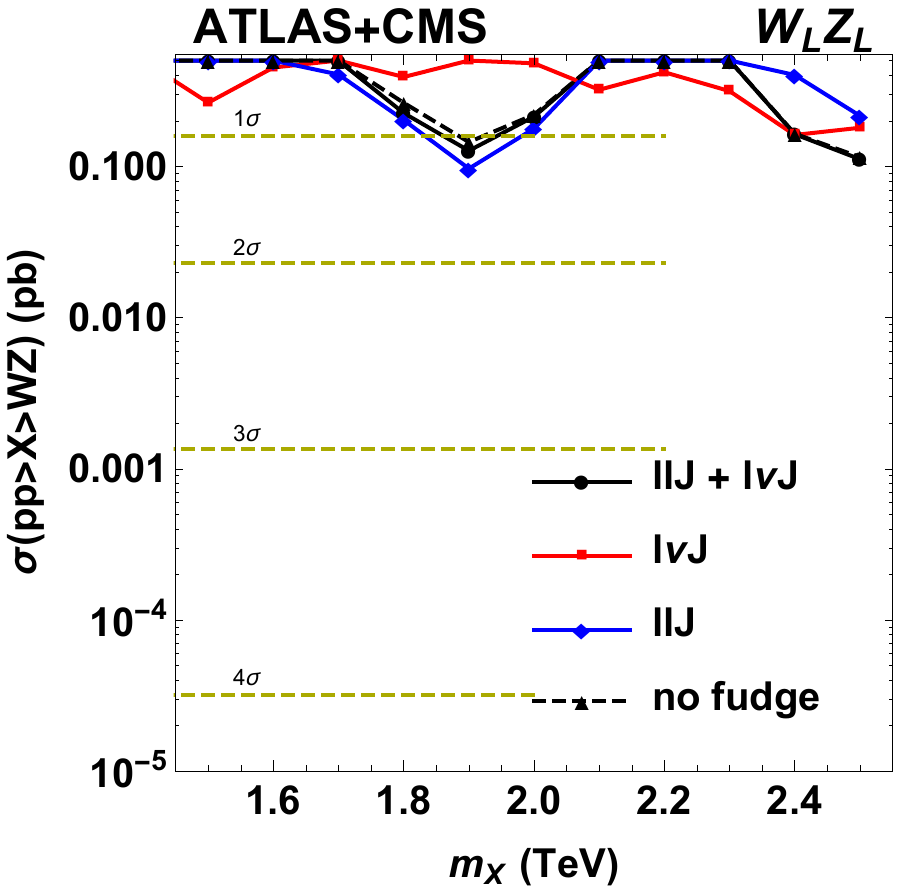}
\caption{\small Combination of semileptonic searches for \GtoZZ (top),
  \GtoWW (middle) and \WptoWZ  (bottom) selections and signal hypotheses,
  and as a function of the resonance mass $\MX$ obtained with the emulation of the ATLAS
(red) and CMS (blue) searches and their  
  combination (black). {\bf Left:} Expected (dashed lines) and observed
  (continuous lines) exclusion limits on 
exotic production cross section. The green and yellow bands
  represent the one and two sigma variations around the median expected
  limits. The results include the correction factors discussed in the text.
 {\bf Right:} Likelihood ratio $p$-values. The dashed black curve
 corresponds to the combined search without the corrections discussed in
 the text. 
\label{fig:semiCombo}}
\end{center}\end{figure*}

Here we discuss the combination of the ATLAS and CMS searches in the
semileptonic channels (\LNUJ and \LLJ) and the interpretation of the
results under different signal hypotheses, with final states including a
leptonic \PW ($\to \ell \nu$) or \PZ ($\to \ell \ell$) decay. The results are
summarised in Fig. \ref{fig:semiCombo}.

Under the hypothesis of a  \ZLZL benchmark model, only the $\LLJ$ searches
are relevant. In this channel, CMS observes a small excess ($\approx 1
\sigma$) between 1.7 and 1.9 TeV, while ATLAS a $< 1 \sigma$ excess
between 1.9 and 2.0 TeV, driven by the presence of one event in the highest bin
of the merged analysis distribution. The combination of the two channels
results in a more stringent limit and a moderate excess of the order of
$1\sigma$ around 1.9 TeV. Above 2 
TeV, ATLAS has not published their search results and the limit considered
here is the one provided by CMS. While the significance of the observed
deviation is too small to
cause any excitement, the sensitivity of this analysis is strongly
reduced. This has implications for the combination result discussed in
Sec.~\ref{sec:combo}. 

On the contrary, under the hypothesis of a \WLWL benchmark model, only the
$\LNUJ$ searches are relevant. An observed upward fluctuation around \MVV =
1.8 TeV in the CMS data spectrum is compensated by a downward 
fluctuation in the same region for the ATLAS data. The two deviations
effectively cancel each other, resulting into observed exclusion limits
which are consistent with the experimental sensitivity and the background-only
hypothesis expectations. 

For the \WLZL benchmark model, we are able to combine the experimental
results in the $\LLJ$ and $\LNUJ$ channels. The sensitivity and the
relative weight of the \LLJ channel is larger than those of the \LNUJ channel in the combination. 
Similar to the interpretation of the search results in the \ZLZL signal
hypothesis, we observe here that the combined results give a small excess
($\approx 1 \sigma$) around $\MVV$ = 1.9 TeV.  

%\clearpage
\section{Combination of hadronic and semi-leptonic channels}
\label{sec:combo}

This section is dedicated to the combination of both hadronic and
semileptonic channels by ATLAS and CMS under different signal hypotheses. The searches in the \JJ and \LNUJ channels contribute to constrain a
hypothetical \GtoWW  production; searches in the \JJ 
and \LLJ channels enter the combination for the interpretation of the
results in a \GtoZZ signal scenario. Finally, all six searches (\ie 
results in three channels by two 
experiments) enter the combination in the \WptoWZ signal hypothesis.

\begin{figure*}[htb]\begin{center}
\includegraphics[width=0.3\textwidth, angle =0 ]{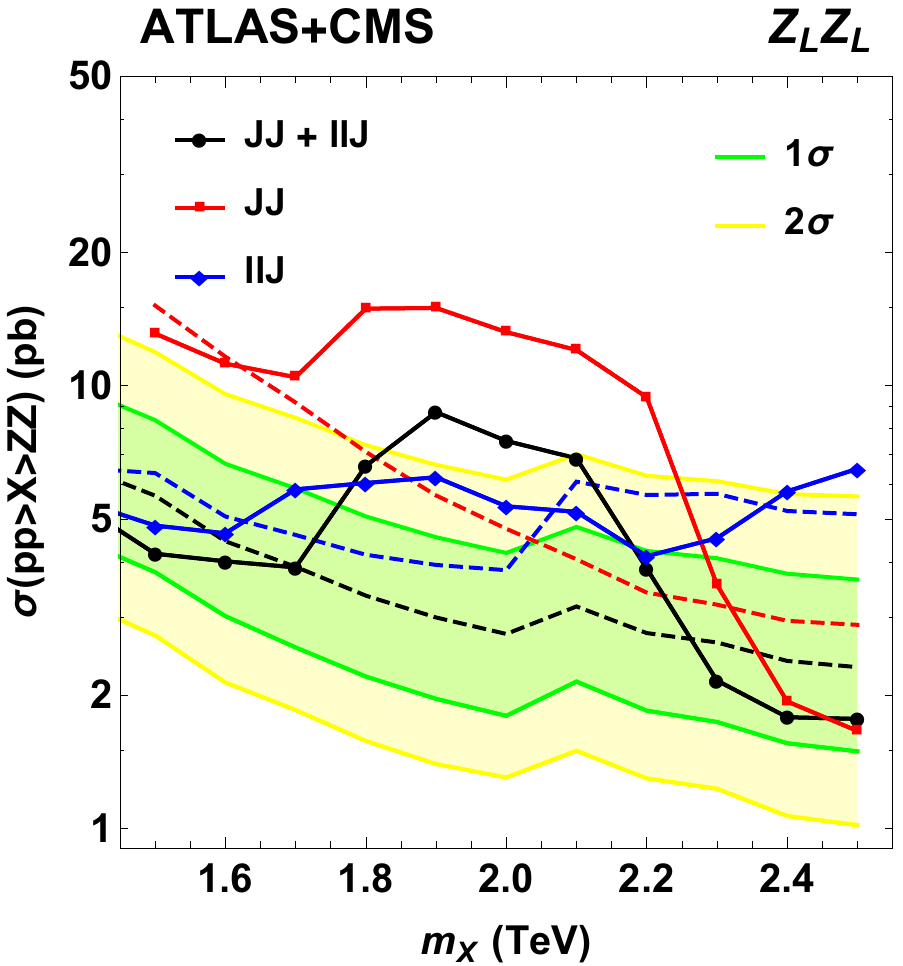}
\includegraphics[width=0.32\textwidth, angle =0 ]{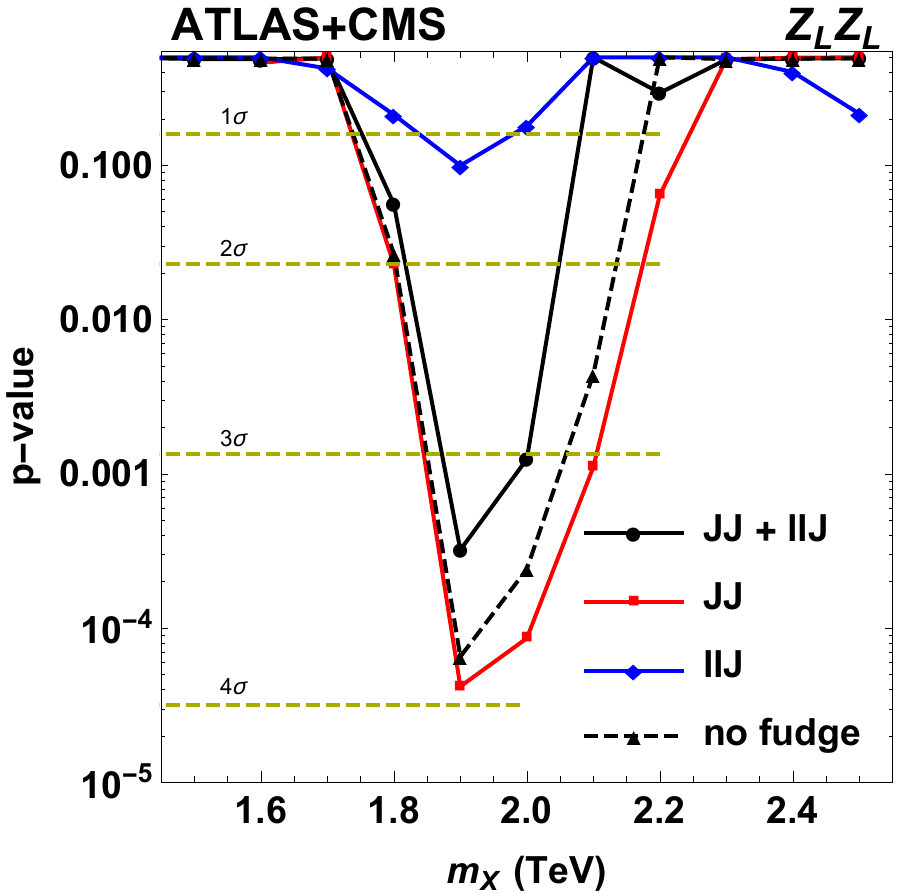}
\includegraphics[width=0.3\textwidth, angle =0 ]{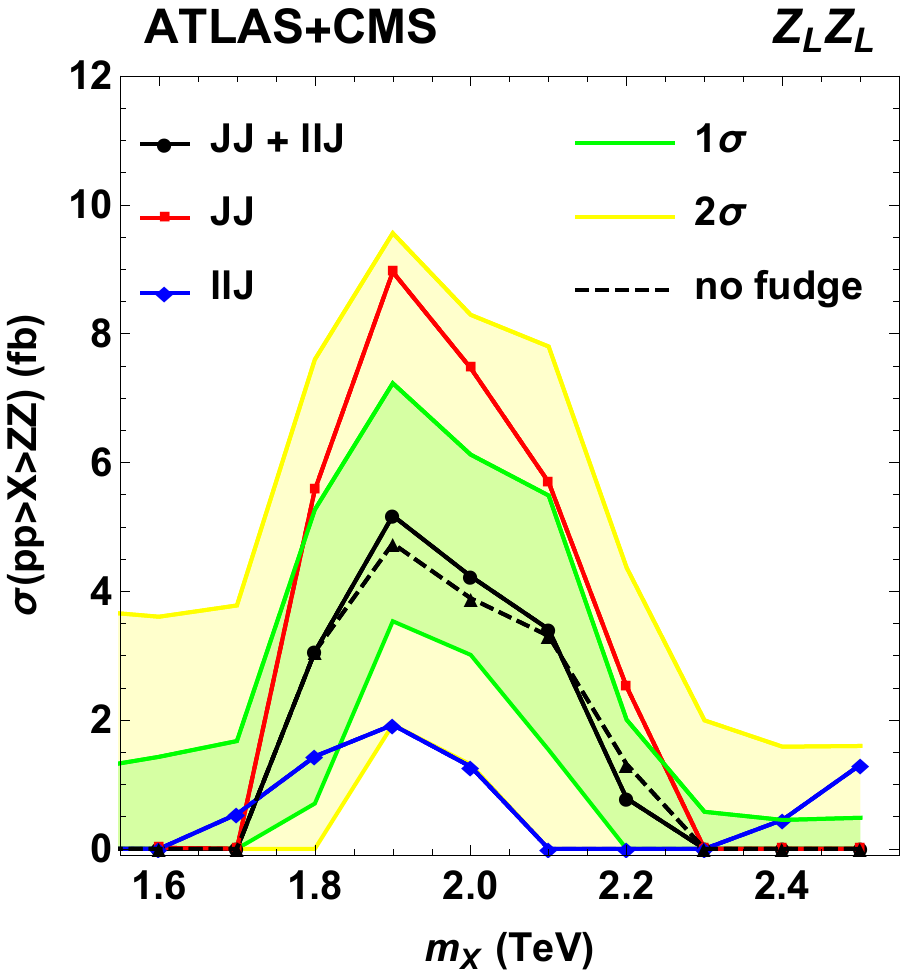}\\
\includegraphics[width=0.3\textwidth, angle =0 ]{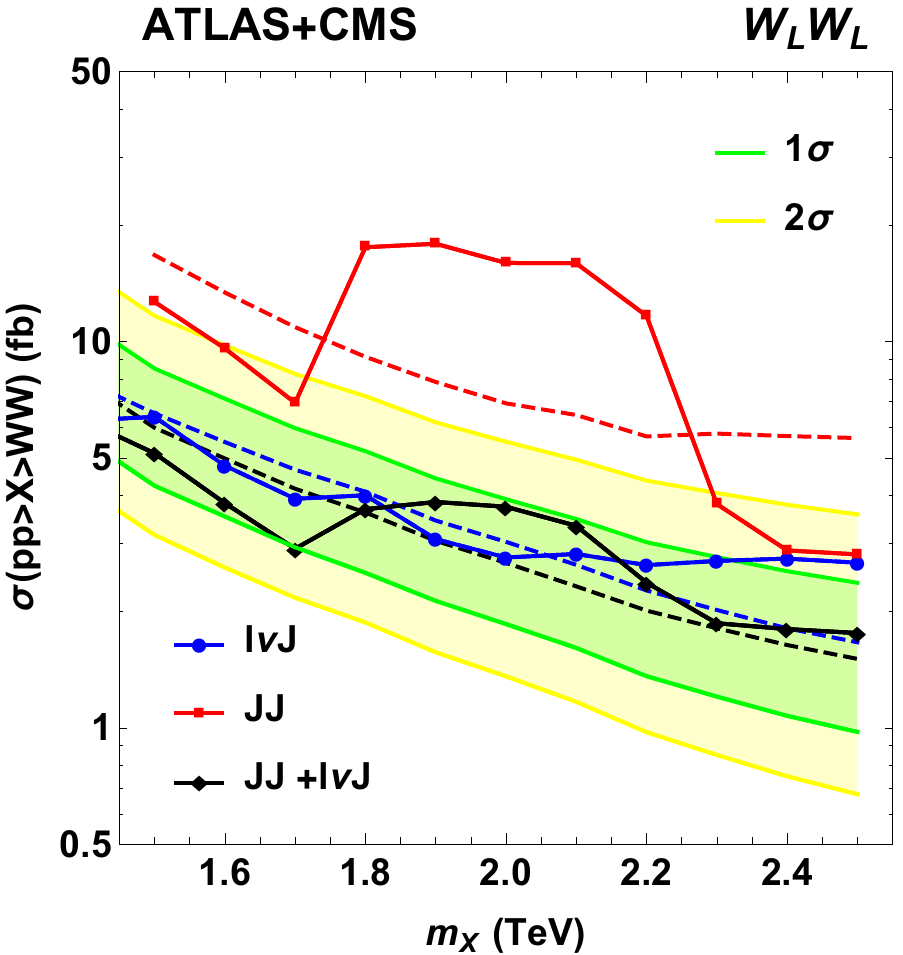}
\includegraphics[width=0.32\textwidth, angle =0 ]{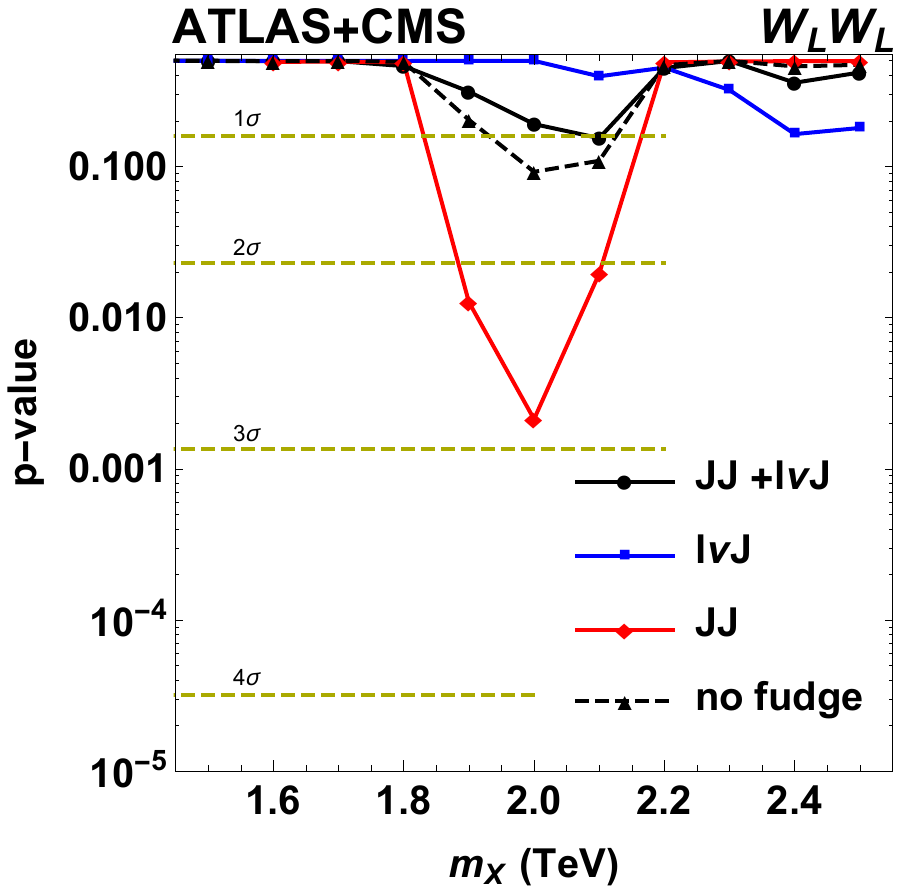}
\includegraphics[width=0.3\textwidth, angle =0 ]{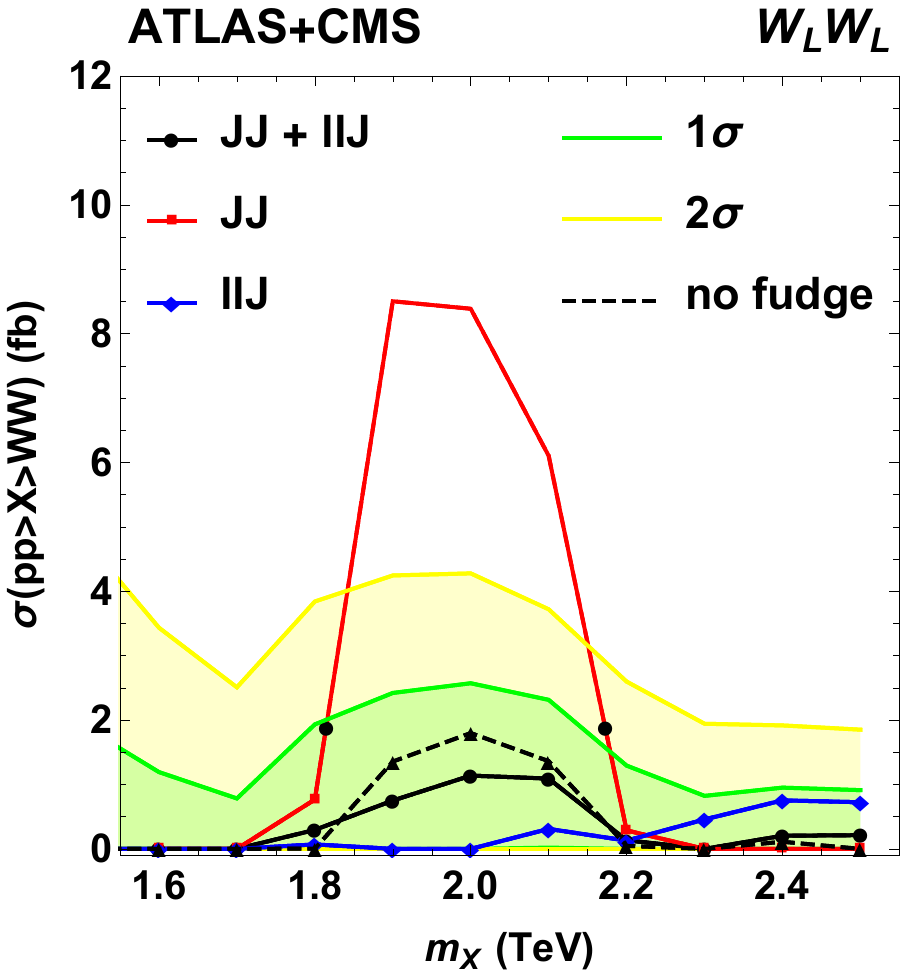}\\
\includegraphics[width=0.3\textwidth, angle =0 ]{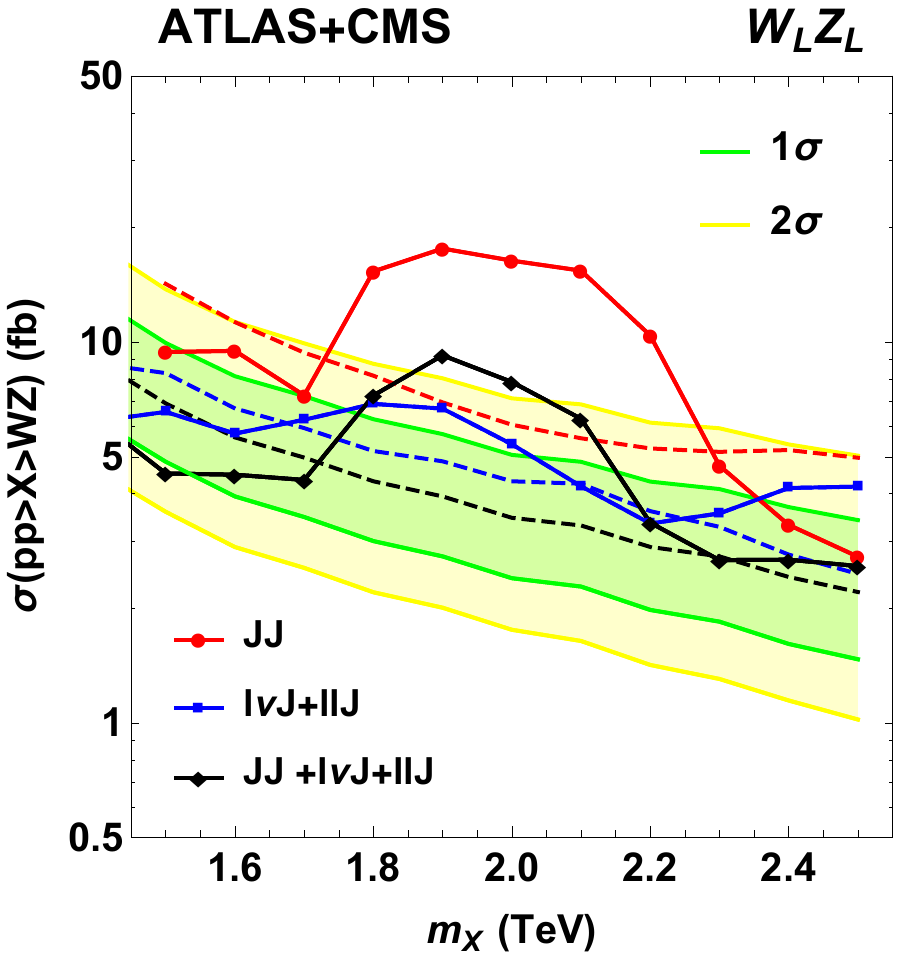}
\includegraphics[width=0.32\textwidth, angle =0 ]{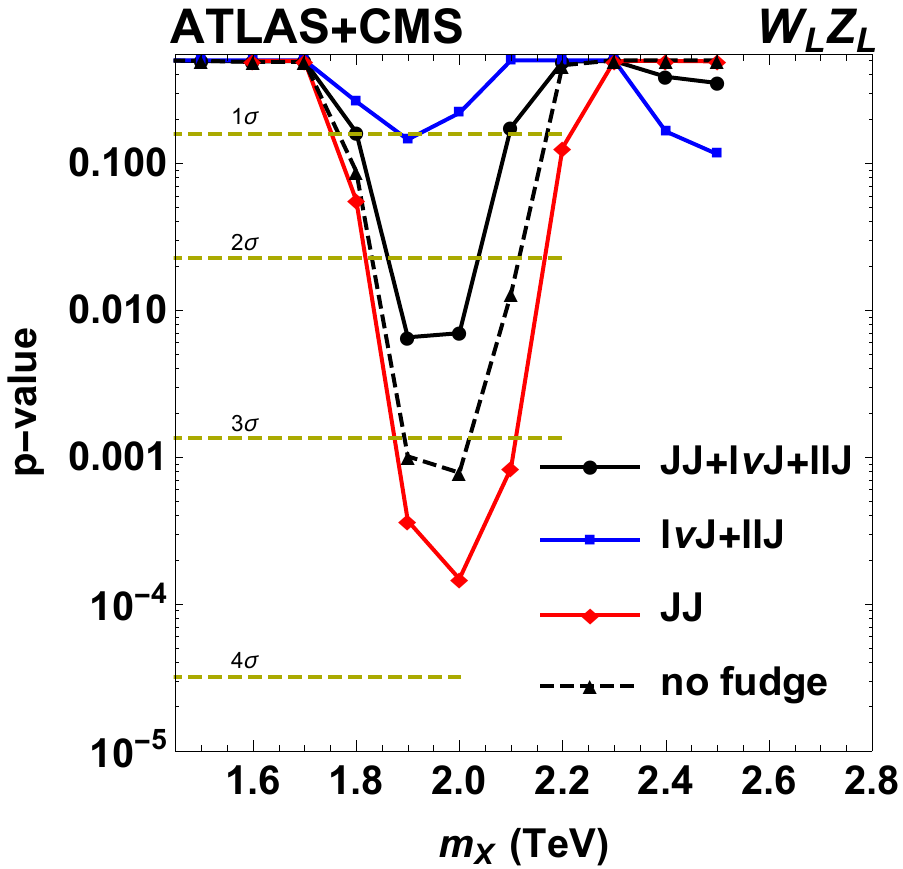}
\includegraphics[width=0.3\textwidth, angle =0 ]{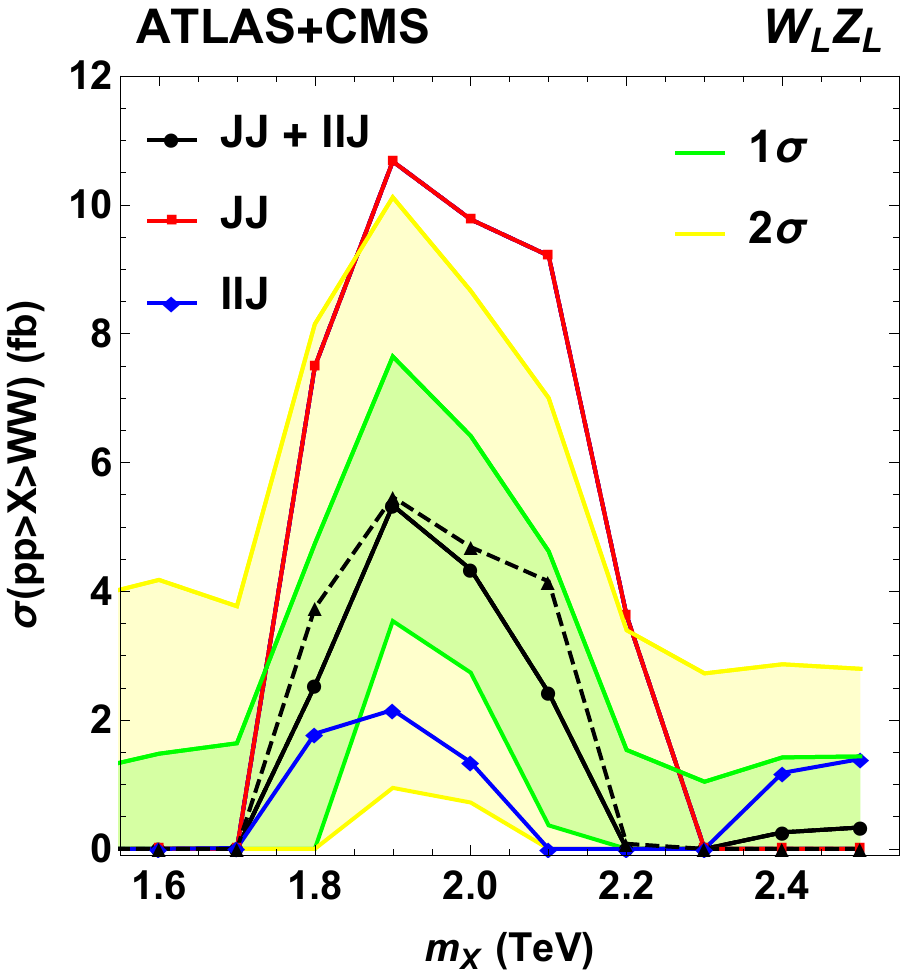}
\caption{\small Combination of all ATLAS and CMS resonance searches for \GtoZZ (top),
  \GtoWW (middle) and \WptoWZ  (bottom) selections and signal hypotheses,
  and as a function of the resonance mass $\MX$ carried out in the hadronic
  (red) and semileptonic (blue) channels and their  
  combination (black). The results include all correction factors discussed
  in the text. {\bf Left:} Expected (dashed lines) and observed
  (continuous lines) exclusion limits on 
exotic production cross section. The green and yellow bands
  represent the one and two sigma variations around the median expected
  limits. 
 {\bf Middle:} Likelihood ratio $p$-values. The dashed black curve
 corresponds to the combined search without the corrections discussed in the text. 
{\bf Right: } Best fitted exotic
  production cross section. The green and yellow bands
  represent the one and two sigma variations around the median 
  values. 
    \label{COMBO_VV_llJ_JJ}}
\end{center}\end{figure*}

\begin{figure*}[htb]\begin{center}
\includegraphics[width=0.32\textwidth, angle =0 ]{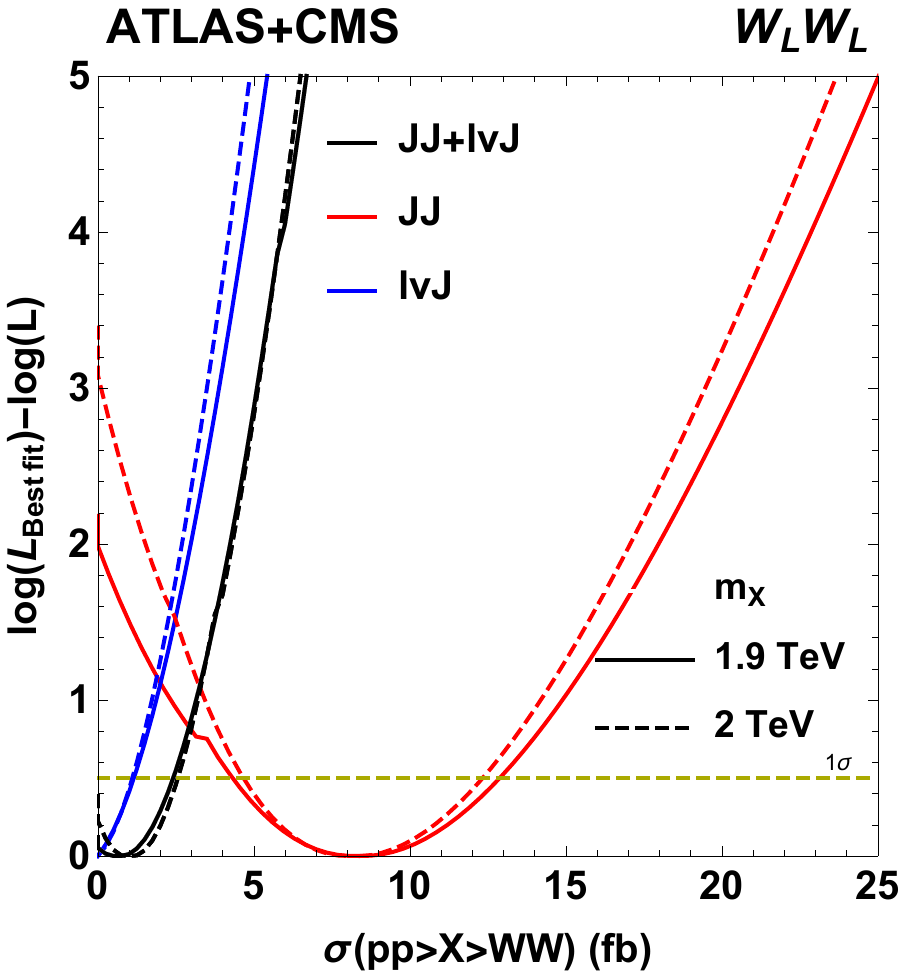}
\includegraphics[width=0.32\textwidth, angle =0 ]{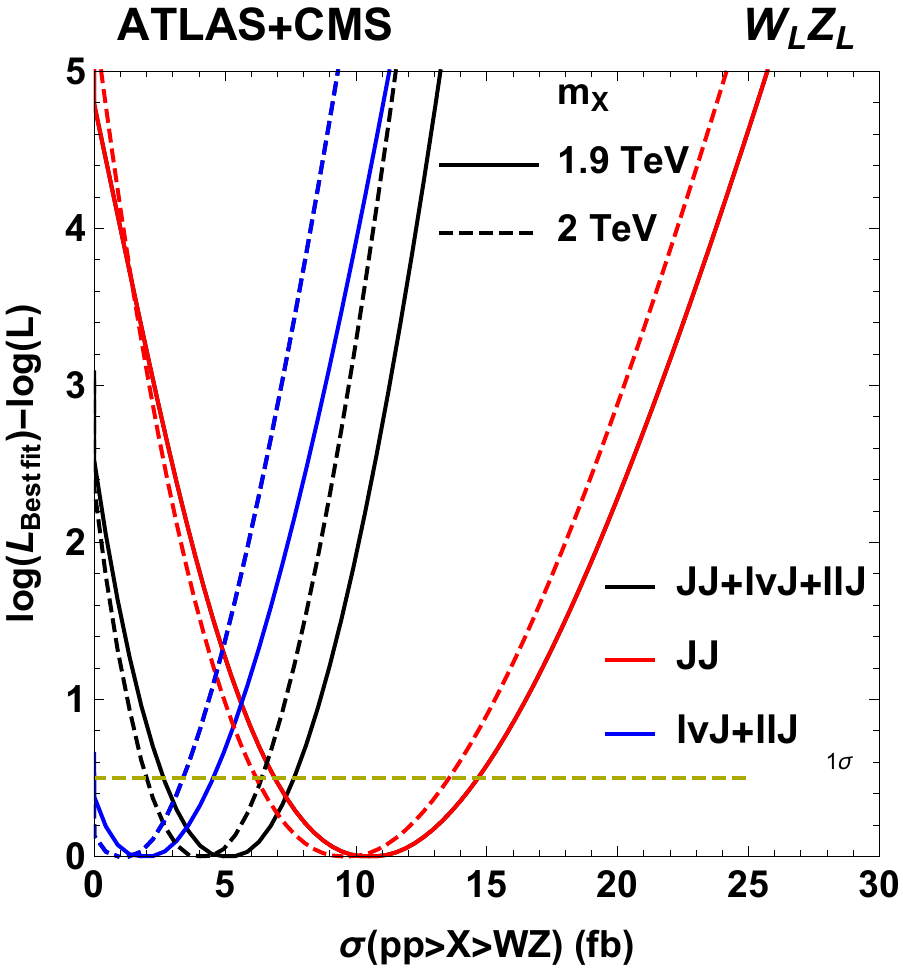}
\includegraphics[width=0.32\textwidth, angle =0 ]{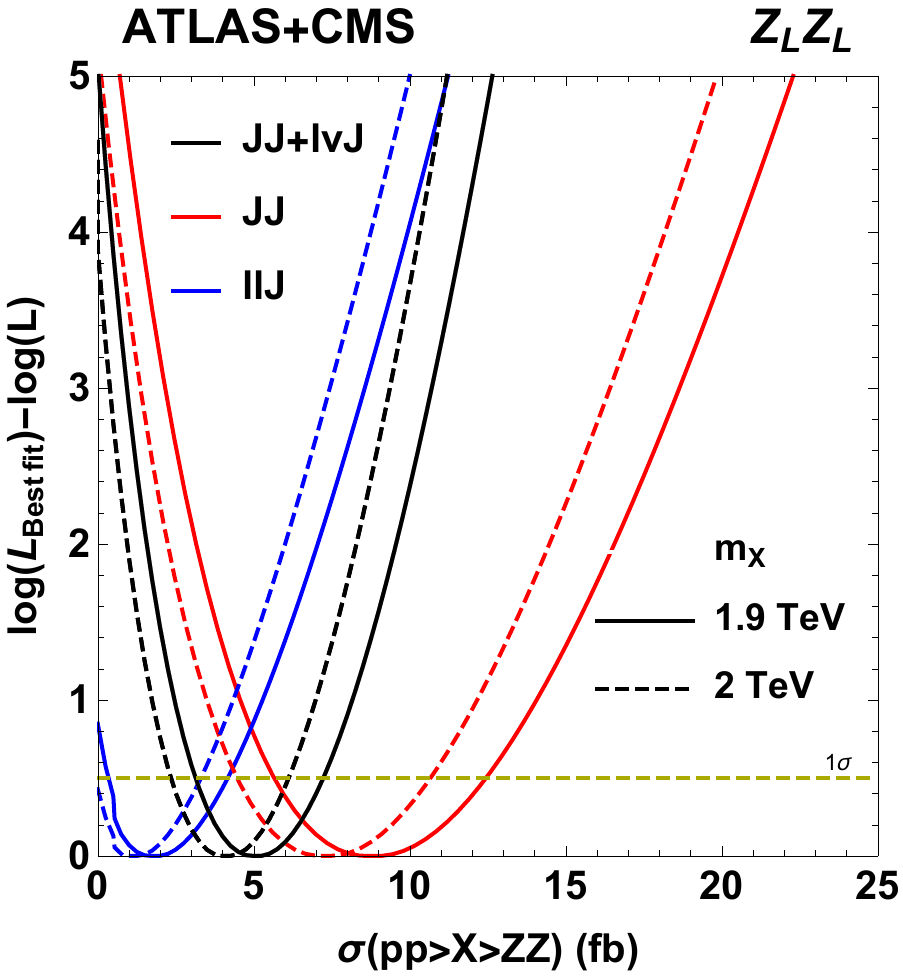}
\caption{\small 
Combination of all ATLAS and CMS resonance searches: Scans of the profile 
  likelihood as a function of  the production cross section for a
  $\MX$ = 2.0 (1.9) TeV signal shown with continuous (dashed) lines
  in the hadronic (red) and semileptonic (blue) channels
  and their combination (black) for \WLWL (left),  \WLZL (middle) and \ZLZL
  (right) selections and signal hypotheses.
\label{COMBO_llJ_JJ_ML}}
\end{center}\end{figure*}

The exclusion limits on production cross section, likelihood-ratio $p$-values, and
best-fit cross sections as a function of a hypothetical resonance mass are
summarised in Fig. \ref{COMBO_VV_llJ_JJ}. Scans of the profile likelihood
as a function of the exotic production cross section for $m_X$ = 1.9 and
2.0 TeV (mass values of largest excesses for the benchmark models considered) 
are given in Fig. \ref{COMBO_llJ_JJ_ML}.
%From one hand the combination is performed under the \PGbulk hypothesis individually for  $\ZLZL$ and $\WLWL$ channels. From the other hand $\WLZL$ hypothesis is considered..
The sensitivity of the search in the \GtoZZ signal hypothesis is dominated
by the semileptonic analyses below 
1.9 - 2.0 TeV and the fully hadronic searches at higher mass ranges.
The largest deviation is observed at $m_X$ = 1.9 TeV, driven by the ATLAS excess
in the \VVJJ channel. The overall significance remains above 3$\sigma$.   
%It is even close to 4 $\sigma$ if one do not apply fudge factors to match our limits to public ones. This may be explained by the large fudge factor applied on ATLAS \JJ limits and that reduces the impact of this channel.
The preferred cross section for a hypothetical \GtoZZ signal as calculated
in the  $\LLJ$ channel is $\approx 2$ fb and increases to 
$\approx 9$ fb for the \JJ channel. When combined, the estimated
cross section is 5 fb. The combination of the two channels reduces the
exotic cross section favoured by the \JJ results, and alleviates the
potential disagreement between different 
channels, without reducing the overall significance of the excess. In other
words, the combination of the two channels leads to a more coherent picture
of the results by the two experiments. 
This is also evident from the profile likelihood scans shown in  Fig.~\ref{COMBO_VV_llJ_JJ}:
given the uncertainty on the best-fit exotic production cross section,
and contrary to what one might expect by considering the individual
exclusion limits, the results obtained in 
different final states are not in tension with each other. In addition, the
combination pushes the excess to mass values below 2
TeV.

The picture is quite different in the \GtoWW signal interpretation. The
lack of a significant excess in the \LNUJ channels is strong enough to reduce the
significance of the \JJ excess below the 1$\sigma$ threshold. The
combination of the ATLAS and CMS results disfavours the hypothesis of a
resonance decaying exclusively to \PWW (an interpretation which in any
case would be difficult to justify phenomenologically).

Finally, the interpretation of the results in the context of a \PWp signal
hypothesis lies between the \GtoZZ and
\GtoWW  scenarios: the \LNUJ analyses are more sensitive than the fully
hadronic ones, but their contribution is not as dominant as in the
\GtoWW case.  Nevertheless, the excess survives above the 3$\sigma$ threshold,
thanks to the presence of a moderate excess in the \LLJ search around the
same mass region. Overall,
the estimated cross section of a hypothetical exotic signal is strongly
reduced: the best-fit value changes
from $\approx 10$~fb (when using the \JJ channel results only) to
$\approx 5$~fb  (when combining the \JJ, \LNUJ and \LLJ
channels). At this smaller cross section value, the outcome of the 
searches in the different channels is quite coherent, as shown in the
profile likelihood scans depicted in
Fig.~\ref{COMBO_VV_llJ_JJ}. The mitigating effect of the \LLJ result is
evident if one compares the \LNUJ-and-\LLJ combined likelihood scan for the \PWp
combination to the likelihood scans in the semileptonic searches. The
$\WLZL$ curve is much more similar to the $\ZLZL$ curve in the \LLJ channel
than to the $\WLWL$ curve in the \LNUJ channel.

In conclusion, a resonance with a production cross section  of $\sim$5~fb and
mass between 1.9 and 2.0 TeV is the scenario most consistent with the
experimental results out of all benchmark models considered in this study,
as long as it does not decay exclusively to a $\WLWL$ final state.

\begin{figure*}[htb]\begin{center}
\includegraphics[width=0.3\textwidth, angle =0 ]{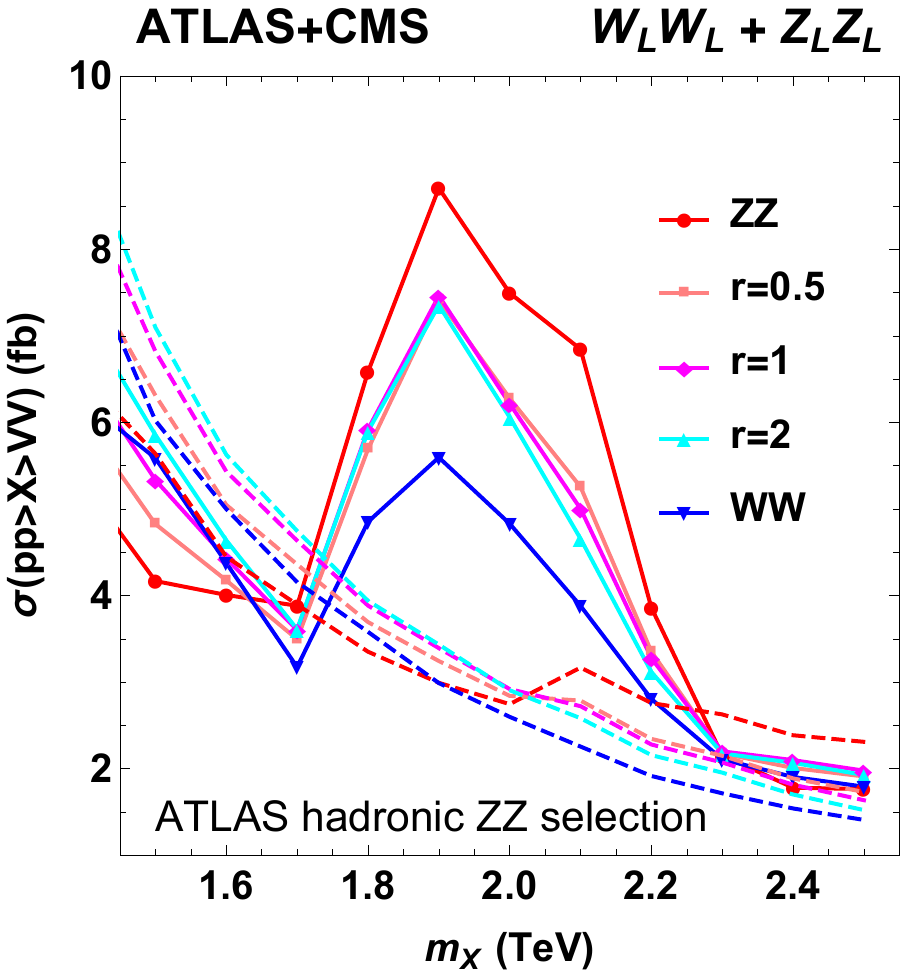}
\includegraphics[width=0.32\textwidth, angle =0 ]{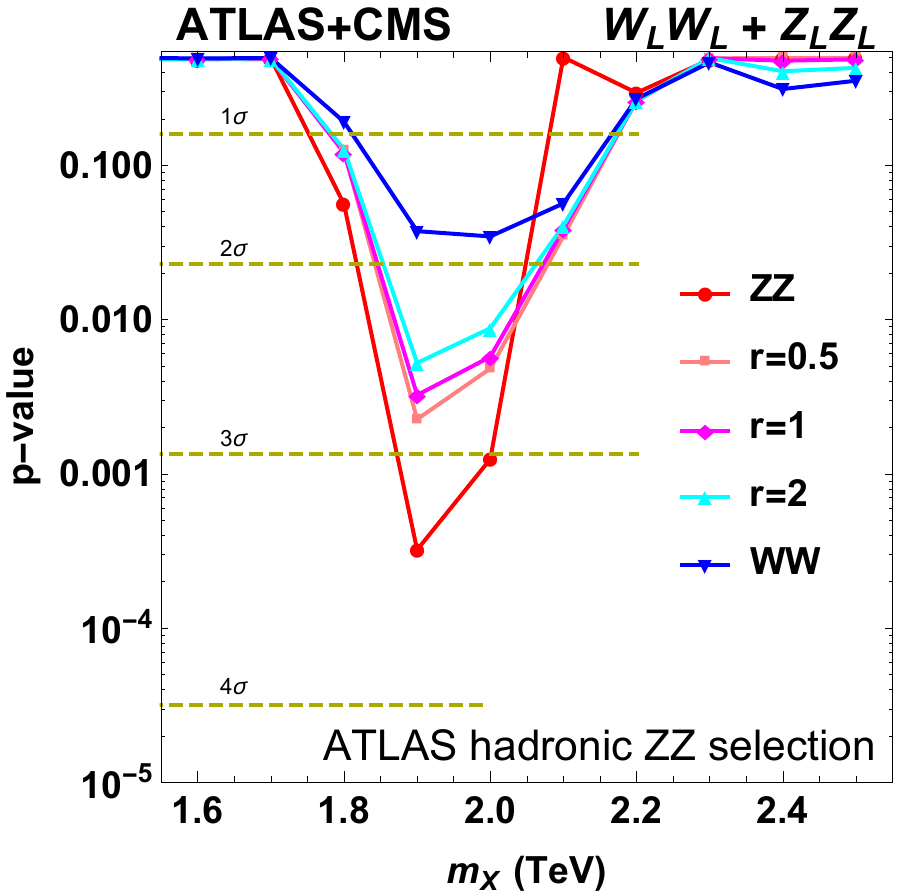}
\includegraphics[width=0.3\textwidth, angle =0 ]{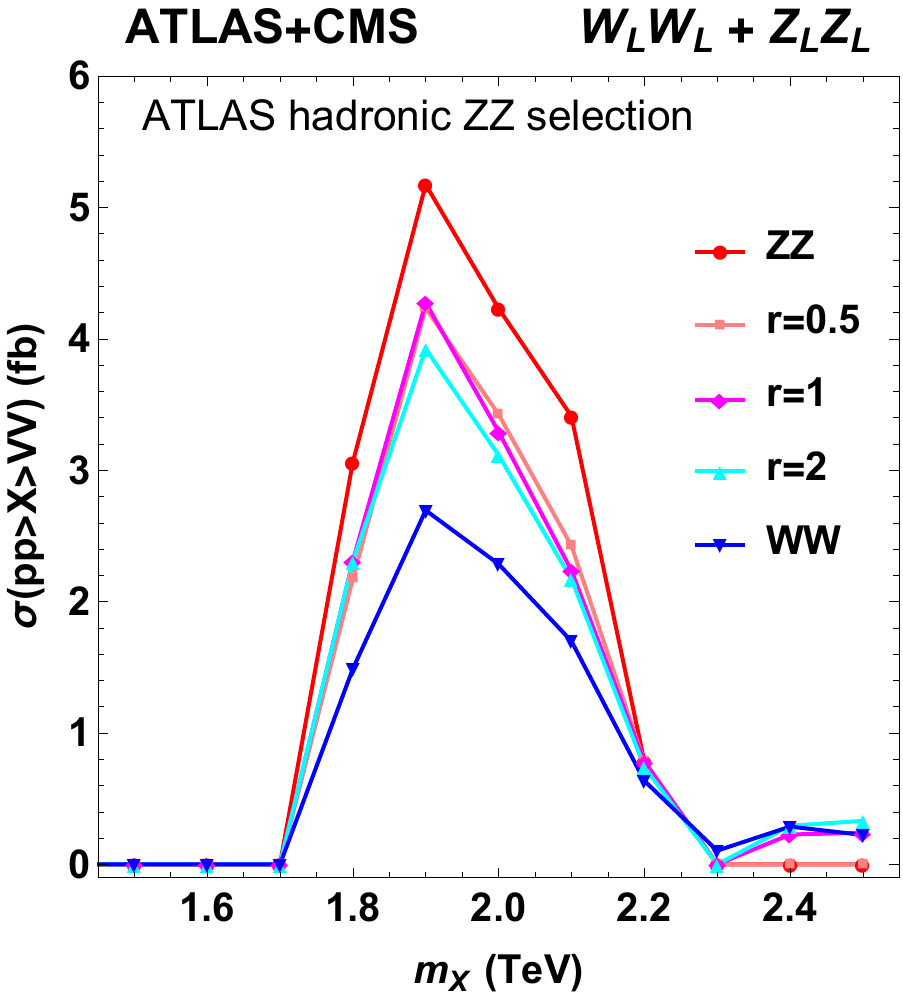}
\caption{\small 
Combination of all ATLAS and CMS resonance searches, 
and dependence of results obtained in this study on
  the $r \equiv {\mathcal{B}(X \to \PW\PW)}/{\mathcal{B} (X \to \PZ\PZ)}$
  parameter for a neutral bulk RS-like spin--2
  particle hypothesis, and as a function of the resonance mass \MX. {\bf Left:}
  expected (dashed lines) and  
  observed (continuous lines) exclusion limits on exotic production cross
  section. {\bf Middle:} likelihood-ratio $p$-values. {\bf
    Right:} best fitted exotic production cross section.
\label{MODEL_INDEPENDANT_COMBO}
}
\end{center}\end{figure*}

An example of the model independent combination of the $\ZLZL$ and $\WLWL$
channels is shown in Fig. \ref{MODEL_INDEPENDANT_COMBO}. In this case, one considers a
resonance that can decay to both $\WLWL$ and $\ZLZL$, with the relative
branching fraction determined by the $r$ parameter introduced in
Eq.(\ref{eq:r}). For $r\to 0$ one recovers the \GtoZZ case, while for $r\to
\infty$ one recovers the \GtoWW limit. It should be noted
that for this combination we use a common mass window for the ATLAS
analyses, namely the one that corresponds to the \PZZ search, giving the
best overall sensitivity (see Sec. \ref{sec:JJ}). Therefore, the results
obtained here on the \WLWL exclusion limits and $p$-values are somewhat
different than the ones presented in Fig.  \ref{COMBO_VV_llJ_JJ}.

The results obtained for generic values of $r$ are similar to the
$\ZLZL$ case, \ie they point to an 
overall excess. The size of the excess is reduced to 2$\sigma$, with a
best-fit exotic production cross section around
4~fb. Particularly interesting is the $r=2$ case,
corresponding to a resonance with universal couplings to the
pseudo-Goldstone bosons. In this case, despite the fact that
${\cal B}(X \to \WLWL)=2{\cal B}(X \to \ZLZL)$, the combined deviation
is found to have a $\approx 
2.4\sigma$ significance for a cross section of $\approx 4$~fb. 

It may be interesting to comment here on how the statistical methods that
we have employed in this study 
%for the combination of the search results across channels and experiments 
compare with the simplified practices used
by the theoretical community. A standard technique employed in many theoretical
papers is to assume
Gaussian likelihoods for the cross section of hypothetical signals, taking
as central value the difference between the observed and expected limits,
and as standard deviation the expected (95\% C.L.) limit divided by 1.96. Then, one can
use the cross sections and uncertainties as estimated 
in the various channels, and calculate a weighted average. This method should, in
principle, work well for cases in which the fitted cross section comes with
a relatively small uncertainty (which is, typically, not 
the case in most searches) and the systematic uncertainties can be considered as
uncorrelated among channels and experiments (which may, or may not be the
case). As an example, we note that the simplified combination of the search
results in the \GtoWW interpretation yields a best-fit cross section of 2.5
$\pm$ 1.6 fb (2.5 $\pm$ 1.4 fb) at $\MX$ = 1.9 TeV (2.0 TeV), to be
compared with our result of $0.75^{+1.67}_{-0.75}$ fb ($1.1^{+1.4}_{-1.1}$
fb). Similarly, the simplified combination in the \GtoZZ interpretation
yields a best-fit cross section of 4.7 $\pm$ 1.9 fb (4.4 $\pm$ 1.8 fb) at
$\MX$ = 1.9 TeV (2.0 TeV), to be compared with our result of
$5.2^{+2.1}_{-1.6}$ fb ($4.2^{+1.9}_{-1.2}$ fb). 

While more data is needed to clarify the situation, the results from the
analysis of the diboson searches is unquestionably one of the
most interesting outcomes of the ATLAS and CMS exotic programmes during
the first LHC run. The situation 
is even more intriguing if one adds to the picture the $\approx 2\sigma$
excess at 1.8-1.9~TeV observed by CMS in a $\PW H$ resonance search. The
\PWp{} results shown in Fig.~\ref{COMBO_VV_llJ_JJ} emerge as the most
promising hint in the quest for a new heavy resonance in the ATLAS and CMS data, as
already pointed out in Ref.~\cite{Brehmer:2015cia}. \\

%\clearpage

\section{Conclusions}
\label{sec:conclusion}
We have performed a combination of the ATLAS and CMS searches for a heavy
resonance decaying to a diboson final state, derived from the public
information available for the six relevant
analyses~\cite{ATLASVV,CMSVV,CMSZVWV,ATLASWV,ATLASZV}. We have developed
a methodology for the combination procedure, which begins with the work to
emulate the public results by ATLAS and CMS for each individual
analysis. This process is adjusted when necessary with correction factors to
account for unknown uncertainties, and has been validated by reproducing
the official results by the two experiments. We have presented combinations
of the ATLAS and CMS 
searches for individual decay modes in various simplified models. At each step, the 95\% CL limits, the likelihood ratio $p$-values, the
profile likelihood scans, and the maximum likelihood fits of the 
production cross section as function of the resonance mass $\MX$ are provided.

The combination is obtained in three scenarios: $\WptoWZ$, $\GtoWW$,
and $\GtoZZ$. We also obtain the full combination results for a \PGbulk resonance
with generic $\WLWL$ and $\ZLZL$ branching fractions.  Out of all benchmark
models considered, the combination
favours the hypothesis of a resonance with mass 1.9-2.0 TeV and a
production cross section $\approx 5$~fb, as long as the
resonance does not decay exclusively to $\WLWL$ final states. Depending
on the details of the resonance model, a signal significance between
2.4 and 3.4$\sigma$ is obtained for notable benchmark scenarios (see
Table~\ref{table:CombinationSummary}). In particular, the possibility of a \PWp{} resonance,
suggested by other searches in different final states, is corroborated
by the diboson searches, with a significance of $\approx3\sigma$ for a resonance
mass of 1.9 TeV.

%%%%%%%%%%%%%% BEG Vic (run-dependant samples)
\begin{table}[htb]
\centering
\topcaption{\small Summary of results obtained in this study: significance,
  $p$-values and best-fit cross sections for 
  different model interpretations at $\MX = 1.9$ and $\MX = 2.0$ TeV, \ie the
  mass values where the largest excesses have been observed for different
  models. Our main results contain corrections that have been introduced
  to account for unknown uncertainties in the official results. (Additional
  results calculated without these correction factors are given inside the parentheses.)
\label{table:CombinationSummary}} 
%\resizebox{\textwidth}{!}
\footnotesize{
%\begin{tabular}{|c|c|c|c|c|}
%\hline
%Signal hypothesis    & $\MX$ (TeV) & Significance    & $p$-value
%& Best-fit cross section (fb) \\\hline
%    & 1.9         & 2.5 (3.1)   & 6.5 (1.0) $\times 10^{-3}$  & $5.3^{+2.3}_{-2.0}$\,\,    ($5.5^{+2.0}_{-1.6}$) \\[1mm] %\cline{3-6}
% \raisebox{1.5ex}[0cm][0cm]{\WptoWZ}               & 2.0           & 2.5 (3.2)   & 7.0 (0.8) $\times 10^{-3}$  & $4.3^{+2.1}_{-1.5}$\,\,    ($4.7^{+1.8}_{-1.3}$) \\[1mm]\hline
%     & 1.9         & 0.49 (0.83) & 0.30 (0.20)                 & $0.75^{+1.67}_{-0.75}$\,\, ($1.4^{+1.7}_{-1.4}$) \\[1mm] %\cline{3-6}
%\raisebox{1.5ex}[0cm][0cm]{\GtoWW}                & 2.0           & 0.88 (1.33) & 0.20 (0.092)                & $1.1^{+1.4}_{-1.1}$\,\,    ($1.8^{+1.8}_{-1.4}$) \\[1mm]\hline
%       & 1.9         & 3.4 (3.8)   & 3.2 (0.65) $\times 10^{-4}$ & $5.2^{+2.1}_{-1.6}$ \,\,    ($4.7^{+1.8}_{-1.2}$)  \\[1mm] %\cline{3-6}
% \raisebox{1.5ex}[0cm][0cm]{\GtoZZ}               & 2.0           & 3.0 (3.5)   & 1.2 (0.24) $\times 10^{-3}$ & $4.2^{+1.9}_{-1.2}$        ($3.9^{+1.6}_{-1.0}$) \\[1mm]\hline \hline
%& 1.9         & 2.6 (3.4)   & 5.2 (0.40) $\times 10^{-3}$ & $3.9^{+2.4}_{-1.5}$        ($4.9^{+2.0}_{-1.7}$) \\[1mm] %\cline{3-6}
%\raisebox{1.5ex}[0cm][0cm]{\PGbulk~(r=2)}                & 2.0           & 2.4 (3.1)   & 8.8 (0.89) $\times 10^{-3}$ & $3.1^{+1.8}_{-1.3}$        ($3.9^{+1.6}_{-1.4}$) \\[1mm]\hline
%\end{tabular}
%
\begin{tabular}{cccccc}
\toprule
Signal hypothesis    & $\MX$ (TeV) & Significance    & $p$-value & \multicolumn{2}{}{}{Best-fit cross section (fb)} \\
\midrule
    	& 1.9         & 2.5 (3.1)   & 6.5 (1.0) $\times 10^{-3}$  & $5.3^{+2.3}_{-2.0}$ &  ($5.5^{+2.0}_{-1.6}$) \\[1mm] %\cline{3-6}
\raisebox{1.5ex}[0cm][0cm]{\WptoWZ}				& 2.0           & 2.5 (3.2)   & 7.0 (0.8) $\times 10^{-3}$  & $4.3^{+2.1}_{-1.5}$ &  ($4.7^{+1.8}_{-1.3}$) \\[1.5mm] 
     	& 1.9         & 0.49 (0.83) & 0.30 (0.20)                 & $0.75^{+1.67}_{-0.75}$ & ($1.4^{+1.7}_{-1.4}$) \\[1mm] %\cline{3-6}
\raisebox{1.5ex}[0cm][0cm]{\GtoWW}			& 2.0           &
0.88 (1.33) & 0.20 (0.092)                & $1.1^{+1.4}_{-1.1}$ &    ($1.8^{+1.8}_{-1.4}$) \\[1.5mm] 
     	& 1.9         & 3.4 (3.8)   & 3.2 (0.65) $\times 10^{-4}$ & $5.2^{+2.1}_{-1.6}$ &  ($4.7^{+1.8}_{-1.2}$) \\[1mm] %\cline{3-6}
\raisebox{1.5ex}[0cm][0cm]{\GtoZZ}				& 2.0           & 3.0 (3.5)   & 1.2 (0.24) $\times 10^{-3}$ & $4.2^{+1.9}_{-1.2}$  &  ($3.9^{+1.6}_{-1.0}$) \\\midrule
	& 1.9         & 2.6 (3.4)   & 5.2 (0.40) $\times 10^{-3}$ & $3.9^{+2.4}_{-1.5}$ & ($4.9^{+2.0}_{-1.7}$) \\[1mm] %\cline{3-6}
\raisebox{1.5ex}[0cm][0cm]{\PGbulk~(r=2)}                & 2.0           & 2.4 (3.1)   & 8.8 (0.89) $\times 10^{-3}$ & $3.1^{+1.8}_{-1.3}$ &  ($3.9^{+1.6}_{-1.4}$) \\
\bottomrule
\end{tabular}
}
\end{table}
%%%%%%%%%%%%%%%%%%%%%
%\clearpage

\vspace{2 mm}
\noindent {\bf Note added in v2 of the paper}\\[2 mm]
\noindent While preparing this
manuscript for submission, ATLAS and CMS presented preliminary results in
searches for diboson resonances with the first $\sqrt{s}$ = 13 TeV $pp$
collision data. They include results in the  $W(\ell \nu)\,V(q \bar{q})$ \cite{ATLAS-CONF-2015-075},
$Z(\ell^+\ell^-)\,V(q\bar{q})$ \cite{ATLAS-CONF-2015-071},
$V(q\bar{q})\,V(q\bar{q})$ \cite{ATLAS-CONF-2015-073} and $Z(\nu
\bar{\nu})\,V(q \bar{q})$ \cite{ATLAS-CONF-2015-068} channels by ATLAS, and
the $W(\ell \nu)\,V(q 
\bar{q})$ and $V(q\bar{q})\,V(q\bar{q})$ channels by CMS \cite{CMS:2015002}. No significant
excess above the SM expectations is observed, however the experimental
sensitivity is, 
in most cases, not comparable with the one from Run-1 yet. The notable
exception is the newly added $Z(\nu 
\bar{\nu})\,V(q \bar{q})$ channel. The most stringent exclusion limits in
the preliminary analysis of Run-2 data are obtained in the following channels:
\begin{itemize}
\item (HVT) \WptoWZ: 25 fb (20 fb) for $m_X$ = 1.9 TeV (2.0 TeV)
 in the $W(q \bar{q})\, Z(\nu \bar{\nu})$ channel by ATLAS, and
  the combination of the two channels considered by CMS.
\item \GtoWW: 15 fb (12 fb) for $m_X$ = 1.9 TeV (2.0 TeV)
 in the $W(\ell \nu)\,W(q \bar{q})$ channel by ATLAS.
\item \GtoZZ: 21 fb (15 fb) for $m_X$ = 1.9 TeV (2.0 TeV)
 in the $Z(\nu \bar{\nu})\,Z(q \bar{q})$ channel by ATLAS.
\end{itemize}
In assessing the compatibility of the Run-2 exclusion limits with the
results obtained in this study (summarised in
Table~\ref{table:CombinationSummary}) we use parton luminosity 
ratio values of 13 (15) for $m_X$ = 1.9 TeV  (2.0 TeV) for $gg$
production (\GtoWW  and \GtoZZ channels) and 8 (8.5) for $m_X$ = 1.9 TeV
(2.0 TeV) for $q\bar{q}$ production (\WptoWZ channels) \cite{Stirling} to
  calculate the 
increase in the exotic signal production cross section from 8 to 13 TeV.
We observe that the absence of a significant deviation in the
Run-2 data
\begin{itemize}
\item creates a $\sim 2-3 \sigma$ tension with the best-fit cross section
  derived in this paper in the \GtoZZ channel, 
\item is consistent (within $1 \sigma$) with the (consistent-with-zero)
  result we obtain in the \GtoWW channel, and
\item is also consistent (within $1 \sigma$) with the best-fit cross
  section that we have derived in the \WptoWZ channel.
\end{itemize}
We, therefore, conclude that the preliminary analysis of the Run-2 data by
ATLAS and CMS does not rule out the small deviation reported in the \WptoWZ
channel of the Run-1 diboson searches. It is widely expected that a clear
picture will emerge with the analysis of the larger 13 TeV datasets.

\acknowledgments

We would like to thank our colleagues at the ATLAS and CMS collaborations
for their exemplary work and publication of a large number of papers on
exotic searches. We thank Andreas Hinzmann for his precious help in the
implementation of the CMS search in the $X \to \VVJJ$ channel. We also
thank Goran Senjanovi\'{c} and Andrea Wulzer for fruitful discussions and
valuable suggestions. A.O. thanks the CERN theory group for their hospitality.

This material is based upon work partially supported by the Cooperation Agreement
(SPRINT Program) between the S\~ao Paulo Research Foundation (FAPESP) and the
University of Edinburgh, under Grant No. 2014/50208-0. A.O. is
supported by the MIURFIRB RBFR12H1MW grant. The work of F. D. and C.L. is
supported by the Science and Technology Facilities Council (STFC) in the UK.

\clearpage
\appendix
\clearpage
\section{Comparison of different approaches to emulate ATLAS \VVJJ analysis}
\label{app-Avvjj}

\par
\begin{figure*}[htb]
\centering
\includegraphics[width=0.32\textwidth, angle =0 ]{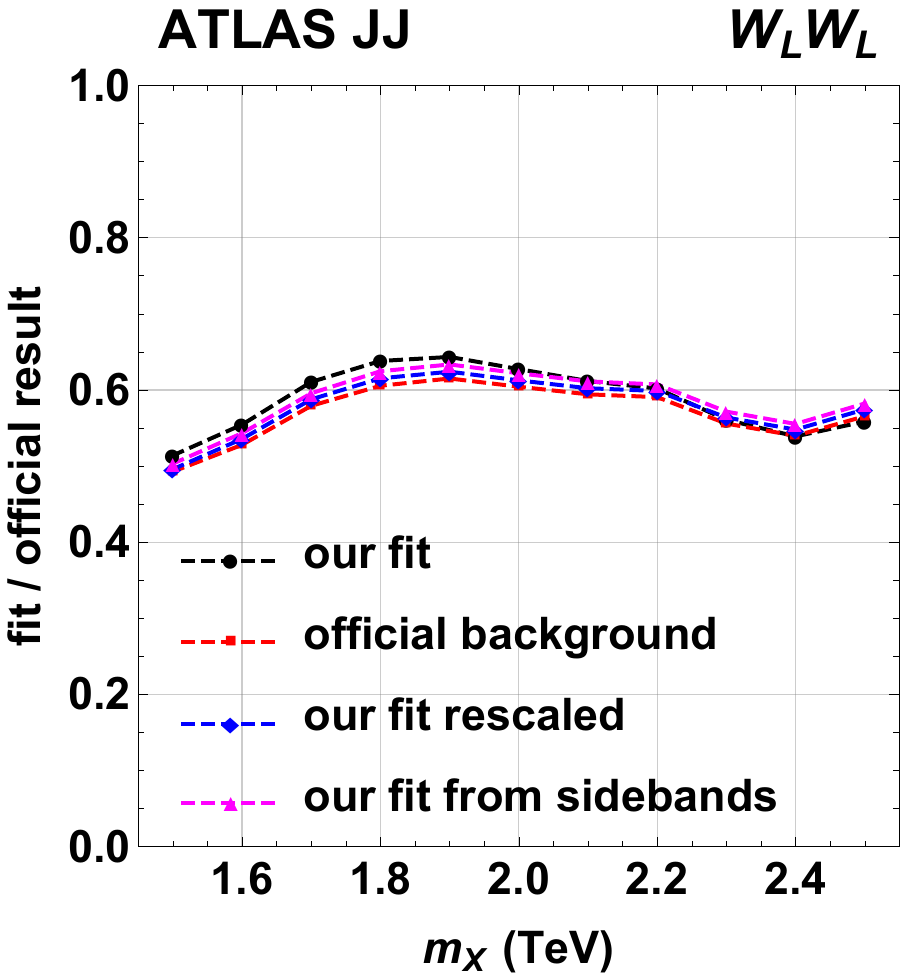}
\includegraphics[width=0.32\textwidth, angle =0 ]{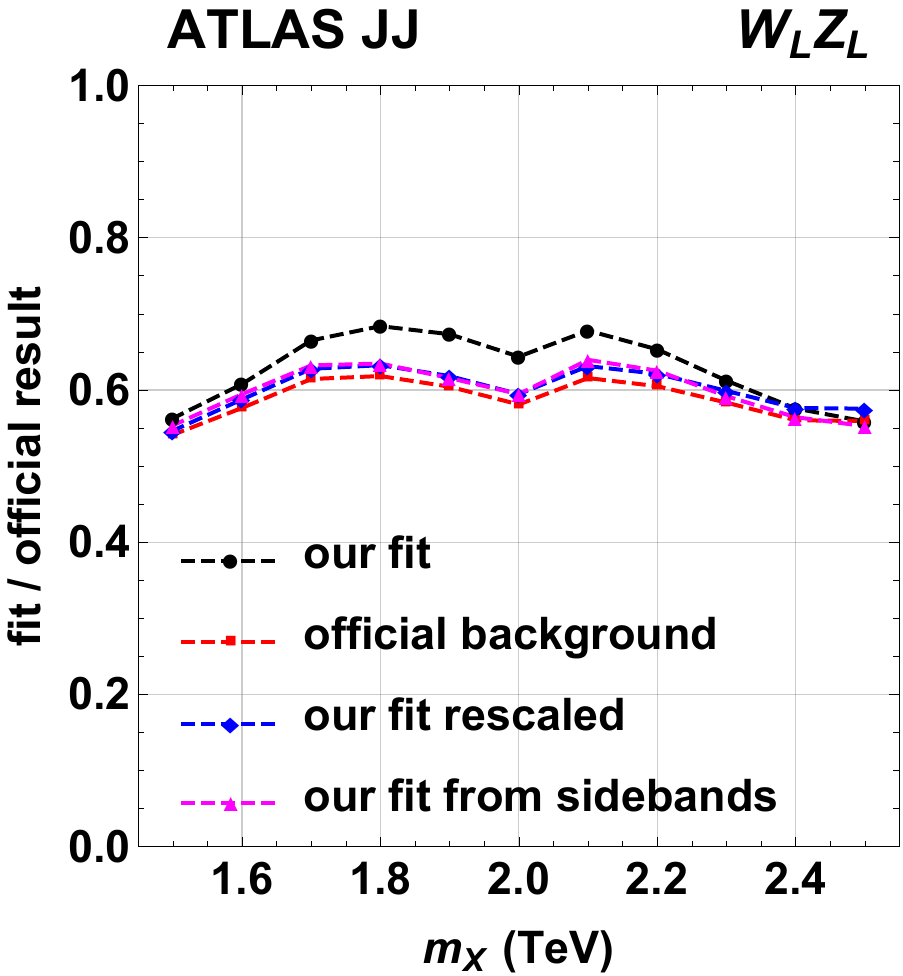}
\includegraphics[width=0.32\textwidth, angle =0 ]{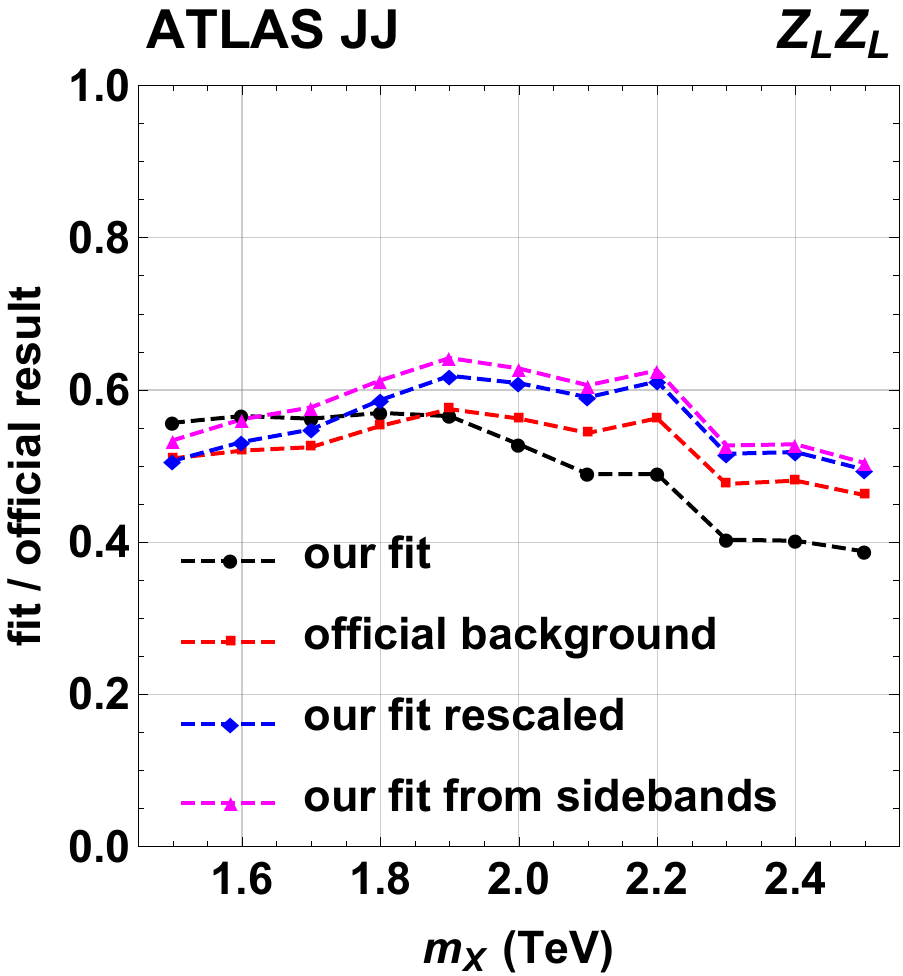}
\caption{Emulation of ATLAS \VVJJ search and comparison of the alternative approaches for the background prediction considered: Fudge factors as a function of the resonance mass $\MX$, determined via the ratio of the expected limits obtained with different background estimation techniques (black: ``pure fitting'', red: ``nominal background'', 
blue: ``rescaling'', magenta: ``sidebands'') over those in the official ATLAS result 
%For comparison, we also show the nominal limits obtained by the ATLAS collaboration in \textbf{grey}, and our final results with the final fudge applied in \text{black}. 
for the $\WLWL$ (left), $\WLZL$ (middle) and $\ZLZL$ channels (right). See text for details.
}
\label{fig:comparisonATLASVVJJ}
\end{figure*}

 The expected limits obtained in the emulation of the ATLAS \VVJJ channel show a 40\% discrepancy with respect to the official results (see Sec. \ref{sec:JJ}). This is the largest discrepancy observed among all the channels considered in this study. We have considered alternative approaches in our strategy and carried out several cross-checks, which are summarised here:
\begin{itemize}
        \item \textbf{Nominal background}: ATLAS publishes a background description with a total background uncertainty. This information can be used directly as an input to our analysis. The disadvantage of this approach is that it combines all systematic uncertainties into a single contribution, implying a correlation model that may not reflect the accuracy of the fit performed by the ATLAS collaboration.
        \item \textbf{Pure fitting}: We have repeated the fit on the data distribution provided by the ATLAS collaboration. The fitting procedure naturally yields a covariance matrix for the shape parameters, which allows to adopt a more realistic correlation model.
        \item \textbf{Rescaling}: This is a mixed approach in which the fit is  performed over the data distribution to obtain the covariance matrix of the fitting function parameters, but the resulting background prediction and the corresponding uncertainties are then rescaled to match those provided by ATLAS. In this approach, the official ATLAS background prediction is used and our fit is only used to model the uncertainties and their correlations. 
        \item \textbf{Sidebands}: In this case we repeat the fit procedure described above, after excluding the region of the largest deviation (1700--2300~GeV) from the fit range, in order to exclude the possibility that it could bias the fit.
\end{itemize}
Fig.~\ref{fig:comparisonATLASVVJJ} shows the ratio of the observed exclusion limits to the ones from the official ATLAS results for the different approaches summarised above. In all cases the differences are very small, which suggests that the explanation for the observed discrepancy should be attributed to a factor other than the background determination procedure. The discrepancy is absorbed in the fudge factor which, when tuned to deliver the official expected exclusion limits, remarkably removes (to a large extent) the differences in the observed limits. One should note that the decision to employ these correction factors in our analysis (for this and other channels) does not change qualitatively the conclusions of this study. This can be seen, for example, in the middle plot of Fig. \ref{COMBO_VV_llJ_JJ}, where it is shown that the two different approaches yield significances that differer typically by 0.5$\sigma$.

%\clearpage
\section{Narrow width approximation}
\label{sec:narrow}

The CMS collaboration assumes a signal with negligible width, whereas the ATLAS collaboration simulates signal distributions with a model-dependent width of $\approx 7\%$ of the resonance mass (see Table 1 of Ref.~\cite{ATLASVV}). In this appendix we estimate the effect of this difference in the final exclusion limits and provide a recipe for obtaining the ATLAS results in the narrow-width approximation.

\begin{figure*}[htb]\begin{center}
\includegraphics[width=0.46\textwidth, angle =0 ]{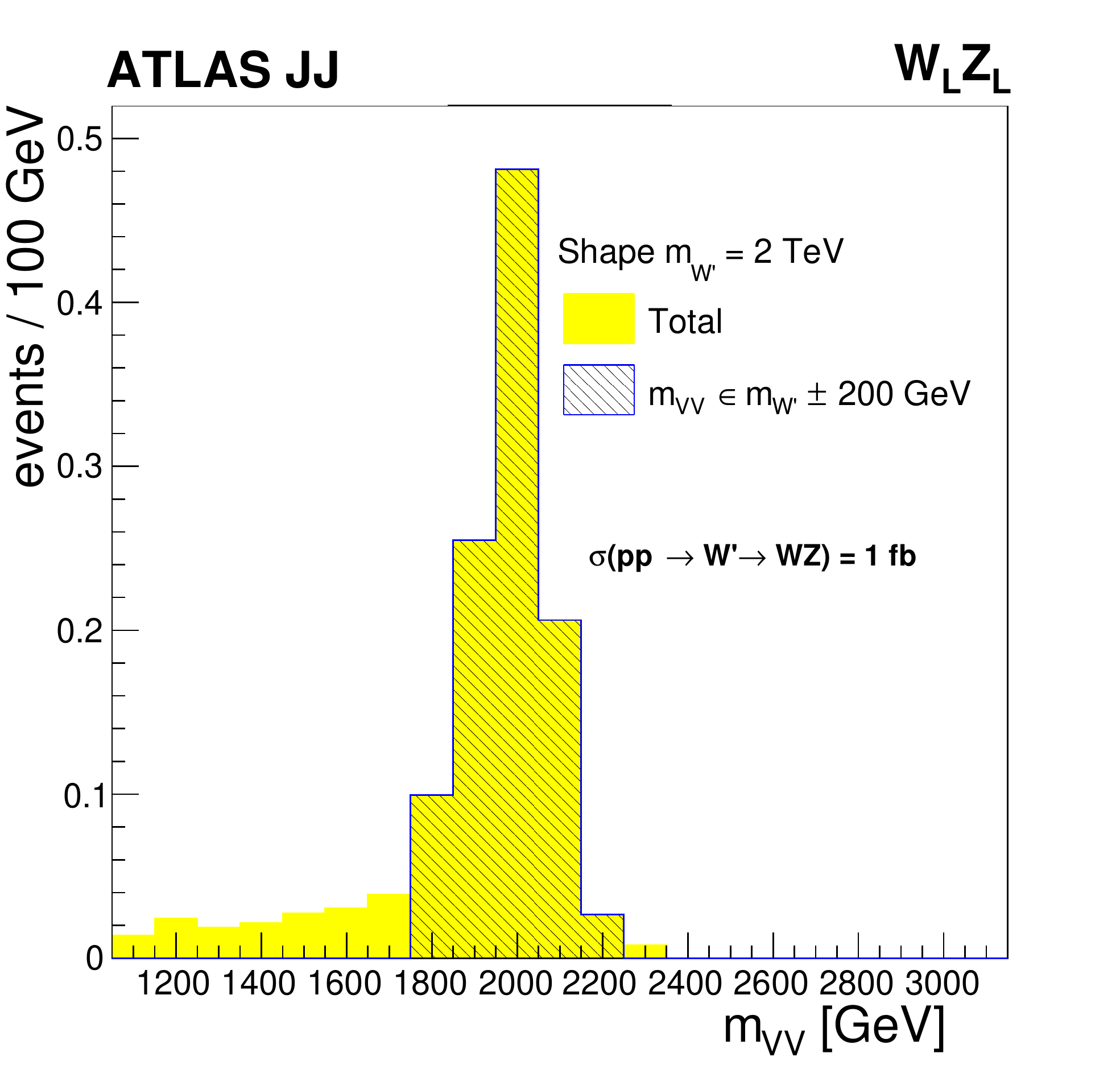}
\includegraphics[width=0.43\textwidth, angle =0 ]{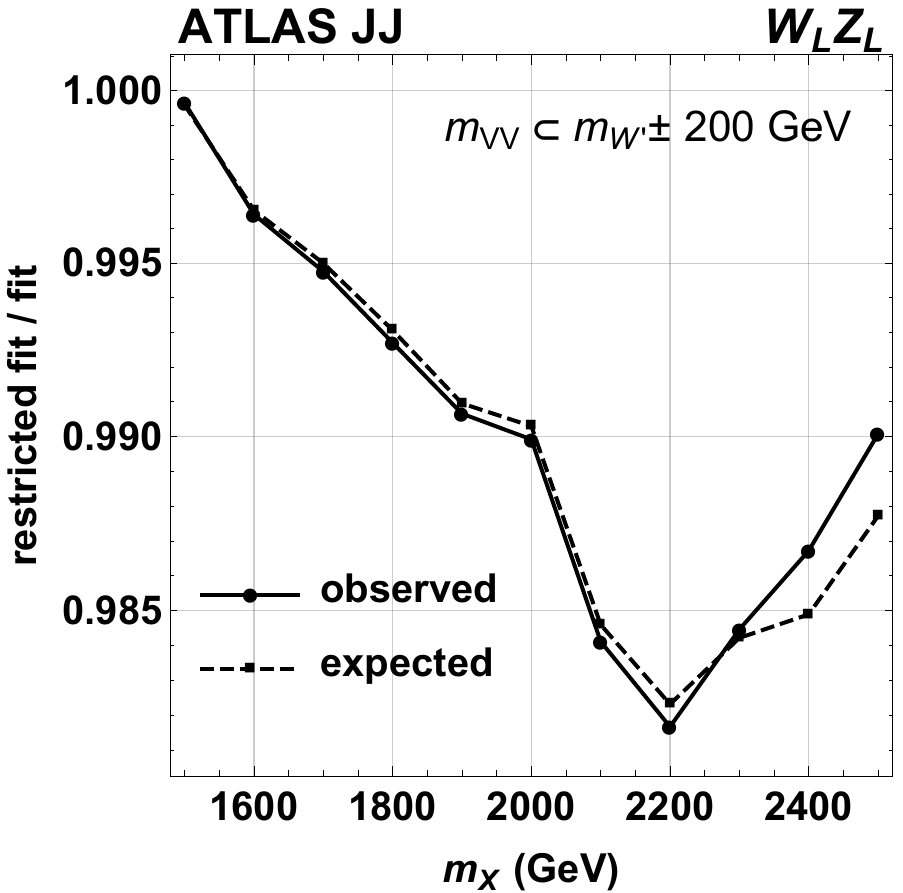}
\caption{\small Narrow-width approximation. {\bf Left:} Signal distribution in the diboson invariant mass for a 2 TeV \PWp signal. The hatched $\pm$200 GeV region around the signal represents the narrow-width approximation. {\bf Right:} Ratio of the expected (dashed lines) and observed (continuous lines) exclusion  limits when constraining the signal width to 10\% of the resonance mass over those obtained with the default shape.  \label{fig:check} }
\end{center}\end{figure*}

The large width hypothesis used by the ATLAS collaboration impacts the limits through the modification of the signal shapes. In the $\JJ$ channel it widens the core for the signal distribution and creates a large left tail due to the interplay between proton PDFs \cite{Harris:2011bh} and the natural width of the resonance, as one can see in the left plot of Fig. \ref{fig:check}. In practice, for a given total cross section we have events \textit{leaking} outside the $\pm 10\%$ window around $\MX$. This value corresponds typically to the experimental resolution of this channel. The amount of this leakage, $f_l$ is provided in Ref.~\cite{ATLASVV} and corresponds typically to 15\% in the region under study in this paper. 

We expect the events in the left tail to have no significant impact on the exclusion limits. A test was performed by truncating the signal to $\MX \pm 200$ GeV and repeating the $\JJ$ limit-setting procedure for the \PWp hypothesis. As one can see in the right plot of Fig. \ref{fig:check}, the difference in the expected exclusion limits does not exceed 2\%. 
To map the ATLAS limits into a narrow width hypothesis we make the following approximation: The main difference between the wide and narrow resonances is the presence of leaking events in the right tail or under the peak. Consequently, by multiplying the signal efficiency of ATLAS by $1/f_l$ we recover most of the properties of the narrow signal. In conclusion, we approximate the narrow signal hypothesis for ATLAS analyses by scaling the fully hadronic and semi-leptonic signals by a factor of 1.1 (\ie by increasing the signal yield by 10\%).

\clearpage
%\section{Per experiment combinations}
%\input{combo_collaborations}

\printbibliography
\end{document}